\documentclass[12pt]{article}

\usepackage{graphicx,color}
\usepackage{amsmath,amssymb,amsfonts}
\usepackage{cite}
\usepackage{bm}
\usepackage{enumerate}
\usepackage{tensor}
\usepackage{mathtools}
\usepackage[colorlinks]{hyperref}
\usepackage[dvipsnames]{xcolor}
\usepackage{subcaption}
\usepackage[capitalise]{cleveref}
\usepackage{dsfont}
\hypersetup{pageanchor=false,linkcolor=NavyBlue,citecolor=NavyBlue,urlcolor=RoyalPurple}
\usepackage{bbm}
\usepackage{setspace}

\makeatletter
\g@addto@macro\bfseries{\boldmath}
\makeatother

\usepackage{cleveref}
\crefname{table}{Table}{Tables}
\crefname{equation}{Eq.}{Eqs.}
\crefname{appendix}{App.}{Apps.}
\crefname{section}{Sec.}{Secs.}
\crefname{figure}{Fig.}{Figs.}

\newcommand{\changelocaltocdepth}[1]{%
  \addtocontents{toc}{\protect\setcounter{tocdepth}{#1}}%
  \setcounter{tocdepth}{#1}%
}

\newcommand{\Cdot}{\!\cdot\!}

\newcommand{\nn}{\nonumber}
\newcommand{\spac}{{\hspace{0.3mm}}}

\newcommand{\np}{\bar{n}}
\newcommand{\nm}{n}
\newcommand{\nip}{\bar{n}_{i}}
\newcommand{\nim}{n_{i}}
\newcommand{\R}[2]{\tensor{\mathcal{R}}{^{#1}_{#2}}}
\newcommand{\dd}{\text{d}}

\newcommand{\cA}{\mathcal{A}}
\newcommand{\cM}{\mathcal{M}}
\newcommand{\Lagr}{\mathcal{L}}

\newcommand{\s}{\hspace{0.8pt}}

\numberwithin{equation}{section}

\begin{document}

\allowdisplaybreaks

\begin{titlepage}

\begin{flushright}
{\small
CERN-TH-2026-175\\
}
\end{flushright}

\vskip80pt
\begin{center}
{\Large \bf 
geoSCET:
Soft Theorems from Power Counting}
\end{center}
  \vspace{0.5cm}
\begin{center}
{\sc Timothy~Cohen,$^{a,b}$ \sc Patrick~Hager,$^{a}$ and Andreas~Helset\,$^{a,c}$} 
\\[6mm]
{\it ${}^a$\! Theoretical Physics Department\\[-1pt] 
CERN, CH-1211 Geneva 23, Switzerland}\\[0.2cm]
{\it ${}^b$\! Institute for Fundamental Science\\[-1pt] 
University of Oregon, Eugene, Oregon 97403, USA}\\[0.2cm]
{\it ${}^c$\! Department of Mathematics and Physics\\[-1pt] 
University of Stavanger, 4036 Stavanger, Norway}\\[0.2cm]
\end{center}
\vskip1cm

\begin{abstract}
\noindent 
We apply the framework of Soft Collinear Effective Theory (SCET) to the theory of the geometric scalar field. The resulting ``geoSCET'' manifests an emergent geometry in the soft sector, mirroring the emergent soft gauge invariance of QCD SCET. We use geoSCET to derive the geometric soft theorems as straightforward consequences of effective-field-theory power counting, including extensions to multiple soft emissions and loop corrections. We demonstrate that theories without a potential satisfy universal geometric soft theorems for any number of soft emissions to all orders in perturbation theory. 
We also show that we can turn on a potential at the soft scale without spoiling the factorization of the soft physics. 
This work demonstrates the underlying field-space diffeomorphism origin of the geometric soft theorem.
\end{abstract}

\end{titlepage}

\setcounter{tocdepth}{2} 
\setcounter{page}{2}
{
\setstretch{1.1}
\tableofcontents
}
\clearpage

\section{Introduction}

The universality of soft dynamics is a hallmark of quantum field theory. 
Manifestations of this property are \emph{factorization theorems}, e.g., the universality of parton distribution functions~\cite{Collins:1984kg,Collins:1989gx}, and effective field theory (EFT) factorization of soft sectors. 
The universality of soft physics has also been leveraged to uncover striking features of scattering amplitudes. Taking low-energy limits of external lines leads to \emph{soft theorems}, which connect higher-point amplitudes with soft legs to lower-point amplitudes without them. 
It is well known that soft factorization in EFTs and soft theorems for amplitudes are fundamentally related, a connection which is central to the arguments presented in this paper.

Classic soft theorems are well known for amplitudes involving photons~\cite{Low:1958sn,Weinberg:1965nx,Burnett:1967km}, gravitons~\cite{Weinberg:1965nx}, and pions~\cite{Adler:1964um}. 
In the case of pions, we now understand that Adler's original insight~\cite{Adler:1964um} is one example of a general feature of Goldstone bosons.
Generalizations for theories of Goldstone bosons beyond the EFT of pions include systems as diverse as fluids~\cite{Cheung:2023qwn}, inflation~\cite{Green:2022slj}, and higher-dimensional branes~\cite{Novotny:2016jkh}. The usual derivations of soft theorems rely on an underlying global/local symmetry. For Goldstone bosons, soft theorems are understood to be a consequence of spontaneously-broken symmetry. For gauge theory and gravity, gauge and diffeomorphism invariance are intricately linked to the corresponding soft theorems. 

Recently, one of the authors derived soft theorems for general EFTs of scalars, called the \emph{geometric soft theorems}~\cite{Cheung:2021yog}, that do not make any reference to an underlying symmetry. The main property of the EFT used in the derivation was the invariance under field redefinitions~\cite{Chisholm:1961tha, Kamefuchi:1961sb, tHooft:1972qbu, tHooft:1973wag, Coleman:1969sm, Callan:1969sn, Deans:1978wn,Politzer:1980me,Arzt:1993gz,Passarino:2016saj, Manohar:2018aog, Criado:2018sdb, Cohen:2024fak}.
Field-redefinition invariance can be made manifest by expressing amplitudes in terms of geometric quantities defined on an underlying field-space manifold~\cite{Volkov:1973vd}.\footnote{The EFT geometry has recently been applied to the Standard Model Higgs sector~\cite{Alonso:2015fsp, Alonso:2016btr, Alonso:2016oah,Helset:2018fgq, Nagai:2019tgi, Helset:2020yio, Cohen:2020xca, Cohen:2021ucp, Alonso:2021rac, Talbert:2022unj, Alonso:2023jsi,Helset:2022pde,Assi:2023zid,Jenkins:2023bls,Assi:2025fsm} and to multifield inflation~\cite{Pinol:2020kvw,Pinol:2026xnl}.} For scalar EFTs with no potential, the geometric soft theorem takes the form
\begin{align}\label{eq:softThm}
    \lim_{k_s\rightarrow 0} \cM_{N+1} = \nabla_{I} \cM_{N} \,.
\end{align}
The soft theorem relates the soft limit $k_s\rightarrow 0$ of a massless scalar with flavor label $I$ in a scattering amplitude with $N+1$ particles to the covariant derivative in field space, $\nabla_I$, acting on the $N$-particle scattering amplitude with no soft scalar. The geometric soft theorem has been extended to theories with fermions and gauge bosons~\cite{Derda:2024jvo} and explicitly verified to hold at one loop~\cite{Cohen:2025dex}.

The possible extensions of the soft theorems fall into three categories: subleading power corrections, loop corrections, and multiple soft emissions. The soft theorem in gauge theory holds up to subleading order in the soft expansion~\cite{Low:1958sn,Burnett:1967km}, while for gravity it has a universal behavior through sub-subleading order~\cite{Cachazo:2014fwa}. The leading-power gauge and gravity soft theorems receive no loop corrections, while at subleading power (and sub-subleading power for gravity) there are loop corrections~\cite{Bern:2014oka}. The loop corrections are well-understood, and they arise only at one-loop order at subleading power in both gauge theory and gravity, and up to two-loop order at sub-subleading power in gravity~\cite{Bern:2014oka,Beneke:2021umj,Beneke:2022pue,Czakon:2023tld,Czakon:2025cxk}.

The case of multiple soft emissions also typically exhibits universal behavior. Several new double soft theorems for scalars and photons were derived at tree level in \cite{Cachazo:2015ksa}. For pions, the single soft limit vanishes, while the double soft limit is nonzero~\cite{Arkani-Hamed:2008owk}. This result was generalized to \textit{any} scalar EFT, where the double soft limit takes the form~\cite{Cheung:2021yog}
\begin{align}
    \lim_{k_I,k_J\rightarrow 0} \cM_{N+2} = \frac{1}{2}\sum_{K}\frac{s_{IK}-s_{JK}}{s_{IK}+s_{JK}} R_{IJK}^{\quad\;\; L} \cM_{N,\dots L \dots} + \nabla_{(I}\nabla_{J)} \cM_{N} \,.
\end{align}
This result, which was derived for tree-level amplitudes, depends on the Riemann curvature tensor in field space, $R_{IJKL}$, as well as the Mandelstam variables $s_{IK} = (k_I + p_K)^2$. Again, we see the appearance of geometric quantities defined on the field-space manifold, further exemplifying the fundamental connection between soft theorems and field-space geometry.

In this paper, we will provide a new perspective on the soft behavior of geometric scalar EFTs by employing the modern EFT approach called Soft-Collinear Effective Theory (SCET)~\cite{Bauer:2000yr,Bauer:2001yt,Bauer:2002nz,Beneke:2002ph,Beneke:2002ni}. The SCET framework is extremely useful because it systematically isolates the dynamics of the soft modes order-by-order in a power-counting expansion. The construction has been extended to include gravity~\cite{Beneke:2021aip,Beneke:2022ehj}. By applying the underlying logic of SCET to the geometric scalar EFT, we derive a new EFT that integrates out hard modes, leading to a factorized theory of soft and collinear scalar modes. This new application of SCET, which we name ``geoSCET,'' provides us with a systematic tool to address the geometric soft theorem. We will use it to analyze subleading power corrections, loop corrections, and multiple soft emissions.

We will see that the geoSCET Lagrangian is factorized into collinear and soft sectors. The soft sector naturally organizes itself geometrically. As we will emphasize in what follows, this is a natural analog of the soft gauge invariance that emerges for QCD SCET. In fact, we will see that parallel transport on the scalar manifold of geoSCET will exactly mirror the construction of QCD SCET using Wilson lines. Exposing this close connection motivates the review of this EFT that is provided in the next section. Once we have established the basics of QCD SCET, the derivation of geoSCET follows in a straightforward way.

We will formulate geoSCET for the case of theories without and with a potential. One interesting conceptual aspect of the construction is the role of soft scalar dressings. We will see that when the theory has no potential, there is a straightforward field redefinition that removes the soft fields from the collinear Lagrangian, following a close analogy with QCD SCET. By contrast, we will see that theories with a potential do not admit an obvious soft dressing to make this decoupling manifest, as was anticipated in \cite{Biswas:2022lsj}. This points to a very interesting underlying difference between these two versions of geoSCET, that we expect will have an impact on future work to develop the connection between geoSCET and the celestial holography program, as reviewed in \cite{Strominger:2017zoo,Pasterski:2021raf,Raclariu:2021zjz,Donnay:2023mrd}.

The paper is organized as follows. After reviewing QCD SCET in \cref{sec:QCDSCET}, we derive geoSCET in \cref{sec:Geometry}. We then apply this EFT to explain the geometric soft theorem in \cref{sec:SoftTh}. We conclude and emphasize a number of exciting future directions in \cref{sec:Conc}. Various technical details are provided in a series of appendices.

\section{Primer: SCET for QCD jets}
\label{sec:QCDSCET}

In this section, we provide a brief primer for the position-space SCET construction. We focus on jet processes, since this is one of its most natural applications, and most closely parallels the applications below. Also, to more directly connect with the new application of SCET presented in what follows, we only consider scalar matter fields coupled to gluons.
The technical details of constructing position space SCET for QCD are given in~\cite{Beneke:2002ph,Beneke:2002ni,Boer:2023yde} and a pedagogical construction of scalar QCD is presented in Section 2 of~\cite{Beneke:2021aip}. The readers who are familiar with SCET can skip this section. For a more thorough pedagogical introduction, see \cite{Becher:2014oda, Becher:2018gno, Cohen:2019wxr}.

\subsection{Kinematics}

For concreteness, we consider the production of two scalar QCD jets, produced by some hard process at partonic center-of-mass energy $\sqrt{s}$, so that the hard scale for the process is $Q\sim \sqrt{s}$. 
The $i^\text{th}$ jet is characterized by a hard momentum $P_i\sim Q$ carried along a light-like direction $\nim^\mu$, and a small jet mass $m_J\ll Q$.
It is convenient to introduce a second light-like vector $\nip^\mu$ such that
\begin{equation}
    \nim^2=\nip^2=0\,,
    \qquad\text{and}\qquad
    \nim\cdot \nip = 2\,.
\end{equation}
If we assume that the $i^\text{th}$ jet points in the $z$-direction, a canonical choice is $n_i^\mu = (1,0,0,1)$ and $\bar{n}_i^\mu = (1,0,0,-1)$.
The collinear momentum of the jet can be decomposed in this basis as
\begin{equation}
    P_i^\mu = \nip\Cdot P_i\spac \frac{\nim^\mu}{2} + \nim\Cdot P_i\spac\frac{\nip^\mu}{2} + P_{i\perp}^\mu\,,
\end{equation}
where the subscript $\perp$ denotes the two transverse directions with respect to $\nim$ and $\nip$.
The natural power-counting parameter $\lambda$ is
\begin{equation}
    \lambda \sim \frac{m_J}{Q} \ll 1\,.
\end{equation}
We are interested in the hierarchy $Q \gg m_J \gg \Lambda_\text{QCD}$, so that perturbation theory applies. 
The collinear momentum is aligned with the $n_i^\mu$ direction, and is characterized by the power counting
\begin{equation}
    (\nip\Cdot P_i,\spac\nim\Cdot P_i,\spac P_{i\perp})\equiv(P_{i-},\spac P_{i+},\spac P_{i\perp})\sim (1,\lambda^2,\lambda)\spac Q\,,
\label{eq:colvirt}
\end{equation}
such that 
\begin{align}
P_i^2\sim\lambda^2Q^2 \sim m_J^2\,.    
\end{align}
The collimated nature of the jets is reflected by the power counting $P_{i\perp}\sim\lambda\spac \nip\cdot P_i$.
In a two-jet process, we often refer to one jet as ``collinear'' and the other as ``anti-collinear.''

It is crucial to note that the sum of two momenta of different collinear sectors is hard, $P_i+P_j\sim Q$ when $i\neq j$.
Therefore, EFT power counting implies that each collinear object must be modelled using a disjoint EFT sector. These sectors can feature interactions among themselves but not with each other.
Physically, we know this picture must be missing some non-trivial dynamics. If this were the whole story, then the two jets would not be able to interact, and thus it would be impossible to describe long-range color correlations due to soft gluon exchanges.
Hence, the EFT must feature an additional sector that facilitates the communication between the independent collinear sectors. This can be accomplished by introducing a soft (sometimes called ``ultrasoft'') sector, whose modes have a characteristic momentum scaling
\begin{equation}
k_{\mathrm{s}}\sim\big(\lambda^2,\lambda^2,\lambda^2\big)\spac Q\,.
\end{equation}
The momentum carried by this isotropic radiation is parametrically small enough so that soft exchanges between different collinear sectors do not change their virtuality as specified in \cref{eq:colvirt}; in other words, $P_i+k_{\mathrm{s}} \sim P_i$. Note that the soft scale is correlated with the collinear scales, since its power counting implies $(P_i^2)^2/Q^2\sim\lambda^4 Q^2$. This setup with soft modes of smaller virtuality than collinear ones is called SCET$_{\mathrm{I}}$. 
This kinematic setup also holds at loop level for a wide range of physically interesting processes.
From here forward, we often follow the usual convention and set $Q=1$.

\subsection{Power counting and degrees of freedom}
Modern EFTs such as SCET are completely analogous to the textbook application of EFT, where one integrates out a field with mass $M$ to derive an EFT valid at scales $E\ll M$, for example Fermi theory. Instead of integrating out a massive field, the generalization relies on splitting a field into a momentum mode expansion, integrating out the hard modes $p\sim M$, and absorbing their effects into local matching coefficients.
Other examples beyond SCET include HQET~\cite{Georgi:1990um}, NRQCD~\cite{Caswell:1985ui}, SdSET~\cite{Cohen:2020php}, EFTs for gravitational waves~\cite{Goldberger:2004jt}.

The expansion into modes can be made manifest by the method of regions~\cite{Beneke:1997zp}, a systematic procedure for obtaining asymptotic expansions of Feynman integrals in the presence of widely-separated scales.
The method identifies momentum regions characterized by distinct scalings, expands the integrands accordingly, and performs the integral over the entire range.
The sum of all regions then reproduces the asymptotic expansion of the integral in powers of a small ratio of external scales.
Obtaining a complete list of all relevant regions to a given observable is challenging and an area of active research~\cite{Gardi:2022khw,Beneke:2023wmt,Ma:2023hrt,Gardi:2024axt,Guan:2024hlf,Ma:2025emu,Ma:2026pjx,Chen:2026dnj}.
The derivation of an EFT can be organized by a method-of-regions analysis as follows:
\begin{enumerate}
    \item[(a)] Identify a power-counting parameter $\lambda$, the relevant momentum regions appearing in the observable of interest and their power counting, and integrate out the short-distance physics by absorbing its dependence into Wilson coefficients of the EFT.
    \item[(b)]  Assign a field to each momentum region, and organize their (self-)interactions in a low-energy effective action organized by a systematic expansion in $\lambda$.
    \item[(c)] Make use of constraints from the UV and symmetries emerging in the infrared to constrain this action.
\end{enumerate}
SCET is a prime example of this methodology, and in the following we provide a high-level conceptual overview of how each step is realized. Let us assume that we want to model an $N$-jet process. 
In this case, (a) tells us that the relevant regions correspond to the $i$-collinear ($i=1,\dots,N)$ modes and the soft mode, while hard modes with momenta $p_h\sim(1,1,1)$ are integrated out.\footnote{In processes with space-like splittings, SCET also features Glauber modes (see~\cite{Rothstein:2016bsq} for a treatment within SCET), whose exchange can generate factorization-breaking contributions~\cite{Catani:2011st}.
These effects have attracted considerable recent attention in the context of superleading logarithms~\cite{Forshaw:2008cq,Becher:2021zkk,Becher:2023mtx,Becher:2024kmk,Becher:2025igg,Dasgupta:2025cgl} and coherence-violating logarithms~\cite{Banfi:2010xy,Gaunt:2014ska,Forshaw:2021fxs,Banfi:2025mra,Becher:2026kbr}.
They also underlie the recently computed factorization-breaking terms in the two-loop splitting amplitude~\cite{Buccioni:2026mfg}, which can be understood from a method-of-regions analysis involving Glauber exchanges~\cite{Becher:2026kbr,Chen:2026dnj,Barcaro:2026dsd}.
We restrict our attention to the soft and collinear sectors and neglect Glauber contributions throughout.}

We now proceed with (b) and introduce the field content, which consists of collinear and soft fluctuations
with power counting (inferred from the two-point functions)
\begin{equation}
    \phi_c\sim\lambda\,,
    \qquad
    \phi_s\sim\lambda^2\,,
    \qquad 
    A_c\sim(1,\lambda^2,\lambda)\,,
    \qquad 
    A_s\sim\lambda^2\,,
\end{equation}
with one collinear field per sector $i=1,\dots,N$.
When constructing the Lagrangian, we will also need the scaling of position arguments, inferred from $p\cdot x\sim 1$ as
\begin{equation}\label{eq:PositionScaling}
    (\nip\Cdot x_c,\spac\nim \Cdot x_c,\spac x_{c\perp})\equiv (x_-,\spac x_+,\spac x_{\perp})\sim (\lambda^{-2},1\spac,\spac\lambda^{-1})\,,
    \qquad \text{and} \qquad
    x_s\sim\lambda^{-2}\,.
\end{equation}
This shows that derivatives of fields scale as momenta, as expected.
The power counting of the measure $\dd^4x$ depends on the field content of the Lagrangian density; it is $\dd^4x\sim\lambda^{-4}$ if collinear fields are present, or $\dd^4x\sim\lambda^{-8}$ for purely soft terms.

Implementing point (c) imposes the gauge symmetry from the UV theory, which we expand upon in the next section, along with constraints due to the UV Lorentz invariance.
Lorentz invariance is broken in the EFT due to the presence of the light-cone reference vectors. The original invariance manifests itself through reparameterization invariance (RPI)~\cite{Manohar:2002fd,Marcantonini:2008qn}, which can be traced back to the fact that the requirements $n^2 = \bar{n}^2 =0$ and $n\cdot \bar{n} =2$ do not uniquely determine these vectors. This leads to three independent transformations (usually called RPI-I, -II, and -III), and we must enforce that the EFT is invariant under all of them. These transformations in turn relate terms at different orders in $\lambda$, and provide non-trivial constraints among the EFT operator coefficients.

\subsection{Gauge symmetry}

In the EFT, the full-theory gauge symmetry is extended into separate collinear and soft gauge transformations which interplay with each other non-trivially.
To see this, consider the EFT gluon field, assuming a single collinear direction:
\begin{equation}\label{eq:GluonDecomp}
    A_\mu(x) = A_{s\mu}(x) + A_{c\mu}(x)\,.
\end{equation}
The two fields on the right-hand side must transform in a way that is consistent with the full gluon transformation on the left-hand side
\begin{equation}
   A \to U A\spac U^\dagger + \frac{i}{g_s}U\big[\partial, U^\dagger\s\big]\,.
\end{equation}
At the same time, the soft field cannot transform under collinear gauge transformations, as this would shuffle collinear modes into the soft field, thereby violating the power counting.
The consistent way to resolve this is to treat the soft field as a dynamical background in the presence of a collinear fluctuation. Following this logic, one can derive the following sets of collinear and soft gauge transformations:
\begin{align}
\label{eq:QCDGaugeTrafoColl}
    &\text{collinear:} & A_c &\to U^{\vphantom{\dagger}}_c \spac A_c^{\vphantom{\dagger}} \spac U_c^\dagger + \frac{i}{g_s}U_c\big[D^{\vphantom{\dagger}}_s, U_c^\dagger \s \big]\,, & \phi_c &\to U_c\spac\phi_c\,, \\[-5pt]
    & & A_s &\to A_s\,, & \phi_s &\to \phi_s \nn\\[10pt]
    \label{eq:QCDGaugeTrafoSoft}
    &\text{soft:} & A^{\vphantom{\dagger}}_c &\to U^{\vphantom{\dagger}}_s\spac A^{\vphantom{\dagger}}_c\spac U_s^\dagger\,, & \phi_c &\to U_s\spac\phi_c\,,\\[-5pt]
    & & A_s &\to U^{\vphantom{\dagger}}_s\spac A^{\vphantom{\dagger}}_s\spac U_s^\dagger + \frac{i}{g_s}U^{\vphantom{\dagger}}_s\big[\partial, U_s^\dagger\s\big]\,, & \phi_s &\to U_s\spac\phi_s\,.\nn
\end{align}
Here, the transformation of the fluctuation $A_c$ is covariant with respect to the soft background $A_s$, which is made manifest by the appearance of the soft-covariant derivative $D_s = \partial - ig_s A_s$ in the transformation, while the background does not transform under these transformations. Under soft transformations, the collinear gluon field transforms as an adjoint matter field, while the background $A_s$ has the usual gauge transformation.
This is the reason why the SCET Lagrangian is naturally formulated through a background-field construction.
For multiple collinear sectors, the gauge group $\mathcal{G}_{\mathrm{SCET}}$ of SCET naturally extends to the semi-direct product as
\begin{equation}
    \mathcal{G}_{\mathrm{SCET}} \simeq SU(3)_s \ltimes \prod_{i=1}^N SU(3)_i\,.
\end{equation}
This is a generic feature of the SCET construction that extends beyond QCD. We will rely on this logic to derive geoSCET below.

\subsection{SCET Lagrangian}
\label{sec:QCDLagrangianConstruction}

The power counting implies that the sum of two momenta from different collinear sectors is hard. Consequently, the EFT Lagrangian factorizes as
\begin{equation}
    \Lagr_{\mathrm{SCET}} = \sum_{i=1}^N\Lagr_i(\phi_{c_i},A_{c_i},\phi_s,A_s) + \Lagr_s + \mathcal{O}_{\text{$N$-jet}}\,,
\end{equation}
where $\Lagr_i$ denotes the soft-collinear Lagrangian that describes the interactions within one collinear sector and with the soft fields (since each $i$-collinear Lagrangian only depends on one direction $\nim$, it is sufficient to construct it once, and so we set $\nim=\nm$ and $P_i = P$ for simplicity in the rest of this section), $\Lagr_s$ is the soft Lagrangian, and $\mathcal{O}_{\text{$N$-jet}}$ represents the collection of $N$-jet operators.
We will introduce the $N$-jet operators in \cref{sec:QCDNJet} below. For now, all we need to know is that they are operators that act as sources, thereby encoding the hard scattering that produces the collinear jets (and soft radiation): thus, only one insertion of an $N$-jet operator is possible for a given amplitude. However, there are infinitely many such operators that differ in power counting, e.g., through the number of collinear fields per sector or the number of soft emissions, and we collectively denote these as $\mathcal{O}_{\text{$N$-jet}}$. 

There is one final essential feature of SCET that we need to introduce: the multipole expansion. The power counting in \cref{eq:PositionScaling} implies that interactions between collinear and soft fields, such as
\begin{equation}
 \Lagr \supset   \int \dd^4x\: \phi_c^2(x)\spac\phi^{\vphantom{2}}_s(x) \,,
\end{equation}
are not homogeneous in $\lambda$, but generate an infinite tower of subleading terms. These terms can be derived by performing a multipole expansion of the soft field around the spacetime point $x_-^\mu = \np\cdot x \spac \frac{\nm^\mu}{2}$. Diagrammatically, this corresponds to the observation that only the combination 
\begin{equation}
    n \Cdot P + n\Cdot  k_{s} \sim \lambda^2
\end{equation}
scales homogeneously in $\lambda$, while $k_{s\perp}$ and $\np \cdot k_s$ are suppressed in $\lambda$ compared to $P_{\perp}$ and $\np\cdot P$; one must therefore Taylor expand accordingly.\footnote{There are two schools of thought to deal with this problem, which are commonly known as ``label SCET'' and ``position-space SCET.'' We adopt the latter here due to its natural lift to geometry.} 
The full multipole expansion is performed at the Lagrangian level by Taylor-expanding soft fields as
\begin{equation}
    \phi_s(x) = \phi_s(x_-) + (x-x_-)^\alpha [\partial_\alpha \phi_s](x_-) + \frac{1}{2}(x-x_-)^\alpha (x-x_-)^\beta [\partial_\alpha \partial_\beta\phi_s](x_-)+\dots\,,
\end{equation}
whenever they appear in interactions with collinear ones.
Here, the square bracket indicates that the field is evaluated at $x_-^\mu$ after acting with the derivatives.

There is a second source of soft-collinear interactions in the Lagrangian; the soft transformations of collinear fields. From~\cref{eq:QCDGaugeTrafoSoft}, we see that $\phi_c(x)\to U_s(x)\phi_c(x)$. Also here, the soft gauge transformation must be multipole expanded as 
\begin{equation}
\label{eq:QCDGaugeMultipoleExpansion}
    U_s(x)\phi_c(x) = U_s(x_-)\phi_c(x) + (x - x_-)^\alpha[\partial_\alpha U_s](x_-)\phi_c(x) + \dots
\end{equation}
in order to have homogeneous power counting in $\lambda$.\footnote{
If the gauge transformation has generators with non-trivial power counting, the expansion is performed to respect the multipole expansion even when the resulting gauge transformation is still inhomogeneous in $\lambda$. This is the case, e.g., for gravity, where momenta generate the transformations, but it will also be true for the geometric scalar field below.}
In fact, it would be preferable to have collinear fields that transform only with the homogeneous gauge transformation $U_s(x_-)$.
Geometrically, this can be achieved by moving the soft gauge transformation of the collinear fields $U_s(x)$ from point $x^\mu$ to $x_-^\mu$ via a parallel transport. To accomplish this, one defines the $\mathcal{R}$ Wilson line~\cite{Beneke:2002ni}, which is a path-ordered exponential
\begin{equation}
\label{eq:QCDRWilsonLine}
    \mathcal{R}(x,x_-) = \mathcal{P}\exp\biggl(ig_s\int_0^1 \dd s\: (x-x_-)^\mu A_{s\mu}\bigl(x_-+s(x-x_-)\bigr)\biggr)\,,
\end{equation}
and transforms as
\begin{subequations}
\begin{align}
\label{eq:QCDRWilsonTransformationColl}
    \mathcal{R}(x,x_-)&\xrightarrow[]{\mathrm{coll\s}} \mathcal{R}(x,x_-)\,,\\[3pt]
    \label{eq:QCDRWilsonTransformationSoft}
    \mathcal{R}(x,x_-)&\xrightarrow[]{\mathrm{soft\s}} U_s^{\vphantom{\dagger}}(x) \mathcal{R}(x,x_-) U_s^\dagger(x_-)\,.
\end{align}
\end{subequations}
The first transformation follows immediately, as the soft fields do not transform under collinear gauge transformations, and the second transformation is a standard property of a Wilson line and can be obtained, e.g., from the parallel-transport equation.
This Wilson line corresponds to the inverse of a soft gauge transformation that fixes \emph{fixed-line gauge}~\cite{Beneke:2002ni}, defined through the condition
\begin{equation}
    (x-x_-)^\mu A_{s\mu}(x) = 0\,.
\label{eq:FLgauge}
\end{equation}
This is a generalization of Fock-Schwinger (FS) gauge ($x^\mu A_\mu(x)=0$) along the light-cone. The gauge choice \cref{eq:FLgauge} has the interpretation as fixing FS gauge along the transverse directions, while leaving the residual gauge field $\nm\cdot A_s(x_-)\frac{\np^\mu}{2}$ unconstrained.
This choice is very natural, as the residual gauge field (also called the emergent soft background field~\cite{Beneke:2021umj}) transforms with the homogeneous transformation $U_s(x_-)$.
The other components of the soft gluon field can be expressed in terms of gauge-covariant objects (such as the field-strength tensor and its covariant derivatives) through the identities~\cite{Beneke:2002ni} (for a derivation, see App.~\ref{sec:App:FLGauge})
\begin{subequations}
\begin{align}
\label{eq:QCDFLIdentity1}
    \nm\Cdot A_s(x) - \nm\Cdot A_s(x_-) &= \int_0^1 \dd s\: (x-x_-)^\mu\spac \nm^\nu F_{s\mu\nu}(y(s))\,,\\
    \np\Cdot A_s(x) &= \int_0^1 \dd s\: s(x-x_-)^\mu\spac \np^\nu F_{s\mu\nu}(y(s))\,,\\
    A_{s\nu_\perp}(x) &=\int_0^1 \dd s\: s(x-x_-)^\mu\spac F_{s\mu\nu_\perp}(y(s))\,.
    \label{eq:QCDFLIdentity3}
\end{align}%
\label{eq:QCDFLIdentity}%
\end{subequations}%
From the transformation~\eqref{eq:QCDRWilsonTransformationSoft}, it is clear that this object moves the gauge transformation of collinear fields from point $x^\mu$ to $x_-^\mu$, and one can redefine the collinear matter fields as
\begin{equation}
\label{eq:QCDCollRedef}
    \phi_c(x) = \mathcal{R}(x,x_-)\hat{\phi}_c(x)\,,
\end{equation}
and likewise for the gluon fields, such that the transformations of the hatted fields are
\begin{align}
    &\text{collinear:} & \hat{A}^{\vphantom{\dagger}}_c &\to U^{\vphantom{\dagger}}_c \spac \hat{A}^{\vphantom{\dagger}}_c \spac U_c^\dagger + \frac{i}{g_s}U^{\vphantom{\dagger}}_c\big[\hat{D}^{\vphantom{\dagger}}_s, U_c^\dagger\big]\,, & \hat{\phi}_c &\to U_c\spac\hat{\phi}_c\,, \\[-5pt]
    & & A_s &\to A_s\,, & \phi_s &\to \phi_s \nn\\[10pt]
    \label{eq:QCDGaugeTrafoSoftNew}
    &\text{soft:} & \hat{A}^{\vphantom{\dagger}}_c &\to U^{\vphantom{\dagger}}_s(x_-)\spac \hat{A}^{\vphantom{\dagger}}_c\spac U_s^\dagger(x_-)\,, & \hat{\phi}_c &\to U_s(x_-)\spac\hat{\phi}_c\,,\\[-5pt]
    & & A^{\vphantom{\dagger}}_s &\to U^{\vphantom{\dagger}}_s\spac A^{\vphantom{\dagger}}_s\spac U_s^\dagger + \frac{i}{g_s}U^{\vphantom{\dagger}}_s\big[\partial, U_s^\dagger\big]\,, & \phi_s &\to U_s\spac\phi_s\,.\nn
\end{align}
Here, $\hat{D}^\mu_s = \partial^\mu - ig_s\, \nm\Cdot A_s(x_-)\frac{\np^\mu}{2}$ is the covariant derivative that contains precisely the residual homogeneous soft background field, and the hatted fields now have homogeneous gauge transformations.
In practice, this redefinition has the benefit of moving all soft fields in the Lagrangian into fixed-line gauge, so that one finds a soft-covariant derivative $\hat{D}_s$, while all subleading soft-gluon terms are expressed through field-strength tensors.
This way, every term in the Lagrangian is then manifestly homogeneous in $\lambda$ and gauge invariant under the homogeneous set of transformations $U_s(x_-)$.

In the previous discussion, we focused on the soft gauge transformations. We now briefly discuss the collinear gauge transformations. From~\cref{eq:QCDRWilsonTransformationColl}, one might wonder how the redefinition~\cref{eq:QCDCollRedef} is consistent with the collinear gauge transformation, and in fact this redefinition is not consistent in general. Instead, one should understand~\cref{eq:QCDCollRedef} as a statement in a physical gauge. The natural choice in SCET is to interpret \cref{eq:QCDCollRedef} in collinear light-cone gauge $\np\cdot A_c(x) = 0$.
Then, to reinstate collinear gauge transformations, one can ``unfix'' the gauge via the collinear Wilson line\footnote{Note that this is the Wilson line of an incoming particle. In factorization theorems for physical observables, one must distinguish in general the Wilson line for incoming and outgoing particles in the gauge-invariant building block to properly account for all phases~\cite{Becher:2026kbr,SchwienbacherThesis}.}
\begin{equation}
\label{eq:QCDCollWilsonLine}
    W_c(x) = \mathcal{P}\exp\biggl(ig_s \int_{-\infty}^0 \dd s\:\np\Cdot A_c(x+s\np) \biggr)\,.
\end{equation}
Under collinear gauge transformations, it transforms as
\begin{equation}
    W^{\vphantom{\dagger}}_c(x) \xrightarrow{\mathrm{coll}} U^{\vphantom{\dagger}}_c(x) W^{\vphantom{\dagger}}_c(x) U_c^\dagger(x-\infty\spac \np)\,,
\end{equation}
and we assume that the transformation falls off at infinity, i.e., we fix $U_c^\dagger(x-\infty\spac \np) = \mathbf{1}$.
One can use the Wilson line to define \emph{collinear gauge-invariant building blocks} as
\begin{subequations}
    \begin{align}
    \label{eq:QCDGaugeInvBB1}
        \hat{\chi}^{\vphantom{\dagger}}_c &= \hat{W}_c^\dagger \hat{\phi}^{\vphantom{\dagger}}_c\,,\\
        \hat{\mathcal{A}}^{\vphantom{\dagger}}_{c\mu} &= \hat{W}_c^\dagger \hat{A}^{\vphantom{\dagger}}_{c\mu} \hat{W}^{\vphantom{\dagger}}_c + \frac{i}{g_s}\hat{W}_c^\dagger \big[\hat{D}^{\vphantom{\dagger}}_{s\mu},\hat{W}^{\vphantom{\dagger}}_c\big]\,,
    \label{eq:QCDGaugeInvBB2}%
    \end{align}%
    \label{eq:QCDGaugeInvBB}%
\end{subequations}%
where the hats on the Wilson line indicate that these are defined with $\hat{A}_c$ in~\cref{eq:QCDCollWilsonLine}.
The covariant version of~\cref{eq:QCDCollRedef} is then obtained as follows by first interpreting both sides of the equation as being in collinear light-cone gauge. The gauge symmetry on the right-hand side can then be reinstated by taking $\hat{\phi}_c \to \hat{\chi}_c$, which is obviously true as $\hat{W}_c=1$ in light-cone gauge. The fully covariant redefinition then reads
\begin{equation}
    \phi^{\vphantom{\dagger}}_c(x) = \mathcal{R}(x,x_-)\hat{W}_c^\dagger(x)\hat{\phi}^{\vphantom{\dagger}}_c(x)\,.
\end{equation}
Beyond this technicality in the redefinition, the collinear Wilson line arises naturally in matching calculations and is essential for the construction of the $N$-jet operator, as we explain below.

The construction of the Lagrangian can then be summarized into a four-step procedure, which will naturally lift to a geometric interpretation, as we explain in the next section.
The steps are:
\begin{enumerate}
    \item[(i)] Insert the decomposition~\cref{eq:GluonDecomp} (and likewise for the scalar field) into the full-theory Lagrangian. Fix collinear light-cone gauge. Expand it as a collinear fluctuation around a dynamical soft background field, using standard background-field methods.
    This Lagrangian is covariant with respect to $A^\mu_s(x)$, but inhomogeneous in $\lambda$.
    \item[(ii)] To render each term homogeneous in $\lambda$, perform the multipole expansion of soft fields around $x_-$ when they appear in interactions with collinear fields. However, collinear fields are still covariant with respect to $A_s(x)$. Thus, gauge transformations break the multipole expansion and mix different orders in $\lambda$.
    \item[(iii)] Identify the emergent \emph{homogeneous background field}, in this case $\nm\Cdot A_s(x_-) \frac{\np^\mu}{2}$, and parallel transport the collinear fields $\phi_c\to\hat{\phi}_c\,, A_c\to\hat{A}_c$, using the $\mathcal{R}$ Wilson line. The hatted fields transform covariantly with respect to this background field, i.e., with $U_s(x_-)$.
    \item[(iv)] The $\mathcal{R}$ Wilson line then dresses the soft fields, such that a homogeneous covariant derivative $\nm\Cdot \hat{D}_s(x_-)$ emerges. All subleading terms are described in terms of field-strength tensors (and their derivatives) only. One can now reinstate collinear gauge transformations using the collinear Wilson line. Then, expand in $\lambda$, and each term is manifestly gauge invariant and homogeneous.
\end{enumerate}
For scalar QCD, this construction is performed in detail in Sec.~2.4 of~\cite{Beneke:2021aip}.
Ultimately, one can organize the Lagrangian as a power expansion
\begin{equation}
    \mathcal{L} = \mathcal{L}^{(0)} + \mathcal{L}^{(1)} + \mathcal{O}(\lambda^2)\,.
\end{equation}
The terms that are particularly relevant for comparison to the geometric scalar field are
\begin{subequations}
    \begin{align}
        \mathcal{L}^{(0)} &= \frac{1}{2} \bigl[ \np\Cdot D_c\spac\hat{\phi}_c \bigr]^\dagger\spac\nm\Cdot D\spac\hat{\phi}_c + \frac{1}{2} \bigl[ \nm\Cdot D\spac\hat{\phi}_c \bigr]^\dagger\spac\np\Cdot D_c\spac\hat{\phi}_c  + \bigl[ D_{c\mu_\perp}\hat{\phi}_c \bigr]^\dagger\spac D_c^{\mu_\perp}\hat{\phi}_c\,,\\
        \mathcal{L}^{(1)}&= \frac{1}{2}\hat{\phi}_c^\dagger x_\perp^\mu \nm^\nu\spac W_c\spac g_s F_{s\mu\nu}\spac W_c^\dagger \spac i\np\Cdot D_c\spac\hat{\phi}_c + \mathrm{h.c.} + \dots\,,
    \end{align}
\end{subequations}
where the dots in the second line denote terms proportional to the soft scalar field, which are not required below. The full Lagrangian is given in~\cite{Beneke:2021aip}.
Importantly, the leading-power Lagrangian features the covariant derivative
\begin{equation}
    D_\mu = \partial_\mu - ig_s A_{c\mu} - ig_s\nm\Cdot A_s(x_-)\frac{\np_\mu}{2}\,,
\end{equation}
which contains the homogeneous soft background field $\nm\Cdot A_s(x_-)\frac{\np_\mu}{2}$. The components $\np\Cdot D$ and $D_\perp$ are purely collinear, indicated in the Lagrangian by the subscript $c$.
The subleading terms contain the field-strength tensor $F_{s\mu\nu}$ and, at higher orders, its covariant derivatives. 
Each term is manifestly gauge invariant under both collinear and the homogeneous soft gauge transformations, ensured by the manifest covariance under soft transformations and by the presence of the collinear Wilson line in the subleading Lagrangians.\footnote{When reinstating collinear gauge transformations via $\hat{\xi}\to\hat{\chi}=W_c^\dagger\hat\xi$, one can make the gauge covariance manifest by employing soft equations of motion. This results in unphysical terms in the Lagrangian where a single collinear mode couples to multiple soft ones. These terms are proportional to the soft equations of motion and drop out in any on-shell matching calculations. For details, we refer to~\cite{Boer:2023yde}.}
Finally, note that the tree-level SCET Lagrangian is exact, i.e., it requires no hard matching: it is obtained from the full theory by the mode decomposition and multipole expansion alone, and its matching coefficient is unity to all orders in the coupling and the $\lambda$-expansion~\cite{Beneke:2002ph}. All non-trivial matching resides in the $N$-jet operator. The Lagrangian still develops the usual anomalous dimensions, and the couplings retain their full-theory running.

\subsection{$N$-jet operator in SCET}
\label{sec:QCDNJet}

The final missing ingredient is the $N$-jet operator, which encodes the hard scattering. For QCD and gravity, the operator basis has been worked out in~\cite{Beneke:2017ztn,Beneke:2018rbh,Beneke:2021aip}, and we state the important results here, following~\cite{Beneke:2021aip}.

First, note that since collinear momenta scale as $\nip\cdot p_i\sim 1$, derivatives acting on collinear fields are unsuppressed, $\np\cdot\partial\sim 1$.
Consequently, operators that differ only by powers of $\nip\cdot\partial$ appear at the same order in power counting. A local operator basis therefore contains infinitely many operators with an arbitrary number of derivatives along the $\nip$-direction at any given order in $\lambda$.
By writing this as
\begin{equation}
    \sum_k \frac{1}{k!}\biggl(\int \dd t\: t^k C(t)\biggr) (\np\Cdot\partial)^k J(x) = \int \dd t\: C(t)\spac J(x+t\s\np) \,,
\end{equation}
one can package all these operators into a single non-local operator.
The dependence on the large momentum is encoded into the light-cone displacement $t$ in the $\np^\mu$ direction and the matching coefficient $C$.

Next, there are constraints from gauge invariance. As the $N$-jet operator encodes a QCD scattering amplitude, it must be gauge invariant under full-theory gauge transformations.
With the factorized realization in SCET, see \cref{eq:QCDGaugeTrafoSoftNew}, this means that any collinear sector must be manifestly gauge invariant on its own, as the collinear transformation is purely within each sector, while the $N$-jet operator as a whole must be invariant under soft gauge transformations.
This can be realized in a simple way:\ any collinear current must be constructed from the gauge-invariant building blocks $\hat{\chi}_i,\,\hat{\cA}_i$ (see Eqs.~\eqref{eq:QCDGaugeInvBB}), which are covariant under soft transformations as
\begin{equation}
    \hat{\chi}_i\xrightarrow{\mathrm{soft}} U_s(x_-)\hat{\chi_i}\,,\qquad\text{and}\qquad \hat{\mathcal{A}} \xrightarrow{\mathrm{soft}} U^{\vphantom{\dagger}}_s(x_-)\hat{\mathcal{A}}^{\vphantom{\dagger}}_s U_s^\dagger(x_-)\,.
\end{equation}
If we place the hard scattering at the spacetime point $X=0$, then this implies $\hat{\chi}_i(t\nip)\to U_s(0)\hat{\chi}_i(t\nip)$. Any soft operator must also be soft covariant (e.g., $\nm\cdot D$ or $F_{s\mu\nu}$) and placed at $X=0$.
Then, if the $N$-jet operator as a whole forms a color singlet, it will also be soft gauge invariant, as all transformation matrices $U_s(0)$ combine into the identity operator.

Therefore, a generic $N$-jet operator, which is an element of $\mathcal{O}_{\text{$N$-jet}}$, takes the form
\begin{equation}
    \mathcal{J} = \int [\dd t]_N\: C(t_{i_1},t_{i_2},\dots)\spac J_s(0) \prod_{i=1}^N J_i(t_{i_1},t_{i_2},\dots)\,,
\end{equation}
with $[\dd t]_N = \prod_{i_k} \dd t_{i_k}$. Here, $C$ denotes the hard matching coefficient, $J_i$ the $i$-collinear operator, and $J_s$ is the soft operator.
This structure again makes manifest that collinear fields can interact within one sector and with the soft background, but an interaction between two sectors requires a hard momentum, which is integrated out.

At leading power, each $J_i$ only contains a single gauge-invariant collinear building block, 
\begin{equation}
    J_i(t_i) = \big\{\hat{\chi}^{\vphantom{\dagger}}_i(t_i\spac\nip),\,\hat{\chi}_i^\dagger(t_i\spac\nip),\,\hat{\mathcal{A}}^{\vphantom{\dagger}}_i(t_i\spac\nip)\big\}\,.
\end{equation}
Power corrections of these collinear operators can be implemented in two ways:
\begin{enumerate}
    \item[(1)] Adding derivatives such as $\partial_\perp$ or $\nm \cdot D_s$. The latter can be removed from the operator basis by applying the collinear equations of motion~\cite{Beneke:2017ztn}. Therefore, one only needs to keep track of the number of $\partial_\perp$ derivatives for these types of operators. We call these $J^{Am}$, where $m$ denotes the power-suppression.
    \item[(2)] Adding more collinear fields in the same direction. Note that $\np\cdot A_c$ is controlled to all orders by the collinear Wilson line $W_c$. Therefore, adding a collinear building block also costs powers of $\lambda$, and thus these operators are suppressed.
\end{enumerate}
There is an additional class of subleading-power operators, namely time-ordered products with the subleading Lagrangian of the form
\begin{equation}
 i\int \dd^4x\: T\Big\{J_i(t_i)\spac\mathcal{L}_i^{(n)}\Big\}\,,
\end{equation}
which can be inserted in the $N$-jet operator.

\subsection{Soft theorem from SCET}
Note that this classification of the operator basis already proves the universality of the soft theorem and its two terms at tree level:
since the covariant derivative $\nm\cdot D$ can be removed by equations of motion, the first allowed building block for the soft gluon field is $F_{s\mu\nu}\sim\lambda^4$.
Consider a non-radiative process matched to a (set of) $N$-jet operators $\mathcal{J}$. To describe soft emission from this process, one can either add an explicit building block, implying process dependence, or use Lagrangian insertions, which are universal.
As the first soft building block is possible only at $\mathcal{O}(\lambda^4)$, the first two terms in the power expansion originate from Lagrangian insertions and are thus generated at $\mathcal{O}(1)$ and $\mathcal{O}(\lambda^2)$. This implies that the leading and subleading terms are \emph{universal}, in that they do not depend on the underlying hard process.
In this way, the soft theorem follows directly from SCET.

One might worry that there is a caveat at subleading power: for the higher terms in the soft theorem, Lagrangian insertions into subleading-power $N$-jet operators are also relevant.
The reason these corrections do not change the conclusions of the previous paragraph is that operators containing insertions of $\partial_\perp$ but no additional collinear or soft building blocks are \emph{completely determined by RPI} and can therefore be related to the leading-order matching coefficient. For details, see App.~\ref{app:RPItower}.
Intuitively, one can understand this as the manifestation of Lorentz invariance of the (hard limit of the) full-theory amplitude, which connects all orders in $\lambda$ for this class of $N$-jet operators.
This completes the argument that the subleading-power terms are universal and completely determined by leading-power matching (i.e., the non-radiative amplitude) and the Lagrangian interactions.

This separation between universal Lagrangian insertions and process-dependent soft building
blocks also accounts for the ``infinite soft theorem''~\cite{Hamada:2018vrw,Li:2018gnc}, a soft theorem first derived from Ward identities of large gauge transformations and later from ordinary gauge invariance. 
While soft emissions are not universal beyond subleading order in gauge theory, the
contributions generated by Lagrangian insertions remain constrained to all orders in the soft expansion. 
As we show in~App.~\ref{app:InfiniteSoft}, the theorem follows directly from this universality. 

\section{Formulating geoSCET}
\label{sec:Geometry}
Building on the framework of QCD SCET reviewed in the previous section, we now introduce geoSCET.\footnote{SCET for scalar field theories has been introduced previously as a pedagogical tool~\cite{Becher:2014oda, Becher:2018gno, Cohen:2019wxr}.} 
The starting point is the geometric scalar field described by the non-linear sigma model Lagrangian
\begin{equation}
\label{eq:GeometricLagrangian}
    \Lagr = \frac{1}{2}g_{IJ}(\phi)\spac\partial_\mu\phi^I\spac\partial^\mu\phi^J - V(\phi)\,,
\end{equation}
where $I,J$ denote flavor indices.
The fields $\phi^I$ are interpreted as coordinates on the field-space manifold. The geometry is encoded in the metric tensor $g_{IJ}(\phi)$, and we will sometimes include a potential $V(\phi)$, which is a scalar function on the manifold.
Formulated in this way, the invariance of $S$-matrix elements under (non-derivative) field redefinitions,
\begin{equation}
\label{eq:GeoScalarSymmetry}
    \phi^I \to f(\phi)^I \,,
\end{equation}
becomes manifest because it can be shown that the $S$-matrix only depends on geometric quantities including the Riemann curvature tensor and covariant derivatives thereof~\cite{Volkov:1973vd}.

The field-redefinition invariance and the geometric formulation are crucial to the construction of the geoSCET Lagrangian.
The case of QCD as summarized in the previous section can naturally be formulated in a covariant language, such that most objects appearing in QCD SCET have direct geometric analogues for the scalar field. In fact, we will see that the construction of the EFT presented here is even simpler than for QCD.

Note that we restrict ourselves to a two-derivative theory. Higher-derivative terms, if added, must be coupled in a covariant way, i.e., through tensors on the field-space manifold to preserve the field-space geometry.
As long as this is respected, it is straightforward to extend the construction outlined here to higher-derivative theories. If the field-redefinition invariance is restricted to be polynomials without derivatives, the construction follows without modification. This is expected since it is symmetry (in this case, field redefinition invariance) and power counting that ultimately constrains the allowed form of the Lagrangian and the $N$-jet operators.

\subsection{Degrees of freedom and field redefinitions}
As explained in \cref{sec:QCDLagrangianConstruction}, it is sufficient to restrict to a single collinear sector for constructing the soft-collinear Lagrangian.
Here, we focus on the conceptual aspects of the construction. Technical details and definitions of the individual geometric objects appearing throughout this section are given in App.~\ref{sec:App:Geometry}.

We start by considering a split of the scalar field into a soft mode $\varphi$ and a collinear mode $\xi$. 
These fields have the power counting
\begin{equation}
    \xi^I\sim\lambda\,,
    \qquad\text{and}\qquad \varphi^I\sim\lambda^2\,.
\end{equation}
Under an infinitesimal transformation $\phi^I \to \phi^I + \varepsilon^I(\phi)$, where $\varepsilon^I$ is some vector field that generates the transformation, the field transforms as
\begin{align}
    \phi^I = \varphi^I+\xi^I&\to \big(\varphi^I+\xi^I\big)+\varepsilon^I(\varphi+\xi)\nn\\
    &= \varphi^I+\xi^I + \varepsilon^I(\varphi) + \xi^J \spac\partial_J \varepsilon^I(\varphi) + \sum_{n=2}^{\infty} \frac{1}{n!}\spac\xi^{J_1}\dots \xi^{J_n}\spac \partial_{J_1}\dots\partial_{J_n} \varepsilon^{I}(\varphi)\,.
\end{align}
This split extends the redefinition-invariance to transformations of the soft and collinear fields.
We assign different transformations to $\varphi^I$ and $\xi^I$ with the condition that the combination must transform like the full scalar field $\phi^I$ as in~\cref{eq:GeoScalarSymmetry}. The counterparts of~\cref{eq:QCDGaugeTrafoColl,eq:QCDGaugeTrafoSoft} are
\begin{enumerate}
    \item[i)] Collinear transformations:
    \begin{subequations}
    \label{eq:GeoCollTrans}%
    \begin{align}
        \xi^I &\xrightarrow[]{\mathrm{coll}} \xi^{I} + \sum_{n=2}^{\infty} \frac{1}{n!}\spac\xi^{J_1}\dots \xi^{J_n}\spac \partial_{J_1}\dots\partial_{J_n} \varepsilon^{I}(\varphi) \,,\\[3pt]
        \varphi^I &\xrightarrow[]{\mathrm{coll}} \varphi^I\,.
    \end{align}%
    \end{subequations}
    The soft scalar cannot transform under collinear transformations while the collinear field transforms nonlinearly.
    This is the analog of the collinear gauge transformations~\cref{eq:QCDGaugeTrafoColl}.
    \item[ii)] Soft transformations:
    \begin{subequations}
    \label{eq:GeoSoftTrans}%
    \begin{align}
        \xi^I &\xrightarrow[]{\mathrm{soft}} \bigl(\delta\indices{^I_J}+\partial_J \varepsilon^I(\varphi)\bigr)\spac\xi^J\,,\\
        \varphi^I &\xrightarrow[]{\mathrm{soft}} \varphi^I + \varepsilon^I(\varphi)\,.
    \end{align}%
    \end{subequations}%
    Here, $\varphi^I$ transforms like an ordinary scalar field with a shift $\varepsilon^I(\varphi)$, while $\xi^I$ transforms as a tangent vector with a Jacobian
    \begin{equation}
    \label{eq:InfinitesimalJacobian}
        U\indices{^I_J} \equiv \frac{\partial f^I(\varphi)}{\partial\varphi^J} = \delta\indices{^I_J}+\partial_J\spac \varepsilon^I(\varphi)\,.
    \end{equation}  
    This transformation of the collinear field is covariant in the sense that $\xi^I$ transforms like a \emph{tangent vector} defined at point $\varphi$, not as an ordinary scalar field.
\end{enumerate}

It is instructive to compare these transformations to their QCD counterparts~\cref{eq:QCDGaugeTrafoColl} and \cref{eq:QCDGaugeTrafoSoft}, respectively.
While the precise form of the transformations is of course different, conceptually we find a very similar structure.
Interestingly, we see that for the geometric scalar field theory, the scalar field itself takes a role similar to the gauge field in QCD, and the collinear field has (by construction) a soft transformation that is covariant with respect to the background $\varphi^I(x)$.

On the other hand, the collinear field does not have a simple transformation property under collinear field redefinitions. However, we will use the freedom of performing collinear field redefinitions to fix a particular field basis where the collinear field has good geometric properties, namely that it is a tangent vector. 
To this end, we decompose the scalar field into a soft (background) mode $\varphi$ and a collinear fluctuation $\xi$ using the exponential map~\cite{Howe:1986vm}
\begin{equation}\label{eq:ExponentialMapRedef}
    \phi^I = \exp_\varphi[\xi]^I\,,
\end{equation}
where $\xi^I$ is defined as a tangent vector at the base-point $\varphi^I$. This is the same decomposition that has been introduced to obtain a covariant formulation of the non-linear sigma model at loop level~\cite{Honerkamp:1971xtx,Honerkamp:1971sh}.
For our purposes, it is useful to have an explicit expression in terms of powers of $\xi$, which reads
\begin{equation}
\label{eq:perturbativeDecomp}
    \phi^I = \varphi^I + \xi^I - \frac{1}{2}\Gamma^I_{JK}(\varphi)\spac\xi^J\xi^K 
    + \mathcal{O}(\xi^3)\,,
\end{equation}
where $\Gamma^I_{JK}$ is the Christoffel symbol.
At linear order in $\xi$, this reduces to the naive decomposition $\phi^I=\varphi^I+\xi^I$,
while the non-linear corrections are enforced by the geometry.
These additional terms modify the Lagrangian to yield a manifestly covariant form with respect to the background $\varphi^I$. This decomposition is analogous to fixing a collinear gauge for gauge theories.

\subsection{geoSCET Lagrangian}
\label{sec:GeoSCETLagrangian}

To construct the effective Lagrangian, one can then follow the same steps (i)-(iv) introduced in~\cref{sec:QCDLagrangianConstruction}.
To start step (i), one simply inserts the covariant field decomposition~\cref{eq:perturbativeDecomp} into the Lagrangian~\cref{eq:GeometricLagrangian} and derives the background-field Lagrangian.
For the potential terms, one finds
\begin{equation}
    V(\phi) = V(\varphi) + \xi^I [\partial_I V](\varphi) + \frac{1}{2}\xi^I \xi^J [\partial_I\partial_J V](\varphi)
    - \frac{1}{2}\Gamma^I_{JK}\spac(\varphi)\xi^J\xi^K [\partial_I V](\varphi)
    + \mathcal{O}(\xi^3)\,,
\end{equation}
where the square bracket indicates that one first takes the derivative and then evaluates the potential at $\varphi^I$. For brevity, we set up the following notation:
whenever a geometric object without explicit argument appears, it is evaluated at $\varphi^I$, e.g.\ $V(\varphi) \equiv V$.
The linear term can be trivially written as $\partial_I \to \nabla_I$ as the covariant derivative of a scalar field is trivial, while the two quadratic terms combine as
\begin{equation}
    \frac{1}{2}\xi^I \xi^J [\partial_I\partial_J V]
    - \frac{1}{2}\Gamma^I_{JK}\spac\xi^J\xi^K [\partial_I V] = \frac{1}{2}\xi^I \xi^J [\nabla_I\nabla_JV]\,,
\end{equation}
where we use the standard definition
\begin{equation}
    \nabla_I V_J = \partial_I V_J - \Gamma^K_{IJ} V_K
\end{equation}
for the covariant derivative of a covector.
This result demonstrates the utility of the exponential map basis choice in~\cref{eq:ExponentialMapRedef}, as it leads to geometric structures that are manifestly covariant with respect to the soft field $\varphi^I$.
Indeed, one can show that to all orders, the potential can simply be written as
\begin{equation}
    V(\phi)
    = \big[\exp(\xi^I\nabla_I)V\big](\varphi)\,.
\end{equation}
The kinetic term is more involved, and the details can be found in App.~\ref{sec:app:BGExpansion}.  

Written as a power series in $\xi$, the Lagrangian then takes the form
\begin{equation}\label{eq:FullLagrangian}
    \mathcal{L} = \mathcal{L}_{\xi^0} + \mathcal{L}_{\xi^1} + \mathcal{L}_{\xi^2} + \mathcal{O}(\xi^3)\,,
\end{equation}
where the individual terms read
\begin{subequations}
\begin{align}
    \mathcal{L}_{\xi^0} &= \frac{1}{2}g_{IJ}(\varphi)\spac\partial_\mu\varphi^I \partial^\mu\varphi^J - V(\varphi)\,,\\[3pt]
    \mathcal{L}_{\xi^1} &= \xi^J\bigl(-g_{IJ}(\varphi)\spac D_{\mu}\partial^{\mu}\varphi^I - \partial_J V(\varphi)\bigr)\,,\\[3pt]
    \mathcal{L}_{\xi^2} &= \frac{1}{2}g_{IJ}(\varphi)\spac D_\mu\xi^I D^\mu \xi^J - \frac{1}{2}R_{IKJL}(\varphi)\spac \partial_\mu\varphi^K \partial^\mu\varphi^L \xi^I\xi^J - \frac{1}{2}[\nabla_I\nabla_JV](\varphi)\spac\xi^I\xi^J\,.
\end{align}
\end{subequations}
Here, we find a set of geometric objects such as the usual field-space covariant derivative~$\nabla$ and the Riemann tensor $R_{IKJL} = g_{IM}R\indices{^M_{KJL}}$, with
    \begin{equation}
    R\indices{^I_{KJL}} = \partial^{\vphantom{I}}_J \Gamma^I_{KL} - \partial^{\vphantom{I}}_L \Gamma^I_{KJ} + \Gamma^I_{JM}\Gamma^M_{KL} - \Gamma^I_{LM}\Gamma^M_{KJ} \,.
\end{equation}
In addition, we define the emergent covariant derivative $D_\mu$ as
\begin{equation}
    D_\mu\xi^I \equiv \partial_\mu \xi^I + \Gamma^I_{KL} \spac\partial_\mu\varphi^K \xi^L\,.
\end{equation}
The covariant derivative preserves the transformation property of $\xi^{I}$ under soft field redefinitions:
\begin{equation}
    D_{\mu}\xi^I(x) \to \frac{\partial\varphi^{\prime I}}{\partial\varphi^J}(x)\spac D_{\mu}\xi^J(x) \equiv U\indices{^I_J}(x)\spac D_{\mu}\xi^J(x)\,.
\end{equation}
A few remarks are in order:
\begin{itemize}
    \item[(a)] The term $\mathcal{L}_{\xi^0}$ describes the soft dynamics and is just the full-theory Lagrangian expressed purely in terms of the soft field $\varphi^I$. This is a generic feature of the SCET construction.
    \item[(b)] The term $\mathcal{L}_{\xi^1}$ describes unphysical processes where a collinear field decays into purely soft fields. Such vertices are not allowed by momentum conservation and should be discarded.
    Integration by parts reveals that this Lagrangian is proportional to the soft equation of motion, so dropping it can also be justified this way.
    \item[(c)] The Lagrangian $\Lagr_{\xi^2}$ as well as all higher terms, which contain the dynamics of the fluctuation field, are entirely formulated in terms of objects that are manifestly covariant with respect to the background geometry encoded by $g_{IJ}(\varphi)$. This makes the invariance under the redefinitions outlined in~Eqs.~\eqref{eq:GeoCollTrans} and \eqref{eq:GeoSoftTrans} manifest term by term.
    \item[(d)] Despite starting only with a single scalar field $\phi^I$, the effective description now contains a soft-covariant derivative 
    \begin{equation}
    \label{eq:GeoCovariantDerivative}
    (D_\mu)\indices{^I_{J}} = \delta\indices{^I_J}\spac\partial_\mu + \Gamma^I_{JK}\spac\partial_\mu\varphi^K \equiv \delta\indices{^I_J}\spac\partial_\mu + (A_\mu)\indices{^I_{J}}\,.
    \end{equation}
    Geometrically, this is the pullback covariant derivative which arises from the pullback of the connection $\nabla$ defined on the field-space manifold to the space-time manifold.
\end{itemize}
Interestingly, the low-energy effective theory features a connection $A_\mu$, just as in the gauge-theory case, and one can therefore follow the same manipulations and geometric intuition as used in \cref{sec:QCDLagrangianConstruction} for the soft gluon $A_s$.
However, unlike in gauge theory, the connection field that appears here scales like $(A_\mu)\indices{^I_{J}}\sim\lambda^4$, so its interactions are power-suppressed. 
In the case of a vanishing potential $V=0$, this will underlie the absence of a leading-power term in the soft theorem since soft emissions in this theory are power-suppressed in $\lambda$.
This also directly implies that there are no soft divergences at leading power in this theory. 
In normal coordinates, the collinear self-interaction starts at $\mathcal{O}(\lambda^2)$, and also collinear divergences are absent at leading power.
Therefore, we can immediately conclude that \emph{the geometric scalar EFT (with $V=0$) is IR finite},\footnote{The equivalence between power-suppressed interactions and IR finiteness holds only in theories without relevant interactions, i.e., in theories where all couplings have non-positive mass dimension. We show below that a counterexample is the cubic coupling in four dimensions, which has positive mass dimension and is IR divergent.} which to our knowledge had never previously been demonstrated.

To summarize, we obtain the manifestly covariant Lagrangian
\begin{equation}\label{eq:CovariantBGLagrangian}
    \mathcal{L} = \mathcal{L}_{\varphi} + \mathcal{L}_\xi\,,
\end{equation}
where
\begin{subequations}
\begin{align}
    \mathcal{L}_\varphi &= \frac{1}{2}g_{IJ}\spac\partial_\mu\varphi^I\spac\partial^\mu\varphi^J - V \,,\\[3pt]
    \mathcal{L}_\xi &= \frac{1}{2}g_{IJ}\spac D_\mu\xi^I D^\mu\xi^J - \frac{1}{2}\Bigl(R_{IKJL}\spac\partial_\mu\varphi^K\spac\partial^\mu\varphi^L + [\nabla_I\nabla_JV]\Bigr)\xi^I \xi^J + \mathcal{O}(\xi^3) \,.
    \label{eq:GeoCollLagrangianPreExp}
\end{align}
\end{subequations}
Note that this Lagrangian is just the ordinary background-field Lagrangian as, e.g., derived for the non-linear sigma model, expanded to second order in $\xi^I$. 

Next, we implement steps (ii) and (iii) simultaneously. First, we perform the multipole expansion of the soft field $\varphi^I(x)$ around $x_-$. Specifically, we expand all soft-collinear interactions using
\begin{equation}
    \varphi^I(x) = \varphi^I(x_-) + (x-x_-)^\mu\spac\big[\partial_\mu\varphi^I\big](x_-) + \dots\,.
\end{equation}
As explained below~\cref{eq:QCDGaugeMultipoleExpansion}, we also need to redefine the collinear fields. Currently, they transform under soft field redefinitions as $\xi^I(x)\to (U_s(x))\indices{^I_J}\spac\xi^J(x)$, which mixes different orders in the multipole expansion. We therefore proceed just as in QCD and redefine the collinear fields $\xi^I \to \hat{\xi}^I$, such that the hatted collinear fields transform with $U_s(x_-)$.
In QCD, the two fields are related by the $\mathcal{R}$ Wilson line~\cref{eq:QCDRWilsonLine}, which moves the soft gauge transformation from point $x_-$ to $x$ along a straight line. It is natural to adapt this to the geometric EFT as
\begin{align}
    \bigl(\mathcal{R}(x,x_-)\bigr)\indices{^I_J} &= \mathcal{P}\exp \biggl(-\int_{0}^1 \dd s\:(x-x_-)^\mu \big(A_\mu(y(s))\big)\indices{^I_J}\biggr)\nn\\
    &= \mathcal{P}\exp \biggl(-\int_{0}^1 \dd s\: (x-x_-)^\mu\spac \Gamma^I_{KJ}\bigl(\varphi(y(s))\bigr)\spac\partial_\mu\varphi^K\bigl(y(s)\bigr)\biggr)\,,
\end{align}
where $y(s)=x_-+s(x-x_-)$.\footnote{As in QCD, this $\mathcal{R}$ Wilson line can be expanded in $\lambda$ as $(x-x_-)^\mu \spac\partial_\mu\varphi^I\sim \lambda^3$.}
Since $A_\mu$ is a bona fide connection with the usual transformation, the $\mathcal{R}$ Wilson line has the same properties as its QCD counterpart, only lifted to the geometric setting.
In particular, it transforms under soft gauge transformations as~\cref{eq:QCDRWilsonTransformationSoft}
\begin{equation}
    \bigl(\mathcal{R}(x,x_-)\bigr)\indices{^I_J} \to \bigl(U(x)\spac \mathcal{R}(x,x_-) \spac U^{-1}(x_-)\bigr)\indices{^I_J}\,,
\end{equation}
and it can be used to relate hatted and unhatted fields in complete analogy to~\cref{eq:QCDCollRedef} as
\begin{equation}
\label{eq:GeoCollRedef}
    \xi^I(x) = \bigl(\mathcal{R}(x,x_-)\bigr)\indices{^I_J}\spac\hat{\xi}^J(x)\,.
\end{equation}
We conclude that $\hat{\xi}^I$ transforms with the homogeneous $U^I_{\phantom{I}J}(x_-)$ as desired.

To complete steps (ii) and (iii), we simply insert the redefinition~\cref{eq:GeoCollRedef} into the (multipole-expanded) collinear Lagrangian~\cref{eq:GeoCollLagrangianPreExp}. Instead of simply expanding everything in $\lambda$, it is useful to express the Lagrangian in terms of the \emph{emergent homogeneous background} contained in $\nm\cdot D(x_-)\frac{\np^\mu}{2}$.

To understand the benefit of organizing the theory in this way, we again appeal to the analog with QCD.
In QCD, the gauge transformations are purely internal, and the generators have no power-counting, i.e.,\ $t^a\sim 1$. 
Therefore, all inhomogeneities in $\lambda$ originate purely from the multipole expansion of the gauge fields, and after the redefinition using the $\mathcal{R}$ Wilson line, the residual Lagrangian is homogeneous term-by-term, and the remaining transformation is $U_s(x_-)$, with the corresponding homogeneous background field $\nm\cdot A_s(x_-)\frac{\np^\mu}{2}$.
This naive picture of the emergent background being homogeneous in $\lambda$ breaks as soon as the generators of the gauge transformation have non-trivial power-counting in $\lambda$.
This happens, e.g., for gravity~\cite{Beneke:2021aip}, and it also happens here for the geometric scalar field, since the soft Jacobian power counts as $U\indices{^I_J} = \delta\indices{^I_J} + \mathcal{O}(\lambda^2)$.
Therefore, the emergent background field is the component that corresponds to the subset of soft transformations that leaves the multipole expansion invariant, i.e., to the residual gauge transformation after parallel-transporting the collinear fields from $x$ to $x_-$.
However, as the gauge transformations are inherently inhomogeneous in $\lambda$, there is now always a trade-off:\ one can either work manifestly covariantly, but then the result is inhomogeneous in $\lambda$, as each term generates an infinite tower of subleading-power corrections. Or, one can expand in $\lambda$, then each term is homogeneous in power counting, but gauge transformations connect terms at different orders in power counting in order to form manifestly covariant objects (covariant derivatives, Riemann tensors, etc.).
For the construction of the Lagrangian, it is advantageous to use the first perspective, as one can leverage the symmetry of the low-energy EFT. We are now ready to proceed with the Lagrangian construction.

\vspace{10pt}
\noindent \textbf{Theories with} $\bm{V=0}$\textbf{:} 
We first discuss the case without a potential.
To see how the $\mathcal{R}$ Wilson line dresses the fields, first consider the insertion of the redefinition~\cref{eq:GeoCollRedef} into the covariant derivative. 
It can be written as\footnote{For brevity, we suppress the arguments $x,x_-$ of the $\mathcal{R}$ Wilson line.}
\begin{align}
    (D_\mu)\indices{^I_J}\spac\xi^J &= (D_\mu)\indices{^I_J}\spac \mathcal{R}\indices{^J_K}\spac\hat{\xi}^K\nn\\
    &= \mathcal{R}\indices{^I_A}\spac \mathcal{R}\indices{_B^A}\spac (D_\mu)\indices{^B_J}\spac \mathcal{R}\indices{^J_K}\spac\hat{\xi}^K\nn\\
    &= \mathcal{R}\indices{^I_A}\Bigl( (\hat{D}_\mu)\indices{^A_K} + \bigl(\mathcal{R}\indices{_B^A}\spac (D_\mu)\indices{^B_J}\spac \mathcal{R}\indices{^J_K} - (\hat{D}_\mu)\indices{^A_K} \bigr)
    \Bigr)\hat{\xi}^K\,,
\end{align}
where $\mathcal{R}\indices{^I_A}\spac \mathcal{R}\indices{_J^A} \equiv \delta\indices{^I_J},$ and we split off the emergent homogeneous derivative
\begin{equation}
    (\hat{D}^\mu)\indices{^I_J} = \delta\indices{^I_J}\spac \partial^\mu + \bigl(\nm\Cdot A(x_-)\bigr)\indices{^I_J}\frac{\np^\mu}{2}\,.
\end{equation}
The combination 
\begin{equation}
\bigl(\mathcal{A}_{s\spac\mu}(x)\bigr)\indices{^A_K} \equiv \mathcal{R}\indices{_B^A}\spac (D_\mu)\indices{^B_J}\spac \mathcal{R}\indices{^J_K} - (\hat{D}_\mu)\indices{^A_K}
\end{equation}
encodes the additional subleading terms of the covariant derivative and is covariant under soft transformations.\footnote{This can be seen by explicit computation, or by noticing that it is a difference of two connections.} It corresponds to the constrained components of the connection in fixed-line gauge and satisfies 
\begin{equation}
    (x-x_-)^\mu \mathcal{A}_{s,\mu}(x) = 0\,.
\end{equation}
In QCD SCET, this object has a corresponding counterpart~\cite{Beneke:2021aip}, which can be expanded in terms of field-strength tensors. The soft-collinear Lagrangian then takes the form 
\begin{align}
\label{eq:GeoLagrangianRInserted}
    \mathcal{L}_\xi &= \frac{1}{2}g_{IJ}(x)\spac \R{I}{A}\R{J}{B}
    \Bigl( (\hat{D}_\mu)\indices{^A_K} + (\mathcal{A}_s(x))\indices{^A_K}
    \Bigr)\hat{\xi}^K
    \Bigl( (\hat{D}^\mu)\indices{^B_L} + (\mathcal{A}_s(x))\indices{^B_L}\Bigr)\hat{\xi}^L\notag\\
    &\quad- \frac{1}{2} \R{I}{A}\R{J}{B}\bigr(R_{IKJL}(x)\spac\partial_\mu\varphi^K(x) \spac\partial^\mu\varphi^L(x)\bigr)\hat{\xi}^A \hat{\xi}^B + \mathcal{O}(\hat{\xi}^3)\,.
\end{align}

The next step is to deal with the dressed soft objects appearing in the Lagrangian, which are still evaluated at $x$.
As in QCD, one can make use of fixed-line gauge to facilitate these expansions. The $\mathcal{R}$ Wilson line is again the gauge transformation that moves a soft configuration to this gauge.
Therefore, the dressed soft objects appearing in~\cref{eq:GeoLagrangianRInserted} are simply the respective objects evaluated in fixed-line gauge.
One can again make use of the identities in~Eqs.~\eqref{eq:QCDFLIdentity}, and ``unfix'' the gauge condition using $F^{\mu\nu}(y(s))\to \mathcal{R}^\dagger(y(s))F^{\mu\nu}(y(s))\mathcal{R}(y(s))$ in the integral. One can evaluate the integral and obtain an infinite power series~\cite{Beneke:2002ni},\footnote{The results obtained therein only depend on the fixed-line gauge condition, and are thus valid not just for QCD, but also for the geometric scalar field.} e.g.,
\begin{equation}
    \nm\cdot\mathcal{A}_s(x) = x_\perp^\mu \nm^\nu F_{\mu\nu} + \frac{1}{2}\nm\Cdot x\,\np^\mu \spac\nm^\nu\spac F_{\mu\nu} + \mathcal{O}(\lambda^9)\,.
\end{equation}
Here, the field-strength tensor is the pullback of the Riemann tensor
\begin{equation}
    (F_{\mu\nu})\indices{^I_J} = [D_\mu,D_\nu]\indices{^I_J} = R^I_{\phantom{I}JKL}\spac\partial_\mu\varphi^K \spac\partial_\nu\varphi^L\,.
\end{equation}

We can then express the Lagrangian as
\begin{equation}
    \Lagr = \frac{1}{2} g_{IJ}(\varphi)\spac \np\Cdot\partial\hat{\xi}^I\spac\nm\Cdot \hat{D}\hat{\xi}^J + \frac{1}{2}g_{IJ}(\varphi)\spac\partial_{\mu_\perp}\hat{\xi}^I \partial^{\mu_\perp}\hat{\xi}^J + \mathcal{O}(\lambda^5)\,,
\end{equation}
where we adopted the new notation that soft geometric objects are evaluated at $\varphi^I(x_-)$, and soft fields are evaluated at $x_-$.
Recall that $\nm\cdot \hat{D} = \nm\cdot\partial + \nm\cdot A\sim \lambda^2 + \lambda^4$, as the connection field itself scales as $\lambda^4$. Therefore, the would-be terms of $\mathcal{L}^{(2)}$ are contained in the covariant derivative of $\mathcal{L}^{(0)}$.
This demonstrates that soft-collinear interactions are power suppressed.
It is interesting to observe that the multipole corrections do not begin at $\mathcal{O}(\lambda^3)$, as one would naively expect, but at $\mathcal{O}(\lambda^5)$, consistent with the scaling of the field-strength tensor. Also, this multipole correction is a quartic interaction proportional to $\varphi^2\hat{\xi}^2$, consistent with the observation that the cubic interactions vanish in normal coordinates ($\Gamma^I_{JK}=0$).
If one were to determine the cubic terms in $\hat{\xi}$, one would also find that the corresponding purely-collinear interaction terms are power suppressed.

For practical calculations, it is more convenient to perform a full $\lambda$-expansion of all objects, and to expand the geometric structures around the vacuum expectation value (vev) $v$ through 
\begin{equation}
\varphi(x_-) = v+\varphi_s(x_-)\,.
\label{eq:varphiAboutVev}
\end{equation} 
This way, every term in the Lagrangian has a definite $\lambda$-counting and a definite multiplicity in soft fields $\varphi_s$.
We then find for $V=0$
\begin{subequations}
\begin{align}
    \mathcal{L}^{(0)} &= \frac{1}{2}g_{IJ}\spac\np\Cdot\partial\hat{\xi}^I\spac \nm\Cdot\partial\hat{\xi}^J + \frac{1}{2}g_{IJ}\spac\partial_{\mu_\perp}\hat
    \xi^I\partial^{\mu_\perp}\hat{\xi}^J\,,\\
    \mathcal{L}^{(2)} &=  \Gamma_{IAJ}\Bigl( \varphi_s^A  \spac\partial_\mu \hat{\xi}^I \partial^\mu \hat{\xi}^J + \frac{1}{2}\nm\Cdot\partial\varphi_s^A\spac \hat{\xi}^J\np\Cdot\partial\hat{\xi}^I\Bigr)\,,
\end{align}
\end{subequations}
where now all geometric objects are evaluated at $v$.
Here, we introduce the Christoffel symbol with all lower indices as
\begin{equation}
    \Gamma_{IAJ} = g_{IK}\Gamma^K_{AJ}\,,
\end{equation}
which is symmetric in the last two indices.

As this manifestly homogeneous form becomes quite cumbersome, it is useful to choose a convenient coordinate system, specifically normal coordinates defined around the vev $v$, such that $\Gamma^I_{JK}(v)=0$.\footnote{With this choice, $\mathcal{L}^{(2)}$ would identically vanish. This is completely consistent, and we comment on this below when we derive the soft theorem.} 
For the double soft theorem, we need to extend the construction to $\mathcal{L}^{(4)}$. In normal coordinates, one obtains
\begin{align}
    \mathcal{L}^{(4)}&= \frac{1}{2}R_{ABIJ}\spac \varphi_s^A \partial_\mu\varphi_s^B\spac\hat{\xi}^J\partial^\mu\hat{\xi}^I \,.
    \label{eq:geoSCETL4}
\end{align}

We restrict the construction to the selected terms in $\mathcal{L}^{(2)}$ and $\mathcal{L}^{(4)}$ that will be relevant for the soft theorem derivation below. The construction can be performed in full generality to any desired order in $\lambda$.

\vspace{10pt}
\noindent \textbf{Theories with} $\bm{V\neq 0}$\textbf{:} 
If a non-vanishing potential is present, it generates an additional tower of terms 
\begin{equation}
    \mathcal{L}_V = \sum_{n=2}\sum_{m=0} \mathcal{L}_{V,\xi^n}^{(m)}\,.
\end{equation}
For our purposes here, we only need the terms bilinear in $\hat{\xi}$ up to $\mathcal{O}(\lambda^2)$:
\begin{subequations}
\begin{align}
    \mathcal{L}_V^{(0)} &= - \frac{1}{2}\big[\nabla_I\nabla_J V\big]\hat{\xi}^I\spac\hat{\xi}^J\,,\\[3pt]
    \mathcal{L}_V^{(1)} &= -\frac{1}{2} \big[x_\perp\Cdot\partial \varphi^K\big] \big[\nabla_K\nabla_I\nabla_J V\big]\hat{\xi}^I\hat{\xi}^J\,,\\[3pt]
    \mathcal{L}_V^{(2)} &= -\frac{1}{4} \big[\nm\Cdot x\, \np\Cdot\partial \varphi^K\big] \big[\nabla_K\nabla_I\nabla_J V\big]\hat{\xi}^I\hat{\xi}^J
    - \frac{1}{4} x_\perp^\mu x_\perp^\nu \big[\hat{D}_\mu \partial_\nu \varphi^K\big]\big[\nabla_K\nabla_I\nabla_J V\big]\hat{\xi}^I\hat{\xi}^J
    \,.\label{eq:L2VUnexpanded}
\end{align}
\end{subequations}
Note one subtlety for the potential terms: the interactions $\phi^2$ and $\phi^3$ in $V(\phi)$ have coupling constants $m^2$ and $g_3$, with mass dimension 2 and 1, respectively.
Their $\lambda$-scaling is not fixed by the EFT but is a physical input, determined by their size relative to the hard scale $Q$.
The power counting that is consistent with SCET$_{\mathrm{I}}$ is $m^2\sim\lambda^4Q^2$ for the bilinear and $g_3\sim\lambda^2Q$ for the cubic, so that the purely-soft Lagrangian has no superleading interactions.
The two scalings are not independent and can be understood as follows: 
a massless scalar with cubic self-coupling radiatively generates $m^2\sim g_3^2$. The assignment $m^2\sim\lambda^4 Q^2$ is thus radiatively stable only for $g_3\lesssim \lambda^2 Q$, which enforces the above scaling.\footnote{If the couplings are larger, one is naturally pushed into SCET$_{\mathrm{II}}$-territory, where $m^2\sim \lambda^2 Q^2$, $g_3\sim \lambda Q$ is consistent.
Larger couplings $g_3\sim Q$ imply $m\sim Q$ and an HQET-like description.
Retaining $g_3\sim Q$ within SCET$_{\mathrm{I}}$ is not an alternative power counting, but the statement that no light modes exist: with $m^2\sim g_3^2\sim Q^2$ the would-be soft and collinear modes are heavier than their own virtualities. We return to this issue in~\cref{sec:loops}, where the same statement manifests as a $\lambda^{-2}$ enhancement per collinear loop.
}
In the above Lagrangian, this was not yet taken into account; we power-counted relative to the leading term. In the expansion $\varphi=v+\varphi_s$, we account for the coupling as well, as seen below.
This means that the purely-soft theory has leading-power interactions, but interactions between collinear fields as well as soft-collinear interactions are power-suppressed when one accounts for the cubic coupling $g_3$ inside $\nabla^3V$.

For the theory with a non-vanishing potential, we focus on the cubic interactions with two collinear and one soft field, which read
\begin{subequations}
\begin{align}
    \mathcal{L}_V^{(2)} &= -\frac{1}{2}\big[\nabla_A\nabla_I\nabla_J V\big]\varphi_s^A\hat{\xi}^I \hat{\xi}^J\,,\\
    \mathcal{L}_V^{(3)} &= -\frac{1}{2}\big[\nabla_A\nabla_I\nabla_J V\big]\big[x_\perp\Cdot\partial\varphi_s^A\big]\hat{\xi}^I \hat{\xi}^J\,,\\
    \mathcal{L}_V^{(4)} &= -\frac{1}{4}\big[\nabla_A\nabla_I\nabla_J V\big]\big[\nm\Cdot x\,\np\Cdot\partial\varphi_s^A\big]\hat{\xi}^I \hat{\xi}^J
    -\frac{1}{4}\big[\nabla_A\nabla_I\nabla_J V\big]\big[x_\perp^\mu x_\perp^\nu\partial_\mu\partial_\nu\varphi_s^A\big]\hat{\xi}^I \hat{\xi}^J\,.
\end{align}
\end{subequations}
These interactions will be relevant for the geometric soft theorem for scalar theories with a non-zero potential.
Again, we stress that, like in QCD, the SCET Lagrangian for the geometric scalar field is not renormalized in the sense that all appearing couplings retain their full-theory evolution, and the tree-level form is exact. In particular, if the metric tensor $g_{IJ}(v)$ or related quantities evolve according to the renormalization group, this will be identical in SCET.

\subsection{$N$-jet operator in geoSCET}
\label{sec:geoNJet}
We now turn to the $N$-jet operator structure, following in close analogy with scalar QCD. A generic $N$-jet operator describing a hard scattering at $X=0$ takes the form
\begin{equation}
    \mathcal{J} = \int [\dd t]_N\: C(\{t_{i_k}\})\spac J_s(0) \prod_{i=1}^N J_i(t_{i_1},t_{i_2},\dots)\,,
    \label{eq:NjetgeoSCET}
\end{equation}
where $C$ is the hard matching coefficient, $J_s$ is a soft operator, and $J_i$ are the collinear operators associated with directions $\nip$.
The rules are the same as outlined in Sec.~\ref{sec:QCDNJet}: all building blocks must be covariant under soft gauge transformations, i.e., background-field redefinitions.
Therefore, $\xi^I$ are good building blocks, as they are tangent vectors, while soft fields $\varphi^I$ are not. Soft covariant objects, such as $R_{IJKL}$ or $\partial_\mu\varphi^I$ are suitable building blocks, but they always contain derivatives of the soft field and are thus power-suppressed.

Let us point out an important difference between gauge theory and geoSCET. In QCD SCET, the fixed-line gauge condition implies $\np\cdot A_s(x_-)=A_{s\perp}(x_-)=0$, so only $\nm\cdot D_s(x_-)$ and $F_{s\mu\nu}(x_-)$ as well as their covariant derivatives are soft operators that transform covariant under soft gauge transformations, e.g.,
\begin{equation}
    F_{s\mu\nu}(0)\xrightarrow{\mathrm{soft}} U_s(0)F_{s\mu\nu}(0)U_s^\dagger(0)\,.
\end{equation}
As described above, one can use the collinear equations of motion to eliminate $\nm\cdot D$ in favor of collinear building blocks. Therefore, the power counting for the first possible soft gluon building block is $F_{s\mu\nu}\sim\lambda^4$. This implies that there are two universal terms that govern the leading soft emissions, corresponding to Lagrangian insertions at $\mathcal{O}(\lambda^0)$ and $\mathcal{O}(\lambda^2)$.

For the geometric scalar, the situation is slightly different. While the same approach holds for the covariant derivative $\nm\cdot D_s$ and Riemann tensors, we can additionally construct a covariant soft building block simply from $\partial_\mu\varphi_s^I$, which transforms like a tensor field.
Therefore, we only have two universal terms in the soft theorem since $\partial_\mu\varphi_s^I\sim\lambda^4$ can be inserted in the $N$-jet operator, while $(F_{s\mu\nu})\indices{^I_J}\sim\lambda^8$, so its insertions are further power suppressed.

In addition, specific to the geometric scalar field, every object appearing inside the $N$-jet operator has a dependence on the background field $\varphi$.
For actual calculations, we perform the expansion around the vev $\varphi = v + \varphi_s$, as in \cref{eq:varphiAboutVev}.
This split is ambiguous, as one can shift
\begin{equation}
    v \to v' = v+a\,,
    \qquad\text{and}\qquad 
    \varphi_s \to \varphi_s - a\,.
    \label{eq:geoRPI}
\end{equation}
The $N$-jet operator must be invariant under this shift. The collinear fluctuations are defined with respect to base-point $v+\varphi_s$, so they do not transform under \cref{eq:geoRPI}. Likewise, any soft-covariant object is invariant, as it depends on derivatives of $\varphi$.
However, the matching coefficient has the form
\begin{equation}
    C(\varphi) = C(v + \varphi_s)\,,
\end{equation}
and performing the expansion in the soft field, one sees that there are an infinite number of subleading terms of higher soft-particle multiplicity that are related to the leading-order hard matching coefficient, as
\begin{equation}\label{eq:RPI}
    C(v+\varphi_s) = C(v) + \varphi_s^A [\partial_A C](v) +\frac{1}{2}\varphi_s^A\varphi_s^B[\partial_A\partial_BC] (v)+\dots\,.
\end{equation}
This relation will be important below, as this precisely leads to the partial-derivative terms of the geometric soft theorem.
This is similar to the RPI relations that we use to relate subleading terms in the non-radiative amplitude to the leading-power one. Here, however, the shift invariance relates coefficients of \emph{different multiplicity} to the non-radiative one.
Combining both constraints allows us to completely determine the soft theorem, as we explain below.

\subsection{Soft decoupling}
We now discuss aspects of decoupling the soft modes for the scalar EFTs without and with a potential.

\vspace{10pt}
\noindent \textbf{Theories with} $\bm{V= 0}$\textbf{:}
One interesting feature of geoSCET in the case of a vanishing potential is that there are no leading-power interactions. The first interaction appears at $\mathcal{L}^{(2)}$ mediated by the covariant derivative.
However, just as in QCD, this interaction corresponds to the eikonal limit of the soft emission (but with an additional power of soft momentum $k_s$ in the numerator), and thus we can construct a field redefinition analogous to the soft-decoupling in QCD.
It takes the standard form
\begin{equation}
    \bigl(S_n(x)\bigr)\indices{^I_J} = \mathcal{P}\exp\biggl( \int_{-\infty}^0 \dd s\: \nm\Cdot A(x+s\nm) \biggr)\indices{^I_J}\,,
\end{equation}
with $(A_\mu)\indices{^I_J} = \Gamma^I_{JK}\spac\partial_\mu\varphi^K$, and satisfies the relation
\begin{equation}
    i\nm\Cdot D\spac(S_n(x_-)\hat{\xi}) = S_n(x_-) i\nm\Cdot\partial \hat{\xi}\,.
\end{equation}
Therefore, the soft-collinear interactions described by the emergent covariant derivative can be removed from the Lagrangian and placed inside the $N$-jet operator, by dressing all collinear fields with soft Wilson lines accordingly.
Conceptually, this is precisely the same feature as in QCD or gravity. However, in practice, these interactions are already power suppressed as $A_s\sim\lambda^4$ for the geometric scalar field. Therefore, there are no soft-collinear interactions at leading power to begin with. The decoupling transformation simply allows one to also remove the interactions at $\mathcal{O}(\lambda^2)$ from the Lagrangian.
Unpacking the Wilson line, one has at leading power
\begin{equation}
    S_n = \delta\indices{^I_J} + \frac{\nm\Cdot A\indices{^I_J}}{\nm\Cdot\partial} + \mathcal{O}(\lambda^4)\,,
\end{equation}
and inserting $\nm\cdot A = \Gamma^I_{KJ}\spac\nm\cdot\partial\varphi^K$ results in
\begin{equation}
    S_n = \delta\indices{^I_J} + \Gamma^I_{KJ}\spac\varphi^K + \mathcal{O}(\lambda^4)\,.
\end{equation}
Note that this Wilson line is equivalent to fixing normal coordinates in the residual geometry. 

\vspace{10pt}
\noindent \textbf{Theories with} $\bm{V\neq 0}$\textbf{:}
In the case of theories with a non-zero potential, the cubic soft-collinear interactions are mediated through 
\begin{equation}
    \mathcal{L}^{(2)} = -\frac{1}{2}\big[\nabla_A\nabla_I\nabla_J V\big]\varphi_s^A\spac\hat{\xi}^I\spac\hat{\xi}^J\,.
\end{equation}
These interactions are not placed inside a covariant derivative, and there is no immediate decoupling transformation that can remove these from the Lagrangian. Aspects of non-decoupling for scalar fields have been studied previously in~\cite{Biswas:2022lsj}.

\section{Soft theorems from geoSCET}
\label{sec:SoftTh}
In this section, we apply the formalism developed above to present a new derivation of the geometric soft theorem using geoSCET. While the existence immediately follows from the operator basis and power counting, the actual form of the soft function must be calculated from the Lagrangian insertions. Note that for QCD and gravity, the interactions in the Lagrangian can be rewritten in a form to manifestly yield the soft theorem~\cite{Beneke:2021umj}. This also holds for the geometric scalar field using the same procedure and makes the calculation straightforward.

In the following, we first perform the non-radiative matching, which determines the hard matching coefficient. 
One can do this quite generically since the details of the coefficient are not important; we only care about the form of the relevant $N$-jet operators.
Next, we consider soft emissions via Lagrangian insertions, performing explicit calculations for single and double soft emissions. This results in the single and double soft theorems.
Finally, we discuss loop corrections to these soft theorems.
For ease of notation, we drop all hats on collinear fields, as is standard in SCET.

\subsection{Non-radiative matching}
We begin with the matching relation between the non-radiative amplitude $\mathcal{M}_N$ appearing in the soft theorem \cref{eq:softThm} and the corresponding $N$-jet operators in SCET.
The final results of this section are conceptually straightforward, even though the explicit derivation is quite intricate and therefore somewhat obtuse.
The complexity reflects the fact that several ingredients must be organized consistently within the EFT expansion. This is somewhat akin to the momentum-conservation discussions in the original derivations of the LBK amplitude~\cite{Low:1958sn,Burnett:1967km}.
The final outcome is that the matching coefficients of the relevant $N$-jet operators simply correspond to the hard, non-radiative amplitude order by order in the $\lambda$-expansion.
A more comprehensive discussion of the soft-theorem derivation from position-space SCET is presented in~\cite{Beneke:2021umj}, and we closely follow this reference.

\begin{figure}[t!]
    \centering
    \includegraphics[scale=0.9]{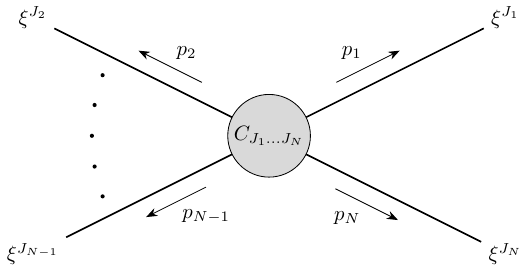}
    \caption{A diagram depicting the non-radiative matching for an $N$-point amplitude. The $\xi^J_1,\dots,\xi^{J_N}$ denote the external collinear building blocks in different collinear sectors with momenta $p_1,\dots, p_N$ (outgoing), and $C_{J_1\dots J_N}$ is the hard matching coefficient.}
    \label{fig:NonRad}
\end{figure}

To begin, consider a full-theory amplitude $\mathcal{M}_{I_1\dots I_N}(p_1,\dots,p_N)$ with $N$ widely-separated energetic legs, schematically depicted in~\cref{fig:NonRad}.\footnote{We fix all-outgoing momenta with $\langle p|\xi|0\rangle = e^{ip\cdot x}$.}
As each collinear sector only contains a single particle, we can align the reference vector $\nim$ with the respective collinear momentum, so that $p_i^\mu = \nip\cdot p_i\, \frac{\nim^\mu}{2}$. We adopt this frame for the \emph{radiative} amplitude, but keep non-zero $p_{i\perp}$ for the non-radiative one.\footnote{This distinction is important, as the radiative scattering features an additional soft momentum $k_s$, so the non-radiative momenta must compensate this via non-vanishing $p_{i\perp}^2/(\nip\cdot p_i)\sim k_s $ if we want to keep the external reference vectors $n_i^\mu$ the \emph{same} for both the radiative and non-radiative amplitude.}

To perform the tree-level matching, we expand the full-theory amplitude in $\lambda$,
\begin{equation}
    \cM^{\vphantom{(0)}}_{I_1\dots I_N} = \cM^{(0)}_{I_1\dots I_N} + \cM^{(1)}_{I_1\dots I_N} + \mathcal{O}(\lambda^2)\,,
\end{equation}
with
\begin{subequations}
\begin{align}
    \cM^{(0)}_{{I_1\dots I_N}} &= \cM^{\vphantom{(0)}}_{I_1 \dots I_N}\bigr\rvert_{p_i^\mu = \nip\cdot p_i\spac \nim^\mu/2}\,,\\
    \cM^{(1)}_{{I_1\dots I_N}} &= \sum_{i=1}^N p_{i\perp}^\mu \biggl[\frac{\partial}{\partial p_{i\perp}^\mu} \cM^{\vphantom{(0)}}_{I_1\dots I_N}\biggr]\biggr\rvert_{p_i^\mu = \nip\cdot p_i\spac \nim^\mu/2}\,.
\end{align}
\end{subequations}
At leading power, the matching equation is obtained by simply equating the hard amplitude with the matrix element of the $N$-jet operator
\begin{equation}
    \mathcal{J}^{(0)} = \int [\text{d}t]_N\: C_{J_1\dots J_N}(t_1,\dots,t_N)\spac \xi^{J_1}(t_1)\dots \xi^{J_N}(t_N)\,,
\end{equation}
which contains a single collinear field in each sector.
It is convenient to work in momentum space, where we find\footnote{For convenience, we express the scattering amplitudes for flavor eigenstates rather than mass eigenstates. The two are related through tetrads~\cite{Cheung:2021yog}.}
\begin{align}
    \cM^{(0)}_{{I_1\dots I_N}} &= \langle p_1,I_1;\dots;p_N,I_N\rvert\, \tilde{\mathcal{J}}^{(0)}\,\lvert 0\rangle\nn\\
    &=\langle p_1,I_1;\dots;p_N,I_N\rvert\,
    \int [\text{d}t]_N\: e^{i \sum_i \nip\cdot p_i\,t_i} C_{J_1,\dots J_N}(t_1,\dots,t_N)\s\xi^{J_1}(t_1)\dots \xi^{J_N}(t_N)\lvert 0\rangle\nn\\
    &= C_{J_1\dots J_N}(\bar{n}_{1}\Cdot p_1,\dots, \bar{n}_{N}\Cdot p_N) \s\delta^{J_1}_{I_1}\dots \delta^{J_N}_{I_N}\nn\\
    &= C_{I_1\dots I_N}(\bar{n}_{1}\Cdot p_1,\dots, \bar{n}_{N}\Cdot p_N)\,.
\label{eq:NjetInsertion}
\end{align}
The leading-power matching coefficient is thus simply the hard limit of the full-theory non-radiative scattering amplitude.

For the subleading soft theorem, one also needs the subleading-power corrections to this non-radiative matching that do not modify the field content, i.e., operators that contain additional $\partial_\perp$ insertions.
We only require the first subleading operator
\begin{equation}
    J^{A1,\mu\spac I} \sim i\partial_\perp^\mu \xi^I\,,
\end{equation}
which we insert \emph{once} into the $N$-jet operator instead of a leading-power insertion of $\xi^I$.
The open indices on $\partial_\perp^\mu$ are contracted with the hard matching coefficient.
This matching coefficient is completely determined by RPI and reads~\cite{Beneke:2019kgv}
\begin{equation}
\label{eq:CA1Coefficient}
    C_i^{A1,\mu}(\np_1\Cdot p_1,\dots,\np_N\Cdot p_N) = -\sum_{k\neq i} \frac{2n^\mu_k}{n_k\Cdot n_i}\frac{\partial}{\partial (\nip \Cdot p_i)} C^{A0}(\np_1\Cdot p_1,\dots,\np_N\Cdot p_N)\,,
\end{equation}
where we suppressed the flavor indices for brevity.
This is a crucial feature: the calculation of the soft theorem requires subleading-power non-radiative $N$-jet operators, but their matching coefficient is constrained by RPI, as we explain in App.~\ref{app:RPItower}. Thus, only the leading-power matching is required.

\subsection{Single soft emission}
\label{sec:SingleSoftTree}
\begin{figure}[t!]
\begin{subfigure}[t]{0.45\textwidth}
    \includegraphics[width=\textwidth]{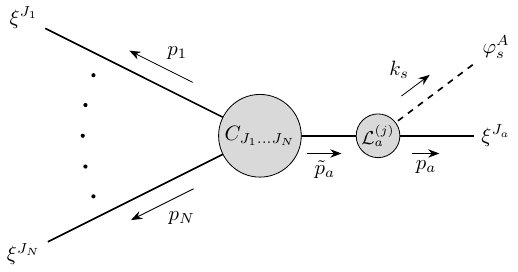}
    \caption{}
    \label{fig:LagrIns}
\end{subfigure}
\hspace{40pt}
\begin{subfigure}[t]{0.45\textwidth}
    \includegraphics[width=\textwidth]{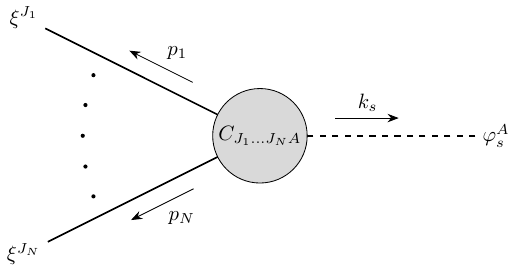}
    \caption{}
    \label{fig:EmFromHard}
\end{subfigure}
\caption{Diagrams depicting (a) soft emissions from external leg $a$ and (b) directly from the hard scattering. The soft scalar is drawn with a dashed line. Diagram (a) contains a subleading Lagrangian insertion $\mathcal{L}^{(j)}_a$ of $\mathcal{O}(\lambda^j)$, intermediate momentum $\tilde{p}_a^\mu = p_a^\mu + \nm_a\cdot k_{s}\,\np_a^\mu/2$, and the non-radiative matching coefficient $C_{J_1\dots J_N}$. Diagram (b) contains the matching coefficient $C_{J_1\dots J_N A}$, which includes an additional soft building block.}
\end{figure}

We now proceed with calculating the impact of radiative effects. 
There are two possible contributions to soft emissions in general: emissions from the external legs via (sub-)leading-power Lagrangian insertions, depicted in~\cref{fig:LagrIns}, and emissions from the hard scattering shown in~\cref{fig:EmFromHard}.
The Lagrangian used in the former is the same for all legs, and thus the emissions are completely independent from the hard matching.
However, in the latter case, if explicit soft building blocks appear that are not constrained by the shift symmetry, they come with their own matching coefficient and are thus process dependent. Thus, they break the universality of the soft emission.
From the discussion of the operator basis in~\cref{sec:geoNJet}, it follows that the first possible shift-invariant soft building block is $\partial_\mu\varphi_{s}^I~\sim\lambda^4$. Consequently, the terms at relative order $\mathcal{O}(\lambda^0)$ and $\mathcal{O}(\lambda^2)$ are universal, in analogy to gauge theory (see~\cite{Beneke:2021umj} for further details and a discussion on the gravitational case).

Let us make a brief remark on the relation between power counting and the expansion in the soft momentum:\ as we insert the soft-collinear Lagrangian into external collinear legs, the intermediate propagator generically takes the form
\begin{equation}
    \frac{i}{\tilde{p}^{\s 2}} = \frac{i}{\nip\Cdot p\,\nim\Cdot k_s}\,,
\end{equation}
and contains the soft pole as $k_s\to 0$ when the propagator goes on shell.
Therefore, the leading-power Lagrangian insertions correspond to $\mathcal{O}(k_s^{-1})$ at the amplitude level and encode the soft divergence.
As $k_s\sim\lambda^2$, the next-to-soft term must arise from $\mathcal{L}^{(2)}$ (or $\mathcal{L}^{(1)}$ inserted into the subleading current containing $J^{A1}$), and thus we see that relative $\mathcal{O}(\lambda^2)$ corresponds to $\mathcal{O}(k_s^0)$ at the amplitude level.

\vspace{10pt}
\noindent \textbf{Theories with} $\bm{V= 0}$\textbf{:} 
First, let us consider a theory with only derivative interactions.
Since there are no leading-power soft-collinear interactions, the first possible soft-collinear interaction term appears at $\mathcal{O}(\lambda^{2})$.
Therefore, the soft theorem features \emph{one universal term} at $\mathcal{O}(\lambda^2)$.
To compute the soft theorem, we restrict the Lagrangian to terms containing one soft and two collinear fields,
\begin{equation}
    \mathcal{L}^{(2)} \supset -\Gamma_{IAJ} \varphi_s^A\xi^J\square\xi^I\,,
\end{equation}
where $\square\xi=\nm\cdot\partial\,\np\cdot\partial\xi$ takes a similar role to the ``universal contraction'' $\np\cdot\partial\xi$ introduced in~\cite{Beneke:2021umj}. In contrast to gauge theory, the additional $\nm\cdot\partial$ cancels the eikonal propagator of the collinear field.

To be concrete, consider the insertion of this Lagrangian into leg $J$ as depicted in~\cref{fig:LagrIns}.
This corresponds to inserting $\mathcal{L}^{(2)}$ into the $N$-jet operator via a time-ordered product and evaluating the matrix element as in~\cref{eq:NjetInsertion}, i.e., computing
\begin{equation}
    \langle p_1,I_1,\dots,p_N,I_N;k_s,A\rvert\, \int [\dd t]_N\:e^{i\sum_i \nip \cdot p_it_i}\tilde{C}_{J_1\dots J_a\dots J_N}\xi^{J_1}\!\dots\! \int \text{d}^4x\:T\{\xi^{J_a}(t_{J_a})\mathcal{L}_a^{(2)}\}\dots \xi^{J_N}\lvert0\rangle\,.
\end{equation}
The internal collinear momentum is $\tilde{p}_a^{\s\mu}=p_a^\mu + \nm_a\cdot k_s\,\np_a^\mu/2$ with on-shell external momenta $p_a^2=k_s^2=0$ and $p^\mu_{a\perp} = 0$.
We find that
\begin{align}
    \bigl(\cM^{(2)}_{\mathrm{rad,\,ext}}\bigr)_{I_1\dots I_a \dots I_N A} &= C_{J_1\dots J_a \dots J_N}\:\delta\indices{^{J_1}_{I_1}}\dots \delta\indices{^{J_N}_{I_N}} \delta\indices{^{\tilde{A}}_{A}} \delta\indices{^{\tilde{J}}_{I}}\, \frac{i g^{\tilde{I}{J_a}}}{\tilde{p}^2}(-i)(-\tilde{p}^2)\Gamma_{\tilde{I}\tilde{A}\tilde{J}}
    \nn\\
    &=-C_{I_1\dots J_a \dots I_N}\spac\Gamma^{J_a}_{AI_a}\,.
\end{align}
By summing all emissions off the external legs, we get 
\begin{equation}\label{eq:EmissionFromLegNoPotential}
    \cM^{(2)}_{\mathrm{rad},\,I_1\dots I_N A} = -\sum_{a=1}^N\Gamma^{J_a}_{AI_a} {C}_{I_1\dots J_a\dots I_N} + \bigl(\cM^{(2)}_{\mathrm{rad},\,\mathrm{hard}}\bigr)_{I_1\dots I_N A}\,.
\end{equation}
The second term in~\cref{eq:EmissionFromLegNoPotential} corresponds to the emission from the hard vertex, as depicted in~\cref{fig:EmFromHard}.
As the process-dependent piece results from an interaction with an explicit factor of $\partial_\mu\varphi_s^I~\sim\lambda^4$, this has no impact on the universal result at $\mathcal{O}(\lambda^2)$. Instead, the piece appearing here is constrained by the shift invariance~\cref{eq:geoRPI}.
Indeed, from the leading-power $N$-jet operator, one has the non-radiative amplitude
\begin{equation}
    \cM^{(0)}_{J_1\dots J_N} = C_{J_1\dots J_N}(v)\,,
\end{equation}
which generates a subleading term contributing to the amplitude with a single extra emission of the form
\begin{align}
    \bigl(\cM^{(2)}_{\mathrm{rad},\,\mathrm{hard}}\bigr)_{J_1\dots J_N A} = \partial_A C_{J_1\dots J_N}(v)\,.
\end{align}
The form of this subleading term is fixed by the shift invariance, see \cref{eq:RPI}.
Inserting this into~\cref{eq:EmissionFromLegNoPotential}, one finds the following relation among amplitudes for a single soft emission\footnote{It is interesting to compare this result to the corresponding term in gauge theory.
Here, the leading-order interaction is
\begin{equation}
    \mathcal{L}^{(0)} = \frac{1}{2}(\nm\Cdot D\xi)^\dagger\,\np\Cdot\partial\xi + \mathrm{h.c.} \supset -\frac{1}{2}g_s\spac\nm\Cdot A_s\, \xi^\dagger i\np\Cdot\partial\xi+\mathrm{h.c.}\,,
\end{equation}
and the emission off the external legs yields
\begin{equation}
    \lim_{k_s\to 0} \cM_{N+1} = -g_s\sum_{a=1}^N t^a \frac{\nm\Cdot\varepsilon^a}{\nm\Cdot k_s} \cM_{N} + \mathcal{O}(k_s^0)\,.
\end{equation}
Identifying $g_s t^a\, \nm\cdot \varepsilon^a \to \Gamma^{J_a}_{A I_a} \nm\cdot k_s$, we see that this is indeed the same structure, arising from the covariant derivative, but suppressed by one power of soft momentum.}
\begin{align}
    \cM_{\mathrm{rad},\,I_1\dots I_N A}^{(2)} &= \partial_A\cM^{(0)}_{I_1\dots I_N} - \sum_{a=1}^N\Gamma^{J_a}_{AI_a} 
    \cM^{(0)}_{I_1\dots J_a\dots I_N}
    = \nabla_A \cM^{(0)}_{I_1\dots I_N} \,,
\end{align}
where in the last equality we see that all the terms assemble into a covariant derivative.

The connection to the soft theorem is now straightforward: we have successfully computed the single soft emission terms up to $\mathcal{O}(k_s^0)$ in the EFT, which reproduces the full-theory amplitude as
\begin{equation}
    \mathcal{M}_{\mathrm{full}} = \mathcal{M}_{\mathrm{rad}}^{(2)} + \mathcal{O}(k_s) \,,
\end{equation}
so one can take the soft limit $k_s\to 0$ (for particle $A$ of an $N+1$-point amplitude) to recover
\begin{equation}
    \lim_{k_s\to 0} \mathcal{M}_{I_1\dots I_NA} = \nabla_A \mathcal{M}_{I_1\dots I_N}\,.
\end{equation}
This successfully reproduces the geometric soft theorem for single emissions found in~\cite{Cheung:2021yog}.

Note that if one fixes normal coordinates $\Gamma^I_{JK}(v)=0$, the contribution due to emission off the external legs vanishes. In this case, $\nabla_I\to\partial_I$ and the soft theorem for a single emission simply reduces to acting with a partial derivative. In SCET, this has the interpretation that the soft theorem is simply due to the shift-invariance of the $N$-jet operator.

\vspace{10pt}
\noindent \textbf{Theories with} $\bm{V\neq 0}$\textbf{:} 
Next, we turn to theories with a non-zero potential. Generically, the presence of a potential results in soft-collinear three-point interactions starting at $\mathcal{O}(\lambda^0)$ for purely soft and $\mathcal{O}(\lambda^2)$ for soft-collinear ones.
However, this power suppression is purely due to the scaling of the dimensionful cubic coupling $g_3\sim\lambda^2$.
As any process-dependent modification to this sector must also be proportional to this cubic coupling, there are still two universal terms, as the first valid building block would be $g_3\,\partial_\mu\varphi_s^I$ (see the discussion below \cref{eq:NjetgeoSCET}).
In this case, the soft theorem thus features \emph{two universal terms}, and we need to consider the leading-order interaction as well as the multipole corrections to obtain the subleading-power term. This is again in close analogy to gauge theory~\cite{Beneke:2021umj}.

Before proceeding, let us briefly comment on the mass term.
With the counting $g_3\sim\lambda^2$ used above, a cubic interaction radiatively generates a mass $m^2\sim g_3^2\sim \lambda^4$, which is leading power in the soft sector, where $k_s^2\sim \lambda^4$.
In the following, we work in a massless theory, imposing the renormalization condition that the physical mass vanishes to all orders, $m_{\mathrm{ph}}^2=0$. We therefore do not consider bilinear interactions but consider a non-vanishing cubic one. We assume the potential is stabilized at quartic order, which is subleading in the infrared and does not affect the analysis below.
While this choice is fine-tuned and not protected from radiative corrections by symmetries, it corresponds to the theory where the actual limit $k_s\to0$ can be taken and the amplitude develops genuine soft and collinear singularities.
We return to the generic case $m\neq 0$, $m\sim\lambda^2$ at the end of this section, where the soft theorem is replaced by a factorization statement for the leading terms in the $\lambda$-expansion.

Focusing on the potential terms, the Lagrangian reads
\begin{subequations}
\begin{align}
    \mathcal{L}_V^{(2)} &= -\frac{1}{2}V_{AIJ}\varphi_s^A\xi^I \xi^J\,,\\[3pt]
    \mathcal{L}_V^{(3)} &= -\frac{1}{2}V_{AIJ}\big[x_\perp\Cdot\partial\varphi_s^A\big]\xi^I \xi^J\,,\\[3pt]
    \mathcal{L}_V^{(4)} &= -\frac{1}{2}V_{AIJ}\biggl(\frac{1}{2}\big[\nm\Cdot x\,\np\Cdot\partial\varphi_s^A\big]
    +\frac{1}{2}\big[x_\perp^\mu x_\perp^\nu\partial_\mu\partial_\nu\varphi_s^A\big]\biggr)\xi^I \xi^J\,,
\end{align}
\end{subequations}
where we have introduced the shorthand notation $V_{IJK}\equiv[\nabla_{(I}\nabla_J\nabla_{K)} V](v)\sim g_3$.
Therefore, if a potential with cubic vertices is present, one can have soft and collinear divergences in the theory, since there are leading-power soft-collinear interactions (relative to the power-counting of the coupling).

The leading-power calculation is straightforward.
For brevity, we introduce the shorthand notation
\begin{equation}
    \np_i\Cdot p \equiv p_-\,,\qquad\text{and}\qquad \nim\Cdot p\equiv p_+\,,
\end{equation}
and likewise for other vectors. One then simply inserts the Lagrangian interaction onto the external legs, obtaining
\begin{align}
    \cM^{(2)}_{\mathrm{rad},\spac V,\, I_1\dots I \dots I_N A} &= C_{J_1\dots J \dots J_N}\:\delta\indices{^{J_1}_{I_1}}\dots \delta\indices{^{J_N}_{I_N}} \delta\indices{^{\tilde{A}}_{A}} \delta\indices{^{\tilde{J}}_{I}}\,\frac{i g^{\tilde{I}J}}{\tilde{p}^{\s 2}}(-i)V_{\tilde{A}\tilde{I}\tilde{J}}\nn\\
    &=C_{I_1\dots J \dots I_N}\frac{V_{AI}^J}{p_-\spac k_{s+}}\,.
\end{align}
Performing the sum over all legs $J=1,\dots,N$, one obtains
\begin{equation}\label{eq:Vsoftfactor}
    \cM^{(2)}_{\mathrm{rad},\spac V,\, I_1\dots I_N A} = \sum_{a=1}^N \frac{V_{A I_a}^{J_a}}{p_-\spac k_{s+}} {C}_{I_1\dots J_a\dots I_N}\,,
\end{equation}
which already reproduces the leading-power correction to the geometric soft theorem due to a potential~\cite{Cheung:2021yog}.

For the subleading-power calculation, one notices that explicit powers of $x^\mu$ appear in the Lagrangian. These result in derivatives of the momentum-conserving $\delta$-functions upon Fourier transforming, and we perform explicit calculations below. Further details are given in App.~A of~\cite{Beneke:2018rbh} and Sec.~4 of~\cite{Jaskiewicz:2021cfw}.

For the subleading-power contribution, first note that there is no $\mathcal{O}(\lambda)$ correction, since we fix $p^\mu_{i\perp}=0$ for the radiative process. Therefore, the contribution from $\mathcal{L}^{(1)}$ vanishes if inserted into the leading-power current. This is consistent with the expansion in soft momenta, as the next term should be suppressed by one order in $k_s\sim\lambda^2$.

The subleading soft term then arises from two contributions,
the insertion of $\mathcal{L}^{(2)}$ into the leading-power current, and the insertion of $\mathcal{L}^{(1)}$ into the subleading-power current, where we change $\xi^J\to i\partial^\mu_\perp\xi^J$ for one leg.
The explicit coordinates appearing in the subleading Lagrangians generate derivatives in momentum space as
\begin{equation}
    X^\mu = \partial^\mu\bigg[(2\pi)^4 \delta^{(4)}\Big(\sum p_{\mathrm{in}}-\sum p_{\mathrm{out}}\Big)\bigg]\,,
\end{equation}
acting on the momentum-conserving $\delta$-function of the vertex.
The easiest strategy to implement the momentum routing is to work inwards from the external legs: assign the external momenta, and perform the momentum-conserving $\delta$ whenever possible, i.e., when no explicit $X$ appears in the vertex. At each subleading-power vertex, introduce a new momentum $\tilde{p}$ that is integrated over, then proceed through the diagram. We explain this strategy explicitly below.

Physically, the insertion of $\mathcal{L}^{(2)}$ is non-zero as $\nm\cdot X$ can hit the $\nm\cdot\tilde{p}=\nm\cdot k_s$ and thus provide a non-zero contribution, while the $\mathcal{L}^{(1)}$ can cancel the explicit $i\partial_\perp \to -\tilde
p_\perp$ of the subleading operator to yield a non-vanishing result.
Both contributions also exist in gauge theory and gravity, where they produce the subleading term~\cite{Beneke:2021umj}.
The calculations are standard, and we present one of them here explicitly. The insertion of $\mathcal{L}^{(1)}$ is completely analogous.

We define the shorthand notation
\begin{equation}
    \int_{\tilde{p}} \equiv \int\frac{\dd^4\tilde{p}}{(2\pi)^4}\,.
\end{equation}
The amplitude for the insertion onto leg $J$ reads
\begin{align}
    \cM^{(4)}_{\mathrm{rad}} &= \frac{1}{2} \int_{\tilde{p}} C_{J_1\dots J \dots J_N}\:\delta\indices{^{J_1}_{I_1}}\dots \delta\indices{^{J_N}_{I_N}} \delta\indices{^{\tilde{A}}_{A}} \delta\indices{^{\tilde{J}}_{I}}\frac{i g^{\tilde{I}J}}{\tilde{p}^2}(-i)V_{\tilde{A}\tilde{I}\tilde{J}}(i\nm\Cdot X\, i\np\Cdot k_s
    - X_\perp^\mu X_\perp^\nu k_{s\mu}k_{s\nu})
    \nn\\
    &= -\frac{1}{2}V_{AI}^J\int_{\tilde{p}} \!C_{I_1\dots J\dots I_N}\frac{1}{\tilde{p}^2}\biggl(\np\Cdot k_s\,\nm^\mu\frac{\partial}{\partial \tilde{p}^\mu} - k_s^\mu k_s^\nu\frac{\partial}{\partial \tilde{p}_\perp^\mu}\frac{\partial}{\partial \tilde{p}_\perp^\nu}\biggr)(2\pi)^4\delta^{(4)}(\tilde{p}-p-k_{s+})\nn\\
    &= \frac{1}{2}\frac{V_{AI}^J}{p_- k_{s+}}\,\np\Cdot k_s\,\nm\Cdot\frac{\partial}{\partial p} C_{I_1\dots J\dots I_N}\,.
\end{align}
To reach the third line, we used the fact that the action of the derivatives on the propagator combines as
\begin{align}
    \int_{\tilde{p}}\frac{1}{\tilde{p}^2}&\biggl(\np\Cdot k_s\,\nm^\mu\frac{\partial}{\partial \tilde{p}^\mu} - k_s^\mu k_s^\nu\frac{\partial}{\partial \tilde{p}_\perp^\mu}\frac{\partial}{\partial \tilde{p}_\perp^\nu}\biggr)(2\pi)^4\delta^{(4)}(\tilde{p}-p-k_{s+})\nn\\
    &= 2\np\Cdot k_s\, \frac{\nm\Cdot \tilde{p}}{\tilde{p}^4} + 2k_s^\mu k_s^\nu\,\frac{g_{\perp\mu\nu}}{\tilde{p}^4}\biggr\rvert_{\tilde{p}\s=\s p+k_{s+}}\nn\\
    &=\frac{2}{(p_-\spac k_{s+})^2}\big(\np\Cdot k_s\,\nm\Cdot k_s + k_{s\perp}^2\big)
\end{align}
which vanishes for $k_s^2=0$.

Summing over all legs, combining with the contribution for the $\mathcal{L}^{(1)}$ insertion,\footnote{The $\mathcal{L}^{(1)}$ insertion yields the transverse pieces $\frac{\partial}{\partial p_{\perp}}$, as in gauge theory. Note that $\frac{\partial}{\partial p_{\perp}}\mathcal{M}\sim 1$, there is no enhancement from the transverse derivative, as this acts on the subleading terms of the non-radiative amplitude. For further details on this, see~\cite{Beneke:2021umj} Secs.~2.5 and 3.4.} and collecting the subleading non-radiative amplitudes into $\cM$, one finally obtains
\begin{equation}
    \cM^{(4)}_{\mathrm{rad}} = \sum_{a=1}^N \frac{V^{J_a}_{AI_a}}{(p_a+k_s)^2} k_s^\mu \frac{\partial}{\partial p_a^\mu} \cM_{I_1\dots J_a\dots I_N}\,,
\end{equation}
with $\cM_{I_1\dots J_a\dots I_N} = \cM^{(0)}_{I_1\dots J_a\dots I_N} + \cM^{(1)}_{I_1\dots J_a\dots I_N} + \mathcal{O}(\lambda^2)$.

We close this section by commenting on the generic case $m\neq 0$, $m\sim\lambda^2$.
Once the soft scalar is massive, the limit $k_s\to 0$ is no longer accessible, since the on-shell condition is $k_s^2=m^2$, and its energy is bounded from below by $k_s^0\geq m$.
Strictly speaking, there is then no soft theorem: the soft factor constructed above corresponds to the leading term in a systematic expansion of the radiative amplitude in $\lambda$.
Concretely, one parametrizes the soft momentum as $k_s^\mu= m\spac v^\mu$ with velocity $v^2=1$, and expands around this configuration rather than $k_s=0$.
The intermediate collinear propagator is unchanged at leading power
\begin{equation}
    (p+k_s)^2 - m^2 = \np\Cdot p\,\nm\Cdot k_s+\mathcal{O}(\lambda^4)\,,
\end{equation}
and the eikonal structure underlying the soft theorem persists.
The covariant form of the leading term is thus unchanged even for a massive theory.
The collinear mass enters the Lagrangian as a bilinear interaction starting at $\mathcal{O}(\lambda^2)$, through
\begin{equation}
    \mathcal{L}^{(2)}_{V,m} = -\frac{1}{2}[\nabla_I\nabla_J V]\xi^I\xi^J\,,
\end{equation}
which is treated as a subleading two-point interaction, as is standard in SCET$_{\mathrm{I}}$ (see, e.g.,~\cite{Boer:2023yde}).
However, the resulting amplitude is now free of infrared divergences, as $m$ serves as an IR regulator for the soft momentum.
This is the familiar form of a soft theorem for massive emissions, and it is a straightforward $\lambda$-expansion following the same rules.
Note that, as we comment on below, the presence of a mass term alters the soft loop corrections.

\subsection{Double soft emission}
Next, we derive the double soft theorem for the case of a vanishing potential $V(\phi)=0$.
Since this calculation becomes quite tedious in a generic coordinate system, we adopt normal coordinates around the vev $v$, i.e., $\Gamma^I_{JK}(v)=0$.
Full covariance can easily be restored at the end of the calculation.

\begin{figure}[t!]
    \centering
    \includegraphics[scale=0.9]{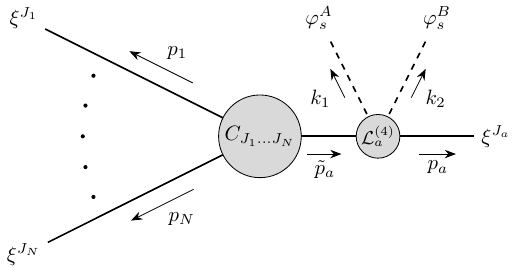}
    \caption{Diagram depicting the non-vanishing topology for double-soft emission in the normal coordinates basis. The subleading Lagrangian $\mathcal{L}^{(4)}_a$ is inserted onto leg $a$.}
    \label{fig:DoubleSoft}
\end{figure}

In this case, the Lagrangian takes the simple form
\begin{equation}
    \mathcal{L}^{(4)} = \frac{1}{2}R_{ABIJ}\spac\varphi_s^A \partial_\mu\varphi_s^B\spac \xi^J \partial^\mu\xi^I\,.
\end{equation}
Inserting this double-emission vertex onto an external leg $J_a$ as shown in \cref{fig:DoubleSoft}, one immediately finds
\begin{align}
    \bigl(\mathcal{M}^{(4)}_{\mathrm{rad.\,ext}}\bigr)_{I_1\dots I_a\dots I_N AB} &= \frac{1}{2}\frac{s_{AI_a}-s_{BI_a}}{s_{AI_a}+s_{BI_a}}R_{ABI_a}^{\phantom{ABI}J_a} C_{I_1\dots J_a\dots I_N}\,,
\end{align}
successfully recovering the Riemann-tensor term of the double soft theorem.
The additional term from the hard matching coefficient, which is fixed by RPI, is
\begin{equation}
    \bigl(\mathcal{M}^{(4)}_{\mathrm{hard}}\bigr)_{I_1\dots I_N AB} =\frac{1}{2}\partial_A\partial_B C_{I_1\dots I_N} \,.
\end{equation}
Since we are working in normal coordinates where $\nabla_I = \partial_I$, we can write this with covariant derivatives. Assembling the pieces, we find that\footnote{To derive this rigorously, one can work in normal coordinates and then reinstate generic coordinates using the soft Wilson lines $S_n$, similar to (un-)fixing light-cone gauge in SCET calculations using soft decoupling. Alternatively, one can use the Lagrangian in generic coordinates and see that the cubic vertices arrange themselves properly.}
\begin{equation}\label{eq:doubleSoft}
    \cM_{\mathrm{rad},\,I_1\dots I_N AB}^{(4)} = \nabla_{(A}\nabla_{B)}\cM_{\mathrm{rad},\,I_1\dots I_N }^{(0)} + \frac{1}{2}\sum_{a=1}^{N} \frac{s_{AI_a}-s_{BI_a}}{s_{AI_a}+s_{BI_a}}R_{ABI_a}^{\phantom{IJI_a}J_a} \cM_{\mathrm{rad},\,I_1\dots J_a \dots I_N }^{(0)}\,.
\end{equation}
This reproduces the tree-level double soft theorem first derived in \cite{Cheung:2021yog}. As we shall see, this result is tree-level exact. In fact, all single soft and multiple soft theorems are tree-level exact for scalar EFTs with a vanishing potential, which we can prove using geoSCET.

In \cite{Cheung:2021yog}, a triple soft theorem was also derived at tree level. One could continue to apply geoSCET to reproduce these multiple soft theorems, which would be a straightforward albeit tedious exercise. Here, we content ourselves with proving the existence and tree-level exactness of these multiple soft theorems through simple power-counting arguments, rather than deriving their explicit forms.

\subsection{Loop corrections}
\label{sec:loops}
We now consider the possible loop corrections to the geometric soft theorems, complementing the one-loop analysis in~\cite{Cohen:2025dex}. (For gravity, this analysis has been performed in~\cite{Beneke:2021umj,Beneke:2022pue}.) The EFT is a powerful tool for such investigations. Since we have a systematic power counting, the different momentum regions of a given loop diagram are already disentangled at the Lagrangian level.
Then, one works order by order in the power counting by writing down all possible topologies that can occur at a given order.
One can distinguish the hard loops, which only affect the matching coefficient, from the soft and collinear loops between the external legs that could modify the soft theorem at the loop level.

In this context, it is worth noting that taking soft or collinear limits of amplitudes is subtle and can give rise to discontinuity functions as discussed in~\cite{Bern:1995ix} (see also~\cite{Cohen:2025dex}). From the SCET perspective, however, the logic is reversed: rather than taking a kinematic limit, one constructs the radiative amplitude from the non-radiative one. The discontinuity functions then arise naturally as loop corrections of matrix elements in the low-energy theory. Further details and a method-of-regions analysis are provided in App.~\ref{sec:app:Discont}.

If such loop corrections arise exclusively from EFT Lagrangian interactions or are constrained by RPI, the resulting corrections will be universal, i.e., they are process-independent. Furthermore, if the loops that arise from Lagrangian interactions are power suppressed compared to the tree-level result, then the soft theorems will be tree-level exact. We first argue that the theories without a potential are indeed tree-level exact, and then we explore the extent to which theories with a potential have universal corrections.
    
\vspace{10pt}
\noindent \textbf{Theories with} $\bm{V= 0}$\textbf{:}   Recall the power counting of the soft theorem for derivatively-coupled theories: there is no contribution at $\mathcal{O}(\lambda^0)$, and the leading soft theorem arises at $\mathcal{O}(\lambda^2)$ corresponding to $\mathcal{O}\big((p\cdot k_s)^{0}\big)$. In this case, the leading terms relevant for the following analysis read in normal coordinates
    \begin{subequations}
        \begin{align}
            \mathcal{L}^{(2)} &\supset -\frac{1}{6}R_{IKJL}\spac \xi^K\xi^L\spac \partial_\mu\xi^I \partial^\mu\xi^J\,,\\
            \mathcal{L}^{(3)} &\supset -\frac{1}{3} R_{KAIJ}\spac \xi^K \partial_\mu\varphi_s^A \spac\xi^I \partial^\mu \xi^J + \mathrm{perm}\,,\\
            \mathcal{L}^{(4)} &\supset -\frac{1}{6}R_{AIBJ}\spac \varphi_s^A \varphi_s^B \partial_\mu\varphi_s^I\spac \partial^\mu \varphi_s^J\,,
        \end{align}
    \end{subequations}
    and crucially, in normal coordinates there are no three-point interactions, and the purely-collinear interaction is at least $\mathcal{O}(\lambda^2)$ suppressed.
    This renders the following analysis trivial.
    Namely, the geometric soft theorem itself has no $\mathcal{O}(\lambda^0)$ term but starts at relative $\mathcal{O}(\lambda^2)$. Any loop correction requires at least another vertex, which is itself suppressed by at least $\mathcal{O}(\lambda^2)$, regardless of whether it is soft or collinear, so it is impossible to construct one-loop topologies that contribute at $\mathcal{O}(\lambda^2)$.
    Therefore, the geometric soft theorem for derivatively-coupled theories is exact to all orders and not modified by loop corrections.
    In practice, this means that it originates purely from the shift-invariance of the hard matching coefficient, and reads to any loop order
    \begin{equation}
        \lim_{k_s\to 0}\mathcal{M}_{I_1\dots I_N A} = \nabla_A \mathcal{M}_{I_1\dots I_N}\,,
    \end{equation}
    where $\mathcal{M}$ can contain an arbitrary number of hard loops, which affect the matching coefficient but are determined by non-radiative matching.
    Let us emphasize that this statement is purely a consequence of power counting in the EFT.

    Moreover, this also holds for higher-emission soft theorems using the same arguments. Any loop one can attach costs at least $\mathcal{O}(\lambda^2)$, and in normal coordinates one can eliminate the cubic vertices. Therefore, there are no topologies at the respective leading power that could modify the soft theorems for any number of soft emissions. The power counting of geoSCET proves that the geometric double soft theorem in \cref{eq:doubleSoft} and the triple soft theorem derived in \cite{Cheung:2021yog}, along with an infinite number of further universal geometric soft theorems, are all tree-level exact.
    
\vspace{10pt}
\noindent \textbf{Theories with} $\bm{V\neq 0}$\textbf{:}
Here we focus on the cubic interaction, which generates the terms that dominate in the infrared. Again, we assume $m_{\mathrm{ph}}^2=0$ and comment on the generic massive case below. The reason for this is twofold. First, we perform the massless analysis to show how the results of~\cite{Cohen:2025dex} are reproduced in geoSCET. Second, the topologies of the loop corrections for the massless and massive theories are the same. Therefore, it will be easy to extend the analysis to the massive case, which is the more physical setup.

With the power counting $g_3\sim\lambda^2 Q$ introduced below~\cref{eq:L2VUnexpanded}, the leading term of the soft theorem arises at $\mathcal{O}(\lambda^2)$ and the subleading one at $\mathcal{O}(\lambda^4)$. 
This corresponds to $\mathcal{O}\bigl((p\cdot k_s)^{-1}\bigr)$ and $\mathcal{O}\bigl((p\cdot k_s)^{0}\bigr)$, respectively. 
The pattern persists in the loop expansion. As $g_3$ is the only quantity of positive mass dimension besides the soft-collinear invariants themselves, dimensional analysis fixes the $n$-loop soft-collinear contribution to be
\begin{equation}\label{eq:cubicPC}
    \biggl(\frac{g_3^2}{p\Cdot k_s}\biggr)^n\sim \lambda^{2n}\,,
\end{equation}
as we explicitly demonstrate below.
Each loop is suppressed by $\lambda^2$ but carries one further inverse power of $p\cdot k_s$. Therefore, the $\lambda$-order of a contribution and the degree of its infrared singularity are no longer locked together, and higher loops are increasingly infrared divergent.

This deserves care, as it spoils the argument of~\cref{sec:GeoSCETLagrangian}, where we concluded that the derivatively-coupled theory is IR finite based on power-counting. The cornerstone of this argument is that power-suppressed terms are less singular in the infrared, 
and it was crucial that all couplings had non-positive mass dimension. In this case, every inverse power of a soft invariant must then be compensated by explicit momenta in the numerator, since nothing else in the theory carries positive mass dimension.
The derivatively-coupled theory satisfies this requirement, as there are even two additional powers of momentum per vertex.
A relevant coupling violates this condition. Specifically, $g_3^2$ supplies the mass dimension on its own, so that further $1/(p\cdot k_s)$ may appear with a coefficient that is $\lambda$-suppressed rather than $\lambda$-enhanced.\footnote{Since $m^2\sim g_3^2$ is the radiatively generated mass, the ratio governing the loop expansion is $m^2/(p\cdot k_s)$, the generated mass measured against the collinear virtuality. If we instead assumed a power counting where $g_3\sim Q$, this ratio becomes $\lambda^{-2}$: each collinear cubic vertex scales as $\mathcal{O}(\lambda^{-1})$ and each collinear loop built from them as $\mathcal{O}(\lambda^{-2})$, indicating the breakdown in the IR.}
This case is specific to the cubic term: already the quartic coupling is dimensionless in $d=4$ and preserves the correspondence.

With this remark aside, we now focus on the leading loop corrections, i.e., the most infrared-divergent terms. The relevant Lagrangian interactions are
\begin{subequations}
    \begin{align}
        \Lagr^{(1)}_\xi &= -\frac{1}{3!}V_{IJK}\spac\xi^I\xi^J\xi^K 
        \,,\\
        \mathcal{L}_\xi^{(2)} &= -\frac{1}{2} V_{AIJ}\spac\varphi_s^A \xi^I \xi^J
        \,,\\
        \Lagr^{(0)}_s &= -\frac{1}{3!}V_{ABC}\spac\varphi_s^A\varphi_s^B\varphi_s^C\,.
    \end{align}
\end{subequations}
Here, the superscript indicates the power counting of the relevant Lagrangian terms.

\vspace{6pt}
\noindent\textbf{Collinear corrections:}
For collinear loops, the loop momentum scales as $\ell\sim(1,\lambda^2,\lambda)Q$. It therefore suffices to consider a single leg. Otherwise, a collinear loop would connect legs of different sectors, and such a topology would include a hard loop momentum, which corresponds to a loop correction of the matching coefficient. As such, it does not exist in the low-energy EFT.

\begin{figure}[t!]
\begin{subfigure}[t]{0.45\textwidth}
    \includegraphics[width=\textwidth]{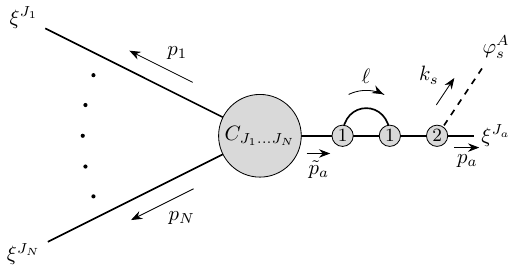}
    \caption{}
    \label{fig:CollBubble}
\end{subfigure}
\hspace{40pt}
\begin{subfigure}[t]{0.45\textwidth}
    \includegraphics[width=\textwidth]{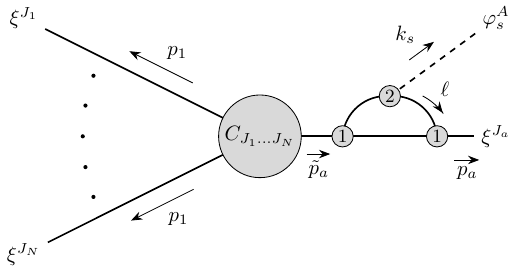}
    \caption{}
    \label{fig:CollTriangle}
\end{subfigure}
\caption{Collinear one-loop topologies modifying the soft theorem with a cubic interaction: the bubble (a) and the triangle topology (b). The numbers $j=1,2$ in the small circles denote the insertion of a subleading Lagrangian $\mathcal{L}_a^{(j)}$. Here, $\tilde{p}_a^\mu = p_a^\mu + \nm_a\cdot k_s\,\np_a^\mu/2$ and the loop momentum $\ell$ scales like an $a$-collinear momentum.}
\label{fig:CollinearTopologies}
\end{figure}

For a single collinear leg, there are two topologies: the bubble and the triangle, as depicted in~\cref{fig:CollinearTopologies}. These factorize and are thus universal.
The bubble topology factorizes into the soft emission attached to the one-loop jet function.
For the insertion into leg $J_a$, one obtains\footnote{Unless otherwise specified, all propagators are understood to include the $+i0$ prescription.}
\begin{align}
    \mathcal{M}_{\text{coll bubble}} &=  iV_{A I_a}^{J_1}\spac V_{J_1 J_2}^{J_3}\spac V_{J_3}^{J_2 J_a} C_{I_1\dots J_a\dots I_N}
    \biggl(\frac{i}{\tilde{p}^2}\biggr)^2
    \frac{1}{2}\tilde{\mu}^{2\varepsilon}\int\frac{\dd^d\ell}{(2\pi)^d}\frac{1}{\ell^2(\ell-\tilde{p})^2}\nn\\
    &= \frac{V_{A I_a}^{J_1}}{2p_a\Cdot k_s}
    \frac{V_{J_1 J_2}^{J_3}\spac V_{J_3}^{J_2 J_a}}{(4\pi)^2}\spac\frac{1}{2p_a\Cdot k_s}\,\frac{1}{2}\mu^{2\varepsilon} e^{\varepsilon\gamma_E} I_2(\tilde{p}^2)\spac C_{I_1\dots J_a\dots I_N}\,,
    \label{eq:CollinearBubble}
    \end{align}
with $\tilde{\mu}^2 = \mu^2 e^{\gamma_E}/(4\pi)$ and  $\mu$ is the factorization scale.
Here, we used $\tilde{p}^\mu = p_a^\mu + \nm_a\cdot k_s\, \np_a^\mu/2$, which is also the scale appearing inside the bubble integral
\begin{equation}
    I_2(p^2) = i\int\frac{\dd^d\ell}{\pi^{d/2}}\frac{1}{\ell^2\spac (\ell - p)^2}\,.
\end{equation}
From the structure in~\cref{eq:CollinearBubble}, we see that this is the leading soft factor $V/(2p_a\cdot k_s)$ multiplying a one-loop correction to the collinear leg, i.e., a one-loop jet function, which depends on the collinear scale $\tilde{p}^2=p_- k_{s+}$. 
As anticipated, the entire contribution scales as $(2p_a\cdot k)^{-2}$, with an additional pole originating from the collinear loop.

Likewise, for the triangle topology as depicted in~\cref{fig:CollTriangle}, one obtains
\begin{align}
    \mathcal{M}_{\text{coll triangle}} &= -iV_{I_a J_1}^{J_2}\spac V_{J_2 A}^{J_3}\spac V_{J_3 J_a}^{J_1}
    \frac{1}{\tilde{p}^2} C_{I_1\dots J_a\dots I_N} \tilde{\mu}^{2\varepsilon}\int\frac{\dd^d \ell}{(2\pi)^d} \frac{1}{\ell^2\spac(\ell-\tilde{p})^2\spac(\ell-\tilde{p}-k_{s+})^2}\nn\\
    &= \frac{1}{2p_a\Cdot k_s} \frac{V_{I_a J_1}^{J_2}\spac V_{J_2 A}^{J_3}\spac V_{J_3}^{J_1 J_a}}{(4\pi)^2}\,\mu^{2\varepsilon}e^{\varepsilon\gamma_E} I^{1m}_3(\tilde{p}^2)\,C_{I_1\dots J_a\dots I_N}\,,
\end{align}
where
\begin{equation}
    I_3^{1m}(p_1^2) = -i\int\frac{\dd^d\ell}{\pi^{d/2}}\frac{1}{\ell^2\spac (\ell - p_2)^2(\ell-p_1-p_2)^2}\,,\quad\text{with}\quad p_2^2 = (p_1+p_2)^2=0
\end{equation}
is the one-mass triangle integral.

Note that the relevant scale appearing in both collinear contributions is the collinear scale $(p+k_{s+})^2 = p_- k_{s+}\sim\lambda^2$, as the multipole expansion in the Lagrangian enforces the momentum-conserving $\delta$-function to be
\begin{equation}
    (2\pi)^d \delta^{(d)}\Bigl(p_{\mathrm{in}} - p_{\mathrm{out}}- \nm\cdot k_s\frac{\np^\mu}{2}\Bigr)
\end{equation}
for incoming collinear momentum $p_{\mathrm{in}}$, outgoing collinear $p_{\mathrm{out}}$, and soft $k_s$.
The bubble and one-mass triangle are single-scale integrals, so the collinear region appearing here covers the entire integral.
One can thus already anticipate that the corresponding soft topologies will be scaleless, as we demonstrate below.

We see that the collinear loop corrections directly reproduce the bubble and triangle topologies identified and computed in~\cite{Cohen:2025dex}.
By power counting, these two topologies already exhaust the leading contributions at one-loop.
For example, take the topologies where the loop is attached to the hard scattering. In this case, we would have an $N$-jet operator with two collinear fields in the same direction, which is suppressed kinematically in $\lambda$, but must also be proportional to $g_3\sim\lambda^2$. Then, connecting the loop and adding a soft emission is an additional suppression of $\lambda^2$, so in total these contributions start at $\mathcal{O}(\lambda^5)$, more suppressed than the one-loop correction we just analyzed, which appears at $\mathcal{O}(\lambda^4)$.
Physically, this configuration cannot produce the IR-leading $1/(p\cdot k)^2$ and is therefore subleading.

\begin{figure}
\begin{subfigure}[t]{0.45\textwidth}
    \includegraphics[width=\textwidth]{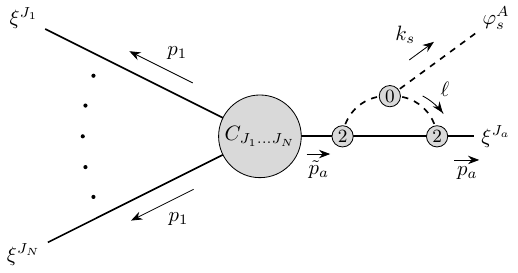}
    \caption{}
    \label{fig:SoftTriangle}
\end{subfigure}
\hspace{40pt}
\begin{subfigure}[t]{0.45\textwidth}
    \includegraphics[width=\textwidth]{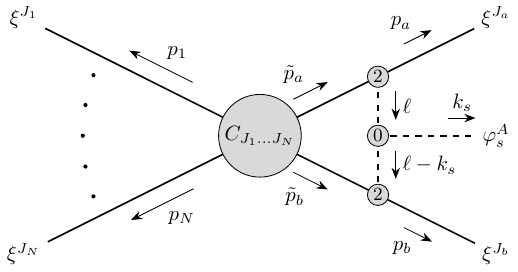}
    \caption{}
    \label{fig:SoftBox}
\end{subfigure}
\caption{Soft one-loop topologies modifying the soft theorem with a cubic interaction: the triangle (a) and the box topology (b). The numbers $j=1,2$ in the small circles denote the insertion of a subleading Lagrangian $\mathcal{L}_{a/b}^{(j)}$. Here, $\tilde{p}_{a/b}^\mu = p_{a/b}^\mu + \nm_{a/b}\cdot k_s\,\np_{a/b}^\mu/2$ and the loop momentum $\ell$ scales like a soft momentum.}
\label{fig:SoftTopologies}
\end{figure}

\vspace{6pt}
\noindent\textbf{Soft corrections:}
Next we consider soft loops, where the scaling of the loop momentum is  $\ell\sim(\lambda^2,\lambda^2,\lambda^2)Q$. The relevant topologies are depicted in~\cref{fig:SoftTopologies}.
The contributions that are only attached to a single leg will be scaleless, as there is no soft scale in the problem due to the eikonal propagators which only carry the single direction $\nim^\mu$.
The only possibility to obtain a non-vanishing result is to have the soft emission directly connected to the soft loop, which must connect two legs of different collinear sectors~\cite{Beneke:2022pue,Ma:2023hrt}. In this case, one has two distinct eikonal propagators which yield $\nim^\mu$ and $\nip^\mu$, while a soft momentum-conserving $\delta$-function transfers the scale $k_{\perp}^2 = k_{-} k_{+}$ into the soft loop.

This is consistent with the regions analysis of the bubble and triangle diagrams: for the kinematics here, the appearing external scale is $\tilde{p}^2=p_- k_+$. In this case, the integrals feature only a single scale, which is completely reproduced by the collinear integrals discussed above. Therefore, the soft region is scaleless.

To make this explicit, consider adding a soft loop to a collinear leg, i.e., the topology of~\cref{fig:CollBubble} with soft loop momentum $\ell$.
The contribution takes the form
\begin{align}
    \mathcal{M} &\sim \int\frac{\dd^d \ell}{(2\pi)^d} \frac{i}{\ell^2}\frac{i}{(\tilde{p}-\ell_+)^2}
    \sim \frac{1}{p_-} \int\frac{\dd^d\ell}{(2\pi)^d} \frac{1}{\ell^2(k_{s+}-\ell_+)}\,.
\end{align}
The resulting transverse integral is scaleless, as one cannot form a Lorentz invariant out of $k_{s+}^\mu = \nm\cdot k_s\,\np^\mu/2$ since $\np^2=0$. The same also holds for the soft triangle topology depicted in~\cref{fig:SoftTriangle}, and again the reason is that the loop is blind to the $n_-^\mu$ direction since only $k_{s+}$ can enter the collinear propagators.

Therefore, the only relevant topology is a soft loop connecting two collinear legs, which we take to be back-to-back for simplicity.
In addition, the soft emission must directly connect to the soft loop, as depicted in~\cref{fig:SoftBox}, so that the full $k_{s\perp}^2$ can enter the soft loop.
In this case, the scale can be built as $k_{s\perp}^2=k_{s-}\,k_{s+}$, where the top eikonal propagator yields $(p_{a-}\,k_{s+})^{-1}$ and the lower one provides $(p_{b+}\,k_{s-})^{-1}$.
From the EFT perspective, we immediately see that this correction factorizes from the hard scattering into a universal soft factor, but crucially it connects two legs and therefore results in non-trivial flavor mixing.

To be concrete, consider now the contribution
\begin{align}
    \mathcal{M}_{\text{soft box}} &= i V_{I_a S_a}^{J_a}\spac V_{I_b S_b}^{J_b}\spac V_A^{S_a S_b}\spac \tilde{\mu}^{2\varepsilon} \int\frac{\dd^d\ell}{(2\pi)^d} \frac{1}{(p_a+\ell_+)^2\spac (p_b-\ell_-+k_{s-})^2\spac \ell^2\spac (\ell-k_s)^2}\nn\\[3pt]
    &=-i\frac{V_{I_a S_a}^{J_a}\spac V_{I_b S_b}^{J_b}\spac V_A^{S_a S_b}}{p_{a-} p_{b+}} \tilde{\mu}^{2\varepsilon}\int\frac{\dd^d\ell}{(2\pi)^d}\frac{1}{\ell^2\spac(\ell-k_s)^2\spac(\ell_- - k_{s-} - i0)(\ell_++i0)}\,.
\end{align}
From dimensional analysis, we see that the remaining soft loop integral must scale as
\begin{equation}
    \int\frac{\dd^d\ell}{(2\pi)^d}\frac{1}{\ell^2\spac(\ell-k)^2\spac(\ell_- - k_{s-} - i0)(\ell_++i0)} = c(\varepsilon)(-k_{s-}\, k_{s+})^{-1-\varepsilon}\,,
\end{equation}
and it is the eikonal limit of the one-mass box integral.
The integration is straightforward, and one obtains
\begin{equation}
    \mathcal{M}_{\text{soft box}} =\frac{1}{8\pi^2}\frac{V_{I_a S_a}^{J_a}\spac V_{I_b S_b}^{J_b}\spac V_A^{S_a S_b}}{p_{a-}\spac p_{b+}} \frac{1}{k_{s-}\spac k_{s+}} \biggl(-\frac{\mu^2}{k_{s-}\spac k_{s+}}\biggr)^{\varepsilon} e^{\varepsilon\gamma_E}\frac{\Gamma^2(1+\varepsilon)\Gamma^3(1-\varepsilon)}{\varepsilon^2\spac\Gamma(1-2\varepsilon)}\,.
\end{equation}
Performing the expansion in $\varepsilon$, one finally finds
\begin{equation}
    \mathcal{M}_{\text{soft box}} = \frac{1}{8\pi^2}\frac{V_{I_a S_a}^{J_a}\spac V_{I_b S_b}^{J_b}\spac V_A^{S_a S_b}}{(2p_{a}\Cdot k_s) (2p_{b}\Cdot k_s)} \biggl(-\frac{\mu^2}{k_{s-}\spac k_{s+}}\biggr)^{\!\varepsilon}
    \biggl(\frac{1}{\varepsilon^2} + \frac{\pi^2}{12}\biggr)\,,
\end{equation}
which precisely corresponds to the box computed in~\cite{Cohen:2025dex}.
It is worth discussing the structure of this contribution, as it is qualitatively different from all other topologies appearing at one loop.
All collinear corrections are proportional to the non-radiative amplitude with at most one flavor index appearing on a single leg. They dress an external line without communicating between sectors.
The soft box is the unique one-loop topology that connects two distinct collinear directions, and correspondingly it is the only one that induces mixing between the flavor indices of different legs.
The mixing arises from the single structure
\begin{equation}
    V_{I_a S_a}^{J_a}\spac V_{I_b S_b}^{J_b}\spac V_A^{S_a S_b}\,,
\end{equation}
a product of three cubic couplings contracted through the soft loop.
This is the direct analog of the color-dipole structure of soft exchange in gauge theory.
There, a soft gluon exchanged between two distinct collinear lines generates the color correlation $\mathbf{T}_a\cdot \mathbf{T}_b$, the unique dipole structure connecting two sectors. In geoSCET, the same role is played by the cubic couplings contracted across the soft loop, with field-space indices replacing color.
In both cases, the correlation arises from the same soft-collinear topology.

To summarize, we have analyzed the most IR-divergent one-loop integrals for the case of a cubic coupling. We have explicitly demonstrated the expansion in $g_3^2/(p\cdot k_s)\sim \lambda^2$, which is increasingly IR divergent, and we have reproduced the statements in~\cite{Cohen:2025dex} from the EFT perspective.
We would like to point out that the question of (non-)factorization is trivial in SCET, as any loop arising purely from Lagrangian insertions is universal from the start.
In addition, this discussion emphasizes the close connection between the different regions of the full-theory Feynman integrals and the sectors of the EFT, see App.~\ref{sec:app:Discont}.

Finally, we comment on how the loop analysis changes for a massive scalar field with $m^2\sim\lambda^4$.
As discussed at the end of~\cref{sec:SingleSoftTree}, the mass term is leading power in the soft sector with $k_s^2\sim\lambda^4$ but a power correction in the collinear one, where $p^2\sim\lambda^2$.
It is therefore retained in the soft propagators while collinear ones remain massless, and the mass enters as an additional bilinear collinear interaction at $\mathcal{O}(\lambda^2)$.

Therefore, the classification of soft topologies changes qualitatively. In the massless theory, soft graphs attached to only a single collinear direction were scaleless: the eikonal propagators only carry the direction $\np^\mu$, and no soft invariant is available.
With a massive soft propagator, this is no longer true, as the mass provides a scale, and integrals of the schematic form
\begin{equation}
    \int\frac{\dd^d\ell}{(2\pi)^d}\frac{1}{\ell^2-m^2}\frac{1}{\nm\cdot \ell}
\end{equation}
are non-zero, even for a single-leg attachment.
Consequently, the soft triangle no longer vanishes, and single-leg topologies must be considered. The mass regulates the infrared limit of this integral, while the UV produces the usual pole in $\varepsilon$ when using dimensional regularization. This UV divergence cancels the IR divergence from the collinear loops. In total, we end up with an IR-finite contribution.
In addition, for diagrams with a soft loop connecting two legs, the soft emission no longer needs to be directly connected to the soft loop to obtain a non-vanishing result (even though these topologies remain power-suppressed in the current analysis).
Computing the systematic one-loop corrections to the massive ``soft theorem'' requires computing all these additional topologies, which we leave for future work. 

Although we do not compute the explicit form of the one-loop contributions for the massive theory, we have shown using the power-counting argument that they are subleading and (in contrast to the IR-divergent massless theory) IR finite. Thus, we can conclude that the factorization of the soft factor in \cref{eq:Vsoftfactor} receives no loop corrections for potentials satisfying the assigned power counting. 

This analysis could be extended to the power counting typical for SCET$_{\mathrm{II}}$, where $m^2\sim \lambda^2 Q^2$ and $g_3\sim \lambda Q$. In this case, the loop corrections from the potential would no longer be power suppressed, and the loop corrections considered here would be genuine corrections to the leading-power soft factorization. We leave for future work the analysis of loop corrections for this setup.

\section{Conclusions}
\label{sec:Conc}
In this paper, we presented geoSCET. This is an EFT that describes the soft and collinear sectors of general scalar field theories, expressed in a way that makes the underlying field-space geometry manifest. We showed that there is an emergent geometry in the soft sector, which has a close analogy with the emergent gauge invariance of the soft sector in QCD. We then put geoSCET to use by demonstrating the origin and extending the validity of the geometric soft theorems using this new construction. We prove the validity of the geometric soft theorems to all orders in perturbation theory in the case where there is no potential, including the extension to multiple soft limits. We provide arguments for the existence of the universality of leading terms for theories with a non-trivial potential. For theories with non-zero masses and cubic couplings on the order of the soft scale, the soft theorem turns into a factorization theorem. In this case, loop corrections are subleading, and the leading-order factorization is robust against loop corrections. Massless theories with cubic couplings, by contrast, are IR divergent and therefore theoretically less well-behaved.

This work provides a jumping-off point for many future directions. One could use this framework to explicitly derive the potential-dependent terms in the soft theorems when assigning different power counting to the masses and cubic couplings. It would be exciting to incorporate fermions, where the soft theorems of~\cite{Derda:2024jvo} could potentially be proven to be tree-level exact.
We anticipate that one could make progress in this direction by combining the approach of expressing field-space geometry in superspace~\cite{Finn:2019aip, Finn:2020nvn, Gattus:2023gep, Gattus:2024ird, Lee:2024xqa, Craig:2026igj} with the framework of collinear superspace~\cite{Cohen:2016jzp, Cohen:2016dcl, Cohen:2018qvn, Cohen:2019gsc}. It would additionally be interesting to incorporate higher-dimensional operators involving gauge bosons to directly explore the relationship between geoSCET and gauge invariance that was introduced in this paper. Furthermore, soft theorems in modern EFTs such as SCET and HQET have been linked to higher-form symmetries~\cite{Berean-Dutcher:2025ohp,Tizzano:2026rgr}. The exactness of the soft theorems hints at a possible underlying connection to higher-form symmetries. We leave for future study the investigation of the connection between the geometric soft theorem and higher-form symmetries.

Another exciting direction would be to develop the generalization of geoSCET within a framework that accommodates field redefinitions with derivatives. This has been an active area of development for the geometric scalar field, with many complementary proposals in the literature already~\cite{Cohen:2022uuw, Cheung:2022vnd, Cohen:2023ekv, Alminawi:2023qtf, Cohen:2024bml, Lee:2024xqa, Cohen:2025prs, Alminawi:2025pwg, Alminawi:2025zij, Delgado:2026zpz, Craig:2026igj}. Currently, all of these approaches are restricted to tree level. One hope is that by generalizing geoSCET to accommodate derivative field redefinitions, one might potentially be able to write down an extended form of the geometric soft theorem that could in turn provide insight into the loop-level generalized geometry. 

Finally, given our results for the $N$-point soft theorem above, it must be possible to connect the framework of geoSCET to the program of asymptotic symmetries \cite{Cheung:2016iub, Campiglia:2017dpg,Campiglia:2017xkp,Henneaux:2018mgn,Biswas:2022lsj} and celestial holography (see~\cite{Strominger:2017zoo,Pasterski:2021raf,Raclariu:2021zjz,Donnay:2023mrd} for reviews) as applied to scalar field theories, especially work exploring the non-linear sigma model~\cite{Melton:2021kkz,Kapec:2022axw,Kapec:2022hih,Kampf:2023elx}. There are hints of the connection between the physics on the celestial sphere and SCET for gauge theories~\cite{Larkoski:2014bxa,Krishna:2023ukw}.
We therefore expect it will prove to be fruitful to explore the analog for scalar field theories and geometric non-linear sigma models. Making these connections precise will be an exciting application of the geoSCET framework introduced here.

\subsection*{Acknowledgments}

We thank Martin Beneke, Dominik Schwienbacher, Michel Stillger, Luigi Tizzano, and Kathryn Zurek for useful discussions.
TC is supported by the U.S.~Department of Energy Grant No.~DE-SC0011640.
AH is supported by the Research Council of Norway under the FRIPRO Young Research Talent grant No.~357307.

\changelocaltocdepth{1}
\appendix
\section*{Appendices}
\addcontentsline{toc}{section}{Appendices}

\section{Details for the geometric Lagrangian construction}
\label{sec:App:DetailsLagr}
\subsection{Conventions and geometric objects}
\label{sec:App:Geometry}

Geometrically, the scalar field can be viewed as a smooth map
\begin{align}
    \phi\colon\, \mathcal{M} &\to \mathcal{N}\,,\\
    x &\mapsto \phi^I(x)\nn
\end{align}
assigning a coordinate $x^\mu$ from Minkowski space $\mathcal{M}$ to a coordinate $\phi^I$ on the manifold of field configurations $\mathcal{N}$, where $I$ denotes the index in the flavor basis.
The geometry of the field-space manifold $\mathcal{N}$ is described by a Riemannian metric tensor defined on the tangent bundle $T\mathcal{N}$ as
\begin{equation}
    g\colon T\mathcal{N}\times T\mathcal{N}\to \mathds{R}\,.
\end{equation}
In turn, this metric induces the usual set of geometric objects, such as the Levi-Civita connection $\nabla$, which acts in local coordinates via
\begin{subequations}
    \begin{align}
        \nabla_I A_J &= \partial_I A_J - \Gamma^L_{IJ}A_L\,,\\
        \nabla_I B^J &= \partial_I B^J + \Gamma^J_{IL} B^L\,,
    \end{align}
\end{subequations}
where $A_J$ ($B^J$) is a covariant (contravariant) vector field.
Here, the Christoffel symbol is defined as
\begin{equation}
     \Gamma^I_{KL} = \frac{1}{2}g^{IM}(\partial_K g_{ML} + \partial_L g_{MK} - \partial_M g_{KL})\,,
\end{equation}
and we define the version with lowered indices as
\begin{equation}
    \Gamma_{IKL} = g_{IM}\Gamma^M_{KL}\,.
\end{equation}
The curvature is encoded in the Riemann tensor as
\begin{equation}
    [\nabla_I,\nabla_J] B^K = R\indices{^K_{L IJ}}B^L\,,
\end{equation}
from which one obtains
\begin{equation}
    R\indices{^K_{LIJ}} = \partial^{\vphantom{K}}_I \Gamma^K_{LJ} - \partial^{\vphantom{K}}_J \Gamma^K_{LI} + \Gamma^K_{IM}\Gamma^M_{LJ} - \Gamma^K_{JM}\Gamma^M_{LI}\,.
\end{equation}

The map $\phi\colon\mathcal{M}\to\mathcal{N}$ induces the pullback, which can be used to transfer smooth functions $V\colon \mathcal{N}\to\mathds{R}$ back to $\mathcal{M}$ via
\begin{align}
    \phi^*V\colon\, \mathcal{M} &\to \mathds{R}\,,\\
    x &\mapsto V(\phi(x))\nn
\end{align}
The pullback can also be extended to vector bundles $E$ such as the tangent bundle $T\mathcal{N}$ and defines a corresponding pullback bundle $\phi^*E$ on $\mathcal{M}$.

One can define the Lagrangian as a scalar function on $\mathcal{M}$ as
\begin{equation}
    \mathcal{L} = \frac{1}{2}g_{IJ}(\phi)\partial_\mu\phi^I\partial^\mu\phi^J - V(\phi)\,,
\end{equation}
where the $x$-dependence is implicit, and the geometric objects are the pulled-back versions.
In coordinates, one then has, e.g.,
\begin{equation}
    (\phi^*g)(\partial_\mu,\partial_\nu)_x = g_{IJ}(\phi(x))\partial_\mu\phi^I(x)\partial_\nu\phi^J(x)\,.
\end{equation}

\subsection{Background-field expansion}
\label{sec:app:BGExpansion}

Naively, one would like to split
\begin{equation}
    \phi(x) = \varphi(x)+\pi(x)\,,
\end{equation}
but this split is ill-defined geometrically
since $\varphi^I$ and $\pi^I$ are points on the manifold, and the addition of points is not a sensible procedure.
In practice, this would manifest through terms breaking the field-redefinition invariance.
The covariant solution is the exponential map~\cite{Howe:1986vm}. 
Since one cannot add the two coordinates, one makes use of the fact that (in a local neighborhood) the points $\varphi^I$ and $\phi^I$ are connected by a unique geodesic, which we denote by $\bm{\Phi}^I(s)$.
This interpolating field satisfies the geodesic equation
\begin{equation}
    \frac{\text{d}^2\bm{\Phi}^I}{\text{d}s^2} + \Gamma^I_{JK}(\bm{\Phi}(s))\frac{\text{d}\bm{\Phi}^J}{\text{d}s}\frac{\text{d}\bm{\Phi}^K}{\text{d}s} = 0
\end{equation}
with boundary conditions
\begin{equation}
    \bm{\Phi}^I(0) = \varphi^I\,,\qquad 
    \frac{\text{d}\bm{\Phi}^I}{\text{d}s}\biggr\rvert_{s=0}=\xi^I\,,\qquad
    \bm{\Phi}^I(1) = \phi^I\,.
\end{equation}
Here $\varphi^I$ is the base-point, and $\xi^I$ the 
tangent vector at $\varphi^I$ pointing in the direction of the geodesic, both of which have a clear geometric meaning.
One has to be careful with notation: while $\varphi^I$ is a coordinate on the manifold $\mathcal{N}$, the perturbation $\xi^I\in T_{\varphi}\s\mathcal{N}$ is a tangent vector. 
The notation is the same as is used in Minkowski space, where e.g.\ $x^\mu$ is a coordinate, while $A^\mu$ is a component of a vector field.
With the pullback, this can again be carried over onto $\mathcal{M}$, and the fluctuation field $\xi^I(x)$ is then a section of the pullback tangent bundle $\phi^*T\mathcal{N}$, i.e., for each coordinate $x^\mu\in\mathcal{M}$ it assigns a tangent vector $\xi^I\in T_{\varphi(x)}\mathcal{N}$.
The proper, covariant background-field decomposition then reads~\cite{Howe:1986vm}
\begin{equation}
    \phi^I = \exp_\varphi[\xi]^I\,,
\end{equation}
where $\exp$ is the standard exponential map.
By solving the geodesic equation perturbatively, one finds $\phi^I = \varphi^I + \pi^I$ with~\cite{Howe:1986vm}
\begin{equation}
\label{eq:FullExpDecomp}
    \pi^I = \xi^I + \chi^I(\varphi,\xi)\,,
    \qquad\text{and}\qquad
    \chi^I = -\sum_{n=2}^{\infty}\frac{1}{n!}
    \Gamma^I_{J_1\dots J_n}\xi^{J_1}\dots\xi^{J_n}\,,
\end{equation}
and
\begin{equation}
    \Gamma^I_{J_1\dots J_n} = 
    \nabla_{(J_1}\dots\nabla_{J_{n-2}}\Gamma^I_{J_{n-1} J_{n})}(\varphi)\,,
\end{equation}
where round brackets denote symmetrization of the indices.

We now present some details relating to the background-field Lagrangian.
These results are known in the literature, and we collect them here as a reference.
To keep the notation concise, we use the same notation as in the main text:
whenever a geometric object without an explicit argument appears, it is evaluated at $\varphi$.

The metric tensor is then expanded as
\begin{equation}
    g_{IJ}(\phi) = g_{IJ} + \xi^K\partial_Kg_{IJ} + \frac{1}{2}\xi^K \xi^L \partial_K\partial_L g_{IJ} - \frac{1}{2}\Gamma^M_{KL}\xi^K\xi^L \partial_M g_{IJ} + \mathcal{O}(\xi^3)\,,
\end{equation}
where the last term arises from the second-order term of the field redefinition,
while for the differential, one finds
\begin{align}
    \partial_\mu\phi^I &= \partial_\mu\varphi^I + \partial_\mu\xi^I - \frac{1}{2}\partial_\mu\bigl[
    \Gamma^I_{JK}\xi^J\xi^K
    \bigr]\nn\\
    &=\partial_\mu\varphi + \partial_\mu\xi^I - \Gamma^I_{JK}\xi^J \partial_\mu\xi^K - \frac{1}{2}\partial_L\Gamma^I_{JK} \partial_\mu\varphi^L\spac\xi^J\xi^K\,.
\end{align}
Inserting this into the kinetic Lagrangian, one finds the purely soft terms at $\mathcal{O}(\xi^0)$
\begin{equation}
    \mathcal{L}_{\xi^0} = \frac{1}{2} g_{IJ}\partial_\mu\varphi^I\partial^\mu\varphi^J - V(\varphi)\,.
\end{equation}
Since this expansion corresponds to a background field expansion with fluctuation $\xi$ and background field $\varphi$, the terms linear in the fluctuation vanish by the background-field equations of motion, in this case by the soft equations of motion. Explicitly, the terms linear in $\xi$ read (adding the potential)
\begin{align}
    \mathcal{L}_{\xi^1} &= g_{IJ}(\varphi)\spac \partial_\mu\varphi^I\spac \partial^\mu\xi^J +\frac 12 \xi^A[\partial_A g_{IJ}](\varphi)\partial_\mu\varphi^I\partial^\mu\varphi^J - \xi^I [\partial_I V](\varphi)\nn\\
    &= \xi^A\Bigl( - \partial^\mu(g_{IA} \partial_\mu\varphi^I) + \frac 12
    [\partial_A g_{IJ}]\partial_\mu\varphi^I \spac\partial^\mu\varphi^J - [\partial_A V]
    \Bigr)\nn\\
    &=\xi^A\Bigl(
    -g_{AI}\square\varphi^I - [\partial_J g_{AI}] \partial_\mu\varphi^I \partial^\mu\varphi^J + \Gamma_{IAJ}\partial_\mu\varphi^I\partial^\mu\varphi^J - [\partial_AV]
    \Bigr)\nn\\
    &=\xi^A\Bigl(-g_{AI}\square\varphi^I-\Gamma_{AJI}\partial_\mu\varphi^I \partial^\mu\varphi^J -[\partial_AV]\Bigr)\nn\\
    &= \xi^A \Bigl(-g_{AI}\nabla^2\varphi^I - [\partial_AV]\Bigr)\,,
\end{align}
and the term in the round brackets in the last line is precisely the equation of motion of $\varphi$. Therefore, one can drop the terms linear in $\xi$.
To obtain this result, we used metric compatibility to rewrite derivatives of the metric tensor in terms of Christoffel symbols
\begin{equation}
    \partial_K g_{IJ} = \Gamma^L_{KI}g_{LJ} + \Gamma^L_{KJ}g_{LI} = \Gamma_{JKI}+\Gamma_{IKJ}\,.
\end{equation}

The quadratic terms are slightly more complicated. One first has
\begin{align}
    \mathcal{L}_{\xi^2} &= \frac 12 g_{IJ}\partial_\mu\xi^I \partial^\mu\xi^J
    +g_{IJ}\partial_\mu\varphi^I\Bigl(-\Gamma^J_{KL}\xi^K \partial^\mu\xi^L - \frac{1}{2}\partial_M \Gamma^J_{KL}\partial^\mu\varphi^M \xi^K\xi^L\Bigr)\\
    &\quad
    +\xi^K \partial_K g_{IJ}\partial_\mu\varphi^I\partial^\mu\xi^J
    +\frac{1}{4}\xi^K \xi^L \partial_K\partial_L g_{IJ} \partial_\mu\varphi^I\partial^\mu\varphi^J
    - \frac{1}{4}\Gamma^M_{KL}\xi^K \xi^L \partial_M g_{IJ}\partial_\mu\varphi^I\partial^\mu\varphi^J\,.\nn
\end{align}
Next, we apply metric compatibility. Throughout this derivation, we can freely symmetrize in $I\leftrightarrow J$ and $K\leftrightarrow L$ as these are contracted with $\partial_\mu\varphi^I \partial^\mu\varphi^J$ and $\xi^K \xi^L$.
In particular, one then has
\begin{align}
    \frac{1}{4}\partial_K\partial_L g_{IJ} = \frac{1}{2}\partial_K \Gamma_{JLI}
    = \frac{1}{2}g_{JM}\partial_K\Gamma^M_{LI} + \frac{1}{2}\Gamma^M_{LI}(\Gamma_{MKJ}+\Gamma_{JKM})\,.
\end{align}
And one obtains
\begin{align}
    \mathcal{L}_{\xi^{2}}&= \frac 12 g_{IJ}\partial_\mu\xi^I \partial^\mu\xi^J
    \textcolor{red}{-\Gamma_{IKJ}\partial_\mu\varphi^I\xi^K \partial^\mu\xi^J}
    -\frac{1}{2}g_{IJ}\partial_\mu\varphi^I\partial_M \Gamma^J_{KL}\partial^\mu\varphi^M \xi^K\xi^L\nn\\
    &\quad
    +\xi^K \xi^L \Bigl(\frac{1}{2}g_{JM}\partial_K\Gamma^M_{LI} + \frac{1}{2}\Gamma^M_{LI}(\textcolor{teal}{\Gamma_{MKJ}}+\Gamma_{JKM})\Bigr) \partial_\mu\varphi^I\partial^\mu\varphi^J\nn\\
    &\quad
    +\xi^K (\textcolor{red}{\Gamma_{IKJ}}+\textcolor{teal}{\Gamma_{JKI}})\partial_\mu\varphi^I\partial^\mu\xi^J
    - \frac{1}{2}\Gamma^M_{KL}\xi^K \xi^L\Gamma_{IMJ}\partial_\mu\varphi^I\partial^\mu\varphi^J\nn\\
    &= \frac 12 g_{IJ}\partial_\mu\xi^I \partial^\mu\xi^J + \textcolor{teal}{\xi^K \Gamma_{JKI}\partial_\mu\varphi^I\partial^\mu\xi^J} + \textcolor{teal}{\frac{1}{2}\Gamma^M_{LI}\Gamma_{MKJ} \xi^K\xi^L\partial_\mu\varphi^I\partial^\mu\varphi^J}\nn\\
    &\quad+\frac{1}{2}\xi^K\xi^L\partial_\mu\varphi^I\partial^\mu\varphi^J\Bigl(
    g_{JM}\partial_K\Gamma^M_{LI} - g_{IM}\partial_J \Gamma^M_{KL}+\Gamma^M_{LI}\Gamma_{JKM} - \Gamma^M_{KL}\Gamma_{IMJ}
    \Bigr)\nn\\
    &= \frac{1}{2}g_{IJ} \Bigl(\partial_\mu\xi^I + \textcolor{teal}{\Gamma^I_{KL}\partial_\mu\varphi^K \xi^L}\Bigr)
    \Bigl(\partial^\mu\xi^J + \textcolor{teal}{\Gamma^J_{MN}\partial^\mu\varphi^M\xi^N}\Bigr) + \frac{1}{2}\xi^K\xi^L\partial_\mu\varphi^I\partial^\mu\varphi^J R_{IKLJ}\nn\\
    &= \frac{1}{2}g_{IJ}D_\mu\xi^I D^\mu \xi^J - \frac{1}{2}R_{IKJL}\partial_\mu\varphi^K\partial^\mu\varphi^L \xi^I\xi^J\,.
\end{align}
To get from the second to the third equality, first note that the terms in the round bracket combine into the Riemann tensor after using the symmetry in $(K,L)$ and $(I,J)$.
In addition, the \textcolor{red}{red} terms in the first equality cancel, while the \textcolor{teal}{teal} ones are used to construct the \emph{pullback covariant derivative}
\begin{equation}
    D_\mu\xi^I \equiv \partial_\mu\xi^I + \Gamma^I_{KL}\partial_\mu\varphi^K \xi^L\,.
\end{equation}
This derivative is required to ensure a covariant transformation of $\partial_\mu \xi^I$ under background-field diffeomorphisms, where one has
\begin{equation}
    \xi^I \to \frac{\partial\varphi^{\prime I}}{\partial\varphi^J}\xi^J \equiv J^I_{\phantom{I}J}\xi^J\,.
\end{equation}
For a generic transformation, the partial derivative is not covariant, as it generates an additional term
\begin{align}
    \partial_\mu\xi^I &\to J^I_{\phantom{I}J} \partial_\mu\xi^J + \xi^J (\partial_\mu J^I_{\phantom{I}J})\nn\\
    &= J^I_{\phantom{I}J}\partial_\mu\xi^J + [\partial_M J^I_{\phantom{I}J}] \partial_\mu\varphi^M \xi^J\,.\label{eq:PartialTransform}
\end{align}
Since the Christoffel symbol transforms as
\begin{equation}
    \Gamma^{I}_{JK} \to \Gamma^L_{MN} J^{I}_L (J^{-1})^M_{\phantom{M}J}(J^{-1})^N_{\phantom{M}K} + J^{I}_{\phantom{I}L} \partial_M(J^{-1})^L_{\phantom{M}K}(J^{-1})^M_{\phantom{M}J}\,,
\end{equation}
the second term in the pullback covariant derivative transforms as
\begin{align}
    \Gamma^{I}_{JK}\partial_\mu\varphi^J \xi^K &\to J^I_{\phantom{I}L} \Gamma^L_{JK} \partial_\mu\varphi^J \xi^K - (\partial_MJ^{I}_{\phantom{I}L} )\partial_\mu\varphi^M \xi^L\,,
\end{align}
which precisely cancels the additional term in~\cref{eq:PartialTransform} ensuring that $D_\mu\xi^I$ has a covariant transformation.

Geometrically, this object arises quite naturally: there is a connection $\nabla$ on the field-space manifold $\mathcal{N}$, and the tangent vector field $\xi$ has a covariant derivative $\nabla\xi$. The covariant derivative appearing in the Lagrangian is, however, an object that eats a vector field $\partial_\mu\in T_x\mathcal{M}$ and a pulled-back vector field $\xi^I \in \Gamma(\varphi^*T\mathcal{N}) $ and yields another (pulled-back) vector field.
Therefore, one can define it rigorously from the pullback of $\nabla$ on $T\mathcal{N}$, defining a map
\begin{equation}
    \varphi^*\nabla\colon \Gamma(\varphi^*T\mathcal{N})\to \Gamma(T^*\mathcal{M}\otimes \varphi^* T\mathcal{N})\,,
\end{equation}
which is an object that eats a vector field on $T\mathcal{M}$ (e.g., $\partial_\mu$) and a pulled-back vector field.
This pullback can then be defined locally by pushing forward the vector field, e.g.
\begin{equation}
    (\varphi^*\nabla)_V\xi = \nabla_{\text{d}\varphi(V)}\xi\,,
\end{equation}
and inserting $V=\partial_\mu$ immediately results in the above form of the covariant derivative.

\subsection{Multipole expansion and fixed-line gauge}
\label{sec:App:FLGauge}

In this appendix, we collect some useful results for the multipole expansion and fixed-line gauge used in both the QCD and geoSCET derivations. First, the identities~Eqs.~\eqref{eq:QCDFLIdentity} 
are generic and valid for an arbitrary connection.
The fixed-line gauge condition $(x-x_-)^\mu A_\mu(x)=0$ reads
\begin{equation}
    x_\perp^\mu A_{\mu}(x) + \frac{1}{2}\nm\cdot x\,\np\cdot A(x) = 0
\end{equation}
and thus corresponds to a partial Fock-Schwinger gauge while keeping $\nm\cdot A(x_-)$ unconstrained.
Therefore, the identities for $A_{s\perp}$ and $\np\cdot A_s$ are related to the standard Fock-Schwinger relation, which expresses the connection $A_\mu$ in terms of the field-strength tensor as
\begin{equation}\label{eq:FixedPointExpansion}
    A_\mu(x) = \int_0^1 \text{d}s\: s\s x^\nu F_{\nu\mu}(sx)\,.
\end{equation}
To prove this relation, write
\begin{align}
    A_\mu(x) &= \int_0^1 \text{d}s\: \frac{\text{d}}{\text{d}s} \bigl( s A_\mu(sx)\bigr)= \int_0^1 \text{d}s\: \bigg( A_\mu(sx) + sx^\nu \frac{\partial}{\partial (sx)^\nu} A_\mu(sx)\bigg)\label{eq:FPProof}\\
    &= \int_0^1 \text{d}s\: \bigg( A_\mu(sx) - sx^\nu F_{\mu\nu}(sx) + sx^\nu \frac{\partial}{\partial (sx)^\mu}A_\nu(sx) + sx^\nu \bigl[ A_\mu\,,A_\nu\bigr](sx)\bigg)\,.\nn
\end{align}
Applying integration by parts to the third term
\begin{equation}
    sx^\nu \frac{\partial}{\partial (sx)^\mu} A_\nu (sx) = \frac{\partial}{\partial(s x)^\mu} (sx^\nu A_\nu(sx)) - A_\mu(sx) = -A_\mu(sx)\, ,
\end{equation}
we see that this cancels the first term in the second line of \cref{eq:FPProof}, and the gauge condition $x^\mu A_\mu(x)=0$ eliminates the commutator term, resulting in~\cref{eq:FixedPointExpansion}.

The corresponding relation for $\nm\cdot A_s$ can be derived from
\begin{equation}
    \nm\Cdot A_s(x) - \nm\Cdot A_s(x_-) = \int_0^1 \text{d}s\: (x-x_-)^\mu \nm^\nu \partial_\mu A_\nu (y(s))\,.
\end{equation}
Next, complete the right-hand side to a field-strength tensor
\begin{align}
    (x-x_-)^\mu \nm^\nu \partial_\mu A_\nu (y(s)) &= (x-x_-)^\mu \nm^\nu F_{\mu\nu}(y(s))
    + (x-x_-)^\mu \nm^\nu \partial_\nu A_\mu (y(s))\nn\\[4pt]
    &\hspace{12pt}+ (x-x_-)^\mu \nm^\nu \bigl[ A_\mu\,, A_\nu\bigr] (y(s))\nn
    \\[4pt]
    &=(x-x_-)^\mu \nm^\nu F_{\mu\nu}(y(s))\,.
\end{align}
Here, the second and third terms vanish by the fixed-line gauge condition as
    \begin{equation}
         (x-x_-)^\mu A_\mu(y(s)) = \frac{1}{s}(y(s) - y_-(s))^\mu A_\mu(y(s)) = 0\,,
    \end{equation}
$(y(s))_- = x_-$, and $[\nm\cdot\partial (x-x_-)] = 0$.
Thus, one obtains the identity
\begin{equation}
    \nm\Cdot A_s(x) - \nm\Cdot A_s(x_-) = \int_0^1 \text{d}s\: (x-x_-)^\mu \nm^\nu F_{\mu\nu}(y(s))\,.
\end{equation}
As one can see, the derivations are completely generic, so they hold for both  QCD and geoSCET. In the latter, one simply replaces
\begin{equation}
    (A_\mu)\indices{^I_J}\to \Gamma^I_{JK}\spac\partial_\mu\varphi^K\,,\qquad\text{and}\qquad (F_{\mu\nu})\indices{^I_J} = R\indices{^I_{JKL}}\spac\partial_\mu \varphi^K \partial_\nu\varphi^L\,,
    \label{eq:AFtoGammaPhi}
\end{equation}
noting that $[D_\mu,D_\nu] = F_{\mu\nu}$ holds with these definitions.

\vspace{6pt}
\noindent\textbf{Parallel transport of soft tensors:}
The $\mathcal{R}$ Wilson line satisfies the parallel-transport equation along the
straight line $y(s)=x_-+s(x-x_-)$,
\begin{equation}
\label{eq:RTransportODE}
    \frac{\dd}{\dd s}\,\R{I}{A}\bigl(y(s),x_-\bigr)
    = -(x-x_-)^\mu \bigl(A_\mu(y(s))\bigr)\indices{^I_J}\,\R{J}{A}\bigl(y(s),x_-\bigr)\,,
\end{equation}
with $\mathcal{R}(x_-,x_-)=\mathbf{1}$.
Consider now an arbitrary field-space tensor $T_{I_1\dots I_n}(\varphi)$ and its
dressed counterpart along the transport path,
\begin{equation}
    \mathcal{T}_{A_1\dots A_n}(s) \equiv
    \R{I_1}{A_1}\dots\R{I_n}{A_n}\, T_{I_1\dots I_n}\bigl(\varphi(y(s))\bigr)\,.
\end{equation}
The derivative of $T$ produces
$(x-x_-)^\mu\partial_\mu\varphi^K\,\partial_K T$, while each instance of $\mathcal{R}$
contributes a Christoffel factor through~Eqs.~\eqref{eq:AFtoGammaPhi} and \eqref{eq:RTransportODE}. Together these
combine into the pullback covariant derivative, so that
\begin{equation}
    \frac{\dd}{\dd s}\,\mathcal{T}_{A_1\dots A_n}(s)
    = (x-x_-)^\mu\, \R{I_1}{A_1}\dots\R{I_n}{A_n}\,
      \bigl[D_\mu T\bigr]_{I_1\dots I_n}\bigl(\varphi(y(s))\bigr)\,,
\end{equation}
with
\begin{equation}
D_\mu T = \partial_\mu\varphi^K\,\nabla_K T\,.    
\end{equation}
Iterating and evaluating at $s=1$ yields the covariant multipole expansion
\begin{equation}
\label{eq:CovariantMultipole}
    \R{I_1}{A_1}\dots\R{I_n}{A_n}\, T_{I_1\dots I_n}\bigl(\varphi(x)\bigr)
    = \sum_{m=0}^{\infty}\frac{1}{m!}\,
      (x-x_-)^{\mu_1}\dots(x-x_-)^{\mu_m}\,
      \bigl[D_{\mu_1}\!\dots D_{\mu_m} T\bigr]_{A_1\dots A_n}(x_-)\,.
\end{equation}
This is the statement that dressing a soft tensor with $\mathcal{R}$ Wilson lines is
the same as parallel transporting it from $x$ to $x_-$, with the multipole tower
generated by covariant derivatives. Each additional order costs one power of
$\partial_\mu\varphi$ and is therefore power suppressed.

Applying this to the metric tensor and using metric compatibility
$\nabla_K g_{IJ}=0$, every term with $m\geq1$ in~\cref{eq:CovariantMultipole}
vanishes, and one obtains the exact identity
\begin{equation}
\label{eq:MetricTransport}
    g_{IJ}\bigl(\varphi(x)\bigr)\,\R{I}{A}(x,x_-)\,\R{J}{B}(x,x_-)
    = g_{AB}\bigl(\varphi(x_-)\bigr)\,.
\end{equation}
Parallel transport is thus an isometry, which is the geometric counterpart of the
unitarity relation $\mathcal{R}\spac\mathcal{R}^\dagger=\mathbf{1}$ satisfied by the
QCD Wilson line. This identity is what allows the dressed metric factors
in~\cref{eq:GeoLagrangianRInserted} to collapse onto the emergent background at
$x_-$, without generating a tower of subleading terms. The same argument applies
to any covariantly constant tensor, while for the Riemann tensor
the expansion~\cref{eq:CovariantMultipole} terminates only order by order in $\lambda$.

\vspace{6pt}
\noindent\textbf{Expansion of the potential:}
The potential terms are treated in the same way. Starting from the covariant
Lagrangian~\cref{eq:GeoCollLagrangianPreExp} and inserting the
redefinition~\cref{eq:GeoCollRedef}, the relevant structure is
$\R{I}{A}\R{J}{B}\bigl[\nabla_I\nabla_J V\bigr](\varphi(x))$, to which
\cref{eq:CovariantMultipole} applies directly with
$T_{IJ}=\nabla_I\nabla_J V$. Using
\begin{subequations}
\begin{align}
    D_\mu\bigl[\nabla_I\nabla_J V\bigr]
    &= \partial_\mu\varphi^K \bigl[\nabla_K\nabla_I\nabla_J V\bigr]\,,\\
    D_\mu D_\nu\bigl[\nabla_I\nabla_J V\bigr]
    &= \bigl[D_\mu\partial_\nu\varphi^K\bigr]\bigl[\nabla_K\nabla_I\nabla_J V\bigr]
    + \partial_\mu\varphi^L\partial_\nu\varphi^K\bigl[\nabla_L\nabla_K\nabla_I\nabla_J V\bigr]\,,
\end{align}
\end{subequations}
and decomposing
$(x-x_-)^\mu = x_\perp^\mu + \nm\Cdot x\,\tfrac{\np^\mu}{2}$
with $x_\perp\sim\lambda^{-1}$ and $\nm\Cdot x\sim1$, the first orders read
\begin{subequations}
\begin{align}
    \mathcal{L}_V^{(0)} &= -\frac{1}{2}\bigl[\nabla_I\nabla_J V\bigr]
        \hat{\xi}^I\hat{\xi}^J\,,\\[3pt]
    \mathcal{L}_V^{(1)} &= -\frac{1}{2}\bigl[x_\perp\Cdot\partial\varphi^K\bigr]
        \bigl[\nabla_K\nabla_I\nabla_J V\bigr]\hat{\xi}^I\hat{\xi}^J\,,\\[3pt]
    \mathcal{L}_V^{(2)} &= -\frac{1}{4}\bigl[\nm\Cdot x\,\np\Cdot\partial\varphi^K\bigr]
        \bigl[\nabla_K\nabla_I\nabla_J V\bigr]\hat{\xi}^I\hat{\xi}^J
        -\frac{1}{4}\,x_\perp^\mu x_\perp^\nu
        \bigl[\hat{D}_\mu\partial_\nu\varphi^K\bigr]
        \bigl[\nabla_K\nabla_I\nabla_J V\bigr]\hat{\xi}^I\hat{\xi}^J\,,
\end{align}
\end{subequations}
reproducing the terms quoted in~\cref{sec:GeoSCETLagrangian}. 
Here, we focused on terms that are at most linear in the soft field and dropped the rest.
As anticipated below~\cref{eq:CovariantMultipole}, each order in the
multipole expansion is accompanied by one further derivative of the soft field, so
that soft-collinear interactions mediated by the potential are systematically power
suppressed relative to the leading term.

\section{Soft limits, discontinuities, and SCET regions}
\label{sec:app:Discont}

In the amplitudes literature, the behavior of loop amplitudes in soft and collinear limits is obtained by computing the $(N+1)$-point amplitude and subsequently taking a kinematic limit.
This procedure is subtle because the limit does not commute with the dimensional regularization expansion in $\varepsilon$. The resulting non-analytic remainders are captured through \emph{discontinuity functions}~\cite{Bern:1995ix}.
The EFT approach reverses the logic: the expansion in the relevant kinematics is performed once and for all at the Lagrangian level, so that loop corrections are single-scale objects that are homogeneous in power counting.
The discontinuities encountered in the asymptotic expansion of the full-theory amplitude then naturally appear as soft and collinear loop corrections in the EFT.
In this appendix, we illustrate this correspondence explicitly using the simplest example in which a discontinuity function appears, and show that the discontinuity is nothing but the contribution of a single momentum region, reproduced in SCET by a low-energy matrix element.

\subsection{The two-mass triangle}

Consider the triangle integral
\begin{equation}
    I_3 = -i\mu^{2\varepsilon}\int\frac{\dd^d \ell}{\pi^{d/2}}\frac{1}{\ell^2\spac(\ell-p_1)^2\spac(\ell-p_1-p_2)^2}\,,
\end{equation}
with the two-mass kinematics
\begin{equation}
    p_1^2 = -Q^2\neq 0\,,\qquad p_2^2=-q^2\neq 0\,,\qquad (p_1+p_2)^2=0\,,
\end{equation}
so that one external leg is exactly light-like. 
The exact result is
\begin{equation}\label{eq:2mExact}
    I^{2m}_3 = 
    \frac{r_\Gamma}{Q^2-q^2}\,\frac{1}{\varepsilon^2}\,\biggl[\biggl(\frac{\mu^2}{Q^2}\biggr)^{\!\varepsilon} - \biggl(\frac{\mu^2}{q^2}\biggr)^{\!\varepsilon}\biggr]\,,\qquad\text{with}\qquad 
     r_\Gamma = 
    \frac{\Gamma(1+\varepsilon)\Gamma^2(1-\varepsilon)}{\Gamma(1-2\varepsilon)}\,.
\end{equation}
We are interested in the limit in which leg $p_2$ approaches the light-cone, $q^2\to 0$, which we implement through the scale hierarchy
\begin{equation}
    q^2 = \lambda^2 Q^2\,,\qquad\text{with}\qquad \lambda\ll 1\,.
\end{equation}
Expanding~\cref{eq:2mExact}, one obtains
\begin{equation}\label{eq:2mExpanded}
    I^{2m}_3 = \frac{r_\Gamma}{Q^2}\frac{1}{\varepsilon^2}\biggl[
    \biggl(\frac{\mu^2}{Q^2}\biggr)^{\!\varepsilon} - \biggl(\frac{\mu^2}{q^2}\biggr)^{\!\varepsilon}
    \biggr] + \mathcal{O}(\lambda^2)\,.
\end{equation}
The first term is the one-mass triangle, i.e., precisely the integral one would have obtained if the limit $q^2\to 0$ had been taken naively at the level of the integrand.
The second term is proportional to the discontinuity function~\cite{Bern:1995ix}
\begin{equation}\label{eq:discontinuity}
    d_1(-q^2) = \frac{1}{\varepsilon^2}\biggl(\frac{\mu^2}{q^2}\biggr)^{\!\varepsilon}\,.
\end{equation}

The origin of the name is worth emphasizing, since it is easily misunderstood: $d_1$ is not an absorptive part or a unitarity cut, but simply a term that is non-analytic in the vanishing invariant.
Keeping the $\varepsilon$-dependence exact, with $\varepsilon<0$, the factor $(q^2)^{-\varepsilon}$ vanishes smoothly as $q^2\to 0$, and there is no discontinuity at all.
Expanded in $\varepsilon$ first, however, one would get a factor $-1/\varepsilon\spac\ln q^2$, which diverges.
The discontinuity function is therefore precisely accounting for the failure of the limit $q^2\to 0$ and the expansion in $\varepsilon$ to commute.

\subsection{The asymptotic expansion from the method of regions}
The same expansion is obtained systematically using the method of regions~\cite{Beneke:1997zp}, which we perform here using the parametric representation. The two-mass triangle is
\begin{equation}
    I^{2m}_3 = \Gamma(1+\varepsilon)\mu^{2\varepsilon}\int\dd\alpha_1\spac\dd\alpha_2\spac\dd\alpha_3\: \delta\Bigl(1-\sum_i\alpha_i\Bigr)\,
    \frac{\mathcal{U}^{-1+2\varepsilon}}{\mathcal{F}^{1+\varepsilon}}\,,
\end{equation}
with the Symanzik polynomials
\begin{equation}\label{eq:TwoMassSymanzik}
    \mathcal{U}=\alpha_1+\alpha_2+\alpha_3\,,\qquad \text{and}\qquad \mathcal{F} = \alpha_2\bigl(\alpha_1\spac Q^2 + \alpha_3\spac q^2\bigr).
\end{equation}
Note that $\alpha_2$ factorizes because the leg $p_1+p_2$ is light-like.

The two scales $Q^2$ and $q^2$ appear in $\mathcal{F}$ in a single bracket, and the expansion is controlled by which of the terms dominates.
This yields two regions.
In general, identifying all regions is a non-trivial task and in particular impeded by so-called ``hidden regions'' (such as Glaubers), see e.g.,~\cite{Jantzen:2012mw,Heinrich:2021dbf,Gardi:2022khw,Beneke:2023wmt,Ma:2023hrt,Guan:2024hlf,Gardi:2024axt,Ma:2025emu,Ma:2026pjx,Chen:2026dnj}.

\vspace{6pt}
\noindent\textbf{Hard region:}
For $\alpha_i\sim 1$, the second term is suppressed in $\lambda$ and
\begin{equation}
    \mathcal{F}_{\mathrm{hard}} = \alpha_1\alpha_2\, Q^2\bigl(1 + \mathcal{O}(\lambda^2)\bigr)\,.
\end{equation}
This is the parametric representation of the one-mass triangle, and one finds
\begin{equation}
    I^{2m,\,\mathrm{hard}}_{3} = \frac{r_\Gamma}{Q^2}\, \frac{1}{\varepsilon^2}\biggl(\frac{\mu^2}{Q^2}\biggr)^{\!\varepsilon}\,.
\end{equation}

\vspace{6pt}
\noindent\textbf{Collinear region:}
The two terms in $\mathcal{F}$ balance for $\alpha_1 = (q^2/Q^2)\hat{\alpha}_1=\lambda^2\hat{\alpha}_1$ with $\hat{\alpha}_1\sim 1$, in which case
\begin{subequations}
    \begin{align}
        \mathcal{U} &= \alpha_2+\alpha_3 + \mathcal{O}(\lambda^2)\,,\\
        \mathcal{F} &= q^2\spac \alpha_2\bigl(\hat{\alpha}_1 + \alpha_3\bigr)\,,
    \end{align}
\end{subequations}
which now only depends on the single scale $q^2$.
As one also expands the $\delta$-function defining the simplex, the rescaled variable is integrated over $\hat{\alpha}_1\in(0,\infty)$, which is the characteristic signature of a region.
Combining the Jacobian with the overall power of $\mathcal{F}$,
\begin{equation}
    \lambda^2\spac (q^2)^{-1-\varepsilon} = \frac{(q^2)^{-\varepsilon}}{Q^2}\,,
\end{equation}
and performing the remaining integrations, one obtains
\begin{equation}
    I^{2m,\,\mathrm{coll}}_3 = -\frac{r_\Gamma}{Q^2}\, \frac{1}{\varepsilon^2}\biggl(\frac{\mu^2}{q^2}\biggr)^{\!\varepsilon}\,.
\end{equation}
The sum of the two regions reproduces~\cref{eq:2mExpanded}. 
Comparing with~\cref{eq:discontinuity}, we see
\begin{equation}
    I^{2m,\,\mathrm{coll}}_3 = -\frac{r_\Gamma}{Q^2}\spac d_1(-q^2)\,.
\end{equation}
We conclude that \emph{the discontinuity function is precisely the contribution of the collinear region}.
There is no third contributing region, as the scaling that retains only the second, $q^2$ term in $\mathcal{F}$ results in a scaleless $\alpha_1$ integral, which vanishes in dimensional regularization.
In particular, there is no region collinear to the leg $p_1+p_2$, since its invariant vanishes identically and therefore provides no scale.

\subsection{Interpretation in momentum space}
It is instructive to translate this into momentum space.
Shifting $\ell' = \ell-p_1$, the propagators read $(\ell'-p_2)^2$, $\ell'^2$, and $(\ell'+p_1)^2$.
Since $p_2$ is hard but has a small virtuality, it necessarily has collinear scaling $p_2\sim(1,\lambda^2,\lambda)Q$.
The two propagators adjacent to the vertex at which $p_2$ attaches, $\ell'^2$ and $(\ell'-p_2)^2$, can then go on shell simultaneously for loop momenta $\ell'$ collinear to $p_2$. This pinch is the collinear region.
In this configuration, the remaining propagator is expanded as
\begin{equation}
    (\ell'+p_1)^2 = \underbrace{\ell'^2}_{\sim\lambda^2 Q^2} + \underbrace{2\spac\ell'\Cdot p_1 + p_1^2}_{\sim Q^2} \to 2\spac\ell'\Cdot p_1 + p_1^2\,,
\end{equation}
which is linear in the loop momentum.
Note that $p_1^2\sim Q^2$ must be retained, since it is of the same order as $2\spac\ell'\cdot p_1 = \np\cdot \ell'\,\nm\cdot p_1 + \mathcal{O}(\lambda\spac Q^2)$.
In the parameter language, this is the term $\hat{\alpha}_1 Q^2$, which after rescaling no longer constitutes an independent scale.
The hard region, in contrast, corresponds to $\ell'\sim(1,1,1)Q$, in which case the expansion amounts to the naive one, setting $p_2$ to a light-like vector with $p_2^2=0$.

\subsection{Interpretation in SCET}

In the effective theory, the two regions are not obtained by expanding a computed integral, but are generated by distinct objects.
Writing the factorization schematically as
\begin{equation}
    I^{2m}_3 = C^{(1)}J^{(0)} + C^{(0)}J^{(1)} + \mathcal{O}(\lambda^2)\,,
\end{equation}
the one-loop matching coefficient $C^{(1)}$ is defined as the full-theory result minus its EFT counterpart, and therefore precisely captures the hard region, i.e., the one-mass triangle.
The collinear region is reproduced by a one-loop low-energy matrix element, here a jet function, computed with the SCET Feynman rules.
The discontinuity function is thus not an obstruction to factorization, but a genuine low-energy contribution.

The virtuality $q^2$ plays the role of an infrared regulator for the collinear sector. If $q^2=0$ exactly, the corresponding collinear integral would be scaleless, and the collinear contribution would be incorrectly attributed to the hard matching coefficient.
Keeping $q^2$ finite results in the collinear region being present on both sides of the matching relation, so that it is consistently subtracted.

Two features are worth emphasizing.
First, the ordering ambiguity that gives rise to the discontinuity function is built into the SCET construction; the Lagrangian is power expanded \emph{before} any loop integration is performed: soft and collinear emissions are attached to the non-radiative interaction through universal Lagrangian insertions, rather than extracted as a limit of a higher-point amplitude.
Second, in the SCET$_\mathrm{I}$ setting relevant here, the soft and collinear modes are separated in virtuality, and each region is individually regulated by dimensional regularization alone.
No rapidity regulator is required, and the assignment of each contribution to the hard or the low-energy sector is unambiguous.
This is the technical reason why the discontinuity functions, which appear as non-trivial structures in the amplitude approach, are automatically accounted for by the EFT.

\subsection{Three-mass triangle}

As a further example, we can also check the three-mass triangle, where one has~\cite{Bern:1995ix}
\begin{equation}\label{eq:ThreeMassDisc}
    I^{3m}_3 \xrightarrow[]{p_3^2\to 0} I^{2m}_3 - \frac{r_\Gamma}{|p_1^2-p_2^2|}\bigl(d_2(p_3^2; p_1^2,p_2^2) - d_2(p_3^2;p_2^2,p_1^2)\bigr)\,,
\end{equation}
with the discontinuity function
\begin{equation}\label{eq:discontinuityd2}
    d_2(p_3^2;p_1^2,p_2^2) = \frac{1}{2\varepsilon^2}(-p_3^2)^{-\varepsilon} - \frac{1}{2\varepsilon^2}\frac{(-p_3^2)^{-\varepsilon}(-p_1^2)^{-\varepsilon}}{(-p_2^2)^{-\varepsilon}} - \mathrm{Li}_2\Bigl(1-\frac{p_1^2}{p_2^2}\Bigr)\,.
\end{equation}
Note that the limit of $p_3^2\to 0$ is finite as $\varepsilon\to 0$ for the three-mass integral.
Again we parametrize the limit $p_3^2\to 0$ by taking $p_3\sim(1,\lambda^2,\lambda)Q$ but $p_1,p_2\sim Q$.

\vspace{6pt}
\noindent\textbf{Hard region:}
If all $\alpha_i\sim 1$, the $\mathcal{F}$ polynomial reduces to
\begin{equation}
    \mathcal{F} = -\alpha_1 \alpha_2\, p_1^2 - \alpha_2\alpha_3\, p_2^2\,,
\end{equation}
which is exactly the polynomial~\cref{eq:TwoMassSymanzik} of the two-mass triangle considered above.
We again see that the hard region is precisely the triangle with one mass set directly to 0:
\begin{equation}
I^{3m,\,\mathrm{hard}}_3(p_1^2,p_2^2,p_3^2\to0) = I_3^{2m}(p_1^2,p_2^2)\,.
\end{equation}

\vspace{6pt}
\noindent\textbf{Collinear region:}
There is a second region, where $\alpha_2\sim\lambda^2\hat{\alpha}_2$ and the polynomials are
\begin{equation}
    \mathcal{U} = \alpha_1+\alpha_3 + \mathcal{O}(\lambda^2)\,,\qquad \mathrm{and}\qquad \mathcal{F} = -\lambda^2 \hat{\alpha}_2(\alpha_1 p_1^2 + \alpha_3 p_2^2) - \alpha_1\alpha_3 p_3^2\,.
\end{equation}
We define $p_1^2 = -s_1$ and $p_2^2=-s_2$, and we perform the parameter integral in $\hat{\alpha}_2$ to find
\begin{equation}
    I_3^{3m,\,\mathrm{coll}} = \frac{\Gamma(1+\varepsilon)}{\varepsilon}\biggl(-\frac{p_3^2}{\mu^2}\biggr)^{-\varepsilon} \int_0^1 \dd \alpha_1\: \frac{\bigl(\alpha_1(1-\alpha_1)\bigr)^{-\varepsilon}}{\alpha_1s_1+(1-\alpha_1)s_2}\,.
\end{equation}
This integral is controlled by the two collinear scales of the endpoints $\alpha_1\to 0,1$
\begin{equation}
    \mu_{12}^2 = \frac{(-p_3^2)(-p_1^2)}{(-p_2^2)}\,,\qquad\text{and}\qquad \mu_{21}^2 = \frac{(-p_3^2)(-p_2^2)}{(-p_1^2)}\,.
\end{equation}
The left-over parameter integral has a closed form in terms of a $_2F_1$.
However, to make contact with the discontinuity function~\cref{eq:discontinuityd2} one can directly expand the integrand in $\varepsilon$ as
\begin{equation}
    J = \int_0^1 \dd x\: \frac{\bigl(x(1-x)\bigr)^{-\varepsilon}}{xs_1+(1-x)s_2} = J_0 + \varepsilon J_1 + \mathcal{O}(\varepsilon^2)\,,
\end{equation}
with
\begin{subequations}
    \begin{align}
        J_0 &= \int_0^1\dd x\: \frac{1}{xs_1 + (1-x)s_2} = \frac{1}{s_1-s_2}\ln\frac{s_1}{s_2}\,,\\[4pt]
        J_1 &= -\int_0^1\dd x\: \frac{\ln x + \ln(1-x)}{x s_1 + (1-x)s_2} = -\frac{1}{s_1 - s_2}\biggl(\mathrm{Li}_2\Bigl(1-\frac{s_1}{s_2}\Bigr) - \mathrm{Li}_2\Bigl(1-\frac{s_2}{s_1}\Bigr)\biggr)\,.
    \end{align}
\end{subequations}
The collinear region then provides
\begin{equation}
    I_3^{3m,\,\mathrm{coll}} = \frac{1}{s_1-s_2}\frac{\Gamma(1+\varepsilon)}{\varepsilon}\biggl(\frac{s_3}{\mu^2}\biggr)^{-\varepsilon}\biggl(
    \ln\frac{s_1}{s_2} - \varepsilon\,\mathrm{Li}_2\Bigl(1-\frac{s_1}{s_2}\Bigr) + \varepsilon\,\mathrm{Li}_2\Bigl(1-\frac{s_2}{s_1}\Bigr)
    \biggr)\,.
\end{equation}
The two regions now directly reproduce~\cref{eq:ThreeMassDisc}.
The hard region directly agrees with $I_{3}^{2m}$. For the collinear region, we pull out $r_\Gamma$ and the denominator to find
\begin{align}
    \frac{s_1-s_2}{r_\Gamma}I_3^{3m,\,\mathrm{coll}} &= \frac{1}{\varepsilon}\ln\frac{s_1}{s_2} -\ln\frac{s_1}{s_2}\ln\frac{s_3}{\mu^2} - \mathrm{Li_2}\Bigl(1-\frac{s_1}{s_2}\Bigr) + \mathrm{Li_2}\Bigl(1-\frac{s_2}{s_1}\Bigr) + \mathcal{O}(\varepsilon)\nn\\[4pt]
    &=d_2(-s_3/\mu^2;-s_1,-s_2) - d_2(-s_3/\mu^2;-s_2,-s_1)\,.
\end{align}
Again, we see that the discontinuity arises from the collinear region.

\section{All-order RPI relations for \texorpdfstring{$A$}{A}-type coefficients}
\label{app:RPItower}

In~\cref{sec:SoftTh}, we used the fact that the subleading non-radiative $N$-jet
operators carrying $i\partial_\perp$ insertions are fixed by RPI in terms of the
leading-power coefficient. Here we demonstrate that this holds for the entire
$A$-type tower, and give the closed form. The relation at first subleading power was obtained
in~\cite{Beneke:2019kgv}. We consider scalar collinear fields. For fields with
spin, RPI acts in addition on the building blocks, and the relations acquire
further terms.

In~\cref{sec:QCDNJet}, we discussed the derivative insertions available for building $A$-type currents from
$\chi_i$. Specifically, the operator $\nip\cdot\partial$ is not independent, as the currents
are non-local along $\nip$ and such insertions are absorbed into the dependence
of the coefficient on $\nip\cdot p_i$, while
$\nim\cdot\partial$ is redundant by the collinear equation of motion~\cite{Beneke:2017ztn}. 
The independent structures at $m^{\text{th}}$ order are therefore the $m$-fold
$\partial_\perp$ insertions, whose on-shell matrix elements are the monomials
$p_{i_1\perp}^{\mu_1}\dots p_{i_m\perp}^{\mu_m}$. As these are linearly
independent, the expansion\footnote{We assume all-outgoing conventions and use $i\partial_\perp \chi_i$ as subleading building blocks.}
\begin{equation}
    \cM = \sum_{m\geq0}\frac{(-1)^m}{m!}\sum_{i_1\dots i_m}
    C^{Am;\,\mu_1\dots\mu_m}_{i_1\dots i_m}\spac
    p_{i_1\perp\mu_1}\dots p_{i_m\perp\mu_m}
    \label{eq:RPIdefC}
\end{equation}
determines the $C^{Am}$ uniquely: they are the Taylor coefficients of the
on-shell amplitude in the transverse momenta, and no further independent
$A$-type coefficient exists.\footnote{Operators with additional collinear building blocks do not contribute to the
on-shell amplitude at tree level, and beyond tree level their contributions are
scaleless (see also~\cite{Beneke:2026ogs}). \Cref{eq:RPIdefC} is thus the matching condition to all orders in the
coupling.}

The amplitude can only depend on Lorentz-invariant combinations of the momenta,
$\cM= F(\{2p_i\cdot p_j\})$, and the leading-power coefficient corresponds to the
restriction to the collapsed configuration $p_i^\mu=p_{i-}^\mu = \nip\cdot p_i\,\nim^\mu/2$. 
Regarding this
restriction as a function of the $N$ light-like vectors $p_{i-}^\mu$,
\begin{equation}
    \mathcal{C}(p_{1-},\dots,p_{N-})\equiv F\bigl(\{2\spac p_{i-}\Cdot p_{j-}\}\bigr)
    = C^{A0}\,,
\end{equation}
the same function evaluated at the physical momenta gives the full amplitude,
\begin{equation}
    \cM(p_1,\dots,p_N)=\mathcal{C}(p_1,\dots,p_N)
    \label{eq:RPImaster}
\end{equation}
to all orders in the power expansion. Therefore, the
leading-power coefficient, viewed as a function of null vectors rather than of
invariants at fixed directions, already encodes the entire tower. 
To see that this is a consequence of RPI, we write
$p_i^\mu=\nip\cdot p_i\,\tilde n_i^\mu/2$ with
$\tilde n_i^\mu=\nim^\mu+z_i^\mu-\tfrac14 z_i^2\nip^\mu$ and
$z_i^\mu=2p_{i\perp}^\mu/(\nip\cdot p_i)$. This is consistent with the requirements $\tilde n_i^2=0$ and
$\nip\cdot\tilde n_i=2$, so $\nim\to\tilde n_i$ is an RPI-I
transformation, with the $z_i^2\nip$ term the non-linear completion required for
it to close.

We can expand~\cref{eq:RPImaster} around $p_i=p_{i-}$ using
\begin{equation}
p_i^\mu=p_{i-}^\mu+\delta p_i^\mu\,,
\qquad\text{with}\qquad
\delta p_i^\mu=p_{i\perp}^\mu-p_{i\perp}^{\s2}\nip^\mu/(2\spac\nip\Cdot p_i)\,,\qquad\text{and}\qquad \nip\Cdot\delta p_i=0\,,
\end{equation}
which gives
\begin{equation}
    \cM=\exp\left[\sum_{i=1}^N \delta p_i\Cdot
    \frac{\partial}{\partial p_{i-}}\right]\mathcal{C}\,,
    \qquad\text{with}\qquad
    \frac{\partial}{\partial p_{i-\mu}}
    =\sum_{k\neq i}2\spac p_{k-}^\mu\spac\frac{\partial}{\partial S_{ik}}\,,
    \label{eq:RPItower}
\end{equation}
where the derivative does not act on the prefactors in the exponent, and $S_{ij}=2p_{i-}\cdot p_{j-}$.
Comparing
with~\cref{eq:RPIdefC}, the first two members are
\begin{subequations}
\begin{align}
    C^{A1\,\mu}_i &= -g_{\perp_i}^{\mu\nu}\spac
    \frac{\partial\mathcal{C}}{\partial p_{i-}^\nu}
    = -\frac{1}{\nip\Cdot p_i}\sum_{k\neq i}\frac{2\spac n_k^{\mu\perp_i}}{\nim\Cdot n_k}\spac
    \frac{\partial C^{A0}}{\partial\ln S_{ik}}\,,
    \label{eq:RPICA1}\\[3pt]
    C^{A2\,\mu\nu}_{ij} &=
    g_{\perp_i}^{\mu\alpha} g_{\perp_j}^{\nu\beta}\spac
    \frac{\partial^2\mathcal{C}}{\partial p_{i-}^\alpha\partial p_{j-}^\beta}
    - \delta_{ij}\spac\frac{g_{\perp_i}^{\mu\nu}}{\nip\Cdot p_i}\spac
    \nip\Cdot\frac{\partial}{\partial p_{i-}}\spac\mathcal{C}\,,
\end{align}
\end{subequations}
where the projectors implement the contraction with sector-transverse momenta, i.e., $g_{\perp_i}$ is transverse to $\nim^\mu$ and $\nip^\mu$.
The $m=1$ result reproduces~\cref{eq:CA1Coefficient} which was obtained in~\cite{Beneke:2019kgv}.

\section{On the infinite soft theorem}
\label{app:InfiniteSoft}

The infinite soft theorem for tree-level, single soft emission was originally derived from the Ward identities of large gauge transformations, and subsequently from ordinary gauge invariance alone~\cite{Hamada:2018vrw,Li:2018gnc}.
The purpose of this appendix is to make the correspondence to SCET explicit. We reproduce the result of~\cite{Li:2018gnc} and show that the infinite soft theorem amounts to the statement that Lagrangian insertions are universal while explicit soft building blocks are not.\footnote{An independent derivation of this result within SCET, together with a systematic classification of the subleading soft building blocks, is in preparation~\cite{Beneke:2026abcd}.}

In scalar QED, one finds for the $\ell^\text{th}$ order term in the soft expansion~\cite{Li:2018gnc}
\begin{equation}
    \mathcal{A}^\mu_{n+1,(\ell)} = \sum_{i=1}^n \frac{1}{(\ell+1)!}\frac{e_i}{p_i\Cdot q} q_{\nu}\spac J_i^{\mu\nu}(q\Cdot \partial_i)^\ell \mathcal{A}_{n} + q_{\alpha_1}\dots q_{\alpha_\ell}A_\ell^{\mu\alpha_1\dots\alpha_\ell}\,,
    \label{eq:InfiniteSoftThm}
\end{equation}
where $\mathcal{A}^\mu_{n+1}$ ($\mathcal{A}^\mu_{n}$) is the (non-)radiative amplitude with polarization vector stripped off, and $A^{\mu\alpha_1\dots\alpha_\ell}$ is a tensor antisymmetric under the exchange of $\mu$ with any $\alpha_i$, encoding the process-dependence. It first arises at $\ell=1$, corresponding to $\mathcal{O}(\lambda^4)$ in the SCET power counting.
A generalization to gravity has also been obtained in~\cite{Hamada:2018vrw,Li:2018gnc}.

The existence of an all-order universal statement is an immediate consequence of the SCET operator basis: while soft building blocks are process-dependent and come with their own matching coefficient, the Lagrangian insertions are universal to all orders. Therefore, a partial universal soft theorem must exist that only results from the Lagrangian insertions on the external legs. In the following, we consider QCD; the Abelian result is obtained by setting $g_s t^a \to e_i$.
With the all-order Lagrangian from~\cite{Beneke:2002ni,Beneke:2021aip}, one finds for the single soft-emission terms of a gluon from scalar matter
\begin{align}
    \mathcal{L} &= \frac{1}{2} g_s\spac \np\cdot j \sum_{k=0}^\infty \frac{1}{(k+1)!}(x-x_-)^\mu (x-x_-)_{\rho_1}\dots (x-x_-)_{\rho_k}\nn\\
    &\hspace{12pt}\times\big[D^{\rho_1}, [D^{\rho_2},\dots, [D^{\rho_k} \nm^\nu F_{s\mu\nu}]\dots]\big] + \dots\,,
\end{align}
with the Noether current $j^{a}_\mu$ defined as in~\cite{Beneke:2021aip}
\begin{equation}
\begin{aligned}
    \np\Cdot j^a = i \hat \chi_c^\dagger t^a\np\Cdot\overset{\leftrightarrow}{\partial}\hat \chi_c\,,\qquad
    \nm \Cdot j^a = i \hat \chi_c^\dagger t^a\nm\Cdot \overset{\leftrightarrow}{\mathcal{D}}\hat \chi_c\,,\qquad
    j_{\mu_\perp}^a = i \hat \chi_c^\dagger t^a \overset{\leftrightarrow}{\mathcal{D}}_{c\mu_\perp} \hat \chi_c\,.
\end{aligned}
\end{equation}
Here,
\begin{equation}
    i\hat{\chi}_c^\dagger t^a \overset{\leftrightarrow}{\mathcal{D}}_{c\mu} \hat{\chi}_c = i\bigl(
    \hat{\chi}_c^\dagger t^a \mathcal{D}_{c\mu} \hat{\chi}_c  - [\mathcal{D}_{c\mu}\hat{\chi}_c]^\dagger t^a  \hat{\chi}_c 
    \bigr)\,,
\end{equation}
and
\begin{equation}
    \mathcal{D}_c^\mu = \hat{D}^\mu - ig_s \mathcal{A}_c^\mu\,.
\end{equation}
We introduce this notation as it is the most compact representation of the trilinear sector of the Lagrangian.

Following~\cite{Beneke:2021umj}, manipulating this Lagrangian as if it were inserted into the matrix element, one finds that the universal contraction and the first $(x-x_-)^\mu$ generate the angular momentum as for the first two terms, while the remaining position arguments reduce to the tower of $(q\cdot \partial)^\ell$ terms.

To see how the tower arises, consider the insertion of this Lagrangian into an $N$-jet operator, evaluated for a single soft emission of momentum $q$ and polarization $\varepsilon$.
First, note that for a single external soft gluon, one has in momentum space
\begin{equation}
    F_{s\mu\nu}\to -i\spac (q_\mu\varepsilon_\nu-q_\nu\varepsilon_\mu)\,,
\end{equation}
while each covariant derivative in the nested commutators contributes one power of the soft momentum, $D^\rho \to -i\spac q^\rho$.

Second, the explicit position arguments are Fourier transformed. As explained in~\cref{sec:SoftTh}, each factor of $(x-x_-)^\rho$ corresponds to a derivative of the momentum-conserving $\delta$ function, which upon integration acts as a derivative with respect to the internal collinear momentum $\tilde{p}$. Contracting with the soft momenta of the commutator, each pair then yields $q\cdot \partial_i$, and the $k^{\mathrm{th}}$ term of the Lagrangian produces $(q\cdot\partial_i)^k$.

Third, the remaining structure arranges into the angular momentum.
Using $\np\cdot j$ together with the eikonal propagator of the collinear line, and
$p_i^\mu = \np\cdot p_i\spac \nm^\mu/2$, the first $(x-x_-)^\mu\spac \nm^\nu F_{s\mu\nu}$ results in
\begin{equation}
    \frac{1}{p_i\Cdot q}
    \biggl(
    p_i\Cdot \varepsilon\, q\Cdot \partial_i - p_i\cdot q\, \varepsilon\cdot \partial_i
    \biggr)
    =
    \frac{\varepsilon_\mu\spac q_\nu}{p_i\Cdot q}\spac
    \bigl(
    p_i^\mu \spac \partial_i^\nu - p_i^\nu \spac \partial_i^\mu
    \bigr)\equiv \frac{\varepsilon_\mu\spac q_\nu}{p_i\cdot q}\spac J_i^{\mu\nu}\,,
\end{equation}
which is precisely the structure appearing in~\cref{eq:InfiniteSoftThm}.
Note that for the scalar field, the orbital piece is the full angular momentum. For fermionic and vector matter, the spin piece arranges naturally as explained in detail in~\cite{Beneke:2021umj}.
Summing over the external legs with Lagrangian insertions then reproduces the sum over $i$.
Note that the leading eikonal emission is not part of this tower. It instead arises separately from $\nm\cdot A_s$ inside the covariant derivative of $\mathcal{L}^{(0)}$.

Finally, note that the combinatorial factors match. The $k^{\mathrm{th}}$ term of the Lagrangian carries $1/(k+1)!$, which corresponds precisely to the $1/(\ell+1)!$ accompanying $(q\cdot \partial_i)^\ell$ in the soft theorem.
The agreement of these factors is a non-trivial check that the universal part of the infinite soft theorem is indeed captured completely by Lagrangian insertions.

The process dependence arises from explicit soft building blocks in the $N$-jet operator, i.e., terms of the form
\begin{equation}
    \mathcal{J} = \int[\text{d}t]_N C(t_1,\dots,t_N) J_s^{(\ell)}(0) \chi_{1}(t_1)\dots\chi_{N}(t_N)
\end{equation}
with $J_s^{(\ell)} = D_{\alpha_1}\dots D_{\alpha_\ell} F_{s\mu\nu}$. For single soft emission, this simplifies to $\partial_{\alpha_1}\dots\partial_{\alpha_\ell}F_{s\mu\nu}$ and the antisymmetry is evident.
A systematic classification of the subleading soft building blocks in QCD is under investigation~\cite{Santana:2026,Beneke:2026abcd}.

From the SCET perspective, the infinite soft theorem is therefore not an independent consequence of gauge invariance, but a direct manifestation of the separation between universal soft interactions encoded in the SCET Lagrangian, and process-dependent soft operators appearing in the hard matching.

The same logic applies to QCD, gravity, and geoSCET, since neither the universality of Lagrangian insertions nor the process dependence of explicit soft building blocks made any reference to the gauge group.
The power counting differs in detail: in gauge theory, the first admissible soft building block is $F_{s\mu\nu}\sim\lambda^4$, in gravity it is $R_{\mu\nu\alpha\beta}\sim\lambda^6,$ and in geoSCET one has $\partial_\mu\varphi_s^I\sim\lambda^4$ and $F_{s\mu\nu}\sim\lambda^8$ appearing later, as discussed in~\cref{sec:geoNJet}.

There is, however, one structural feature in geoSCET with no gauge theory analogue: the hard matching coefficient depends on the soft field only through the invariant combination $C(v+\varphi_s)$, so that the shift invariance~\cref{eq:geoRPI} relates certain coefficients of arbitrarily high soft multiplicity to the non-radiative one, cf.~\cref{eq:RPI}.
Emissions from the hard vertex, which in gauge theory are genuinely process-dependent beyond subleading power, are therefore partially constrained in the geometric case.
We expect that combining this constraint with the Lagrangian tower derived above yields an all-order soft theorem for the geometric scalar field with a larger universal sector than its gauge-theory counterpart. We leave a systematic treatment to future work.

\phantomsection
\addcontentsline{toc}{section}{References}

\pdfbookmark[1]{References}{Refs}
\bibliography{references.bib}

\end{document}